\RequirePackage{lineno} 

\documentclass[twocolumn,showpacs,preprintnumbers,amsmath,amssymb,nofootinbib]{revtex4}
\usepackage{graphicx}% Include figure files
\usepackage{dcolumn}% Align table columns on decimal point
\usepackage{bm}% bold math
 
\usepackage{epstopdf}
\def\bea{\begin{eqnarray}}
\def\eea{\end{eqnarray}}

\def\pp{\mbox{$p$-$p$}}

\def\pa{\mbox{$p$-A}}

\def\pbpb{\mbox{Pb-Pb}}
\def\ppb{\mbox{$p$-Pb}}

\def\pn{\mbox{$p$-N}}
\def\aa{\mbox{A-A}}
\def\nn{\mbox{N-N}}

\def\pt{$p_t$}
\def\mt{$m_t$}
\def\yt{$y_t$}

\def\nch{$n_{ch}$}
\def\mmpt{$\bar p_t$}

\begin{document} 

\setlength{\pdfpagewidth}{8.5in}
\setlength{\pdfpageheight}{11in}

\setpagewiselinenumbers
\modulolinenumbers[5]
%\linenumbers

\preprint{version 1.1}

\title{Precision identified-hadron spectrum analysis for 5 TeV $\bf p$-$\bf Pb$ collisions -- Part I
}

\author{Thomas A.\ Trainor}\affiliation{University of Washington, Seattle, WA 98195}

%%%%%%%%%%%%%%%%%%%%%%%%%%%%%%%

\date{\today}

\begin{abstract}
The $p$-Pb collision system occupies a unique position regarding physical interpretation of high-energy particle data. More-peripheral $p$-Pb is indistinguishable from $p$-$p$ collision while more-central $p$-Pb overlaps an interval of Pb-Pb centrality wherein it has been claimed that quark-gluon plasma (QGP) formation is achieved. One basis for such claims is certain features and centrality trends of identified-hadron (PID) $p_t$ spectra, including similarities between $p$-Pb and Pb-Pb spectra. In order to verify or falsify such claims it is essential that PID spectra for $p$-Pb collisions (as a control experiment) be understood in terms of fundamental QCD principles. This article (Part I of a two-part report) presents application of a two-component (soft + hard) model (TCM) to PID spectra from 5 TeV $p$-Pb collisions. Certain issues remaining from a preliminary PID TCM analysis (e.g.\ proton detection inefficiency) are resolved. Coefficients for TCM model functions previously assumed independent of 
$p$-Pb centrality are obtained directly from spectra. Jet-related spectrum hard components are precisely isolated and their shape evolution with centrality determined relative to a fixed TCM as reference. In Part II the TCM is further elaborated to describe varying spectrum hard components (and therefore entire spectra) within data statistical uncertainties. The completed PID TCM is then used to investigate properties of spectrum and yield ratios (e.g.\ $p/\pi$) and $\bar p_t$ data.
\end{abstract}


\maketitle

%%%%%%%%%
\section{Introduction}

This article reports a follow-up to a previous analysis of identified-hadron (PID) \pt\ spectra from 5 TeV \ppb\ collisions~\cite{ppbpid}. In the previous study a newly-formulated two-component (soft+hard) model (TCM) of  PID hadron production near mid-rapidity was applied to \ppb\ PID spectrum data from Ref.~\cite{aliceppbpid}. (A subsequent article Ref.~\cite{aliceppbpidnew} reports \ppb\ PID spectra over a larger \pt\ acceptance but only for pions, charged kaons and protons.) Jet-related spectrum hard components for pions, kaons, protons and Lambdas were obtained from \ppb\ spectra. Predictions for PID data trends employing the resulting TCM model functions were also compared with measured PID ensemble-mean \mmpt\ data and with spectrum ratios (e.g.\ proton/pion and Lambda/kaon ratios).

One motivation for such a study is the tendency in recent years to interpret certain data features appearing in small collision systems (e.g.\ \pp, \pa) as evidence for ``collectivity'' (i.e.\ hydrodynamic flows) further interpreted to indicate formation of quark-gluon plasma (QGP) in such systems~\cite{ppbridge,ppcms,dusling}. To the extent that such interpretations are based on \pt\ spectrum features it is imperative to establish the fullest possible understanding of PID spectrum composition and its response to A-B centrality variation. That is the main goal of the present study.

While the general framework of the PID TCM developed in Ref.~\cite{ppbpid} appeared to be quite successful several issues remained unresolved. (a) While the TCM otherwise generally described PID spectra satisfactorily that was not the case for identified protons. The substantial systematic discrepancy suggested a problem with proton detection efficiency. (b) To simplify the initial model implementation two critical PID parameters were assumed independent of \ppb\ centrality. That assumption remains to be justified or abandoned. (c) Preliminary choices in formulating the sequence used to isolate spectrum hard components were invoked pending more complete information, also requiring a careful follow-up study.

The present analysis addresses the issues above as follows: (a) A proton inefficiency function is estimated based on variation of the low-\pt\ structure of proton spectra with centrality (justification explained in the text). (b) Modification of the pion soft-component model to accommodate a resonance contribution is explicitly described. (c) Centrality-dependent PID model parameters $z_{si}(n_s)$ and $z_{hi}(n_s)$ are obtained directly from spectrum data and examined for self consistency. (d) PID data spectrum hard components are obtained by a modified procedure that minimizes bias. The result is arguably the most accurate differential description of PID hadron spectra from \ppb\ collisions that can be formulated.

This study demonstrates that within the statistical limits of \ppb\ spectrum data the PID TCM is a {\em necessary and sufficient} data description. All aspects of spectrum data are described by a model including only longitudinal projectile-nucleon dissociation and transverse fragmentation to jets. There is no need for alternative description elements, and spectrum models that have been invoked in the past to support claims of collectivity are rejected on the basis of recent \pp\ spectrum analysis~\cite{tomnewppspec,tommodeltests}.

This article, denoted Part I, introduces the basic elements of a revised PID TCM applied to \ppb\ spectrum data, isolates minimum-bias spectrum hard components and reveals their shape variation with multiplicity \nch\ relative to a fixed-TCM reference. Follow-up Part II defines a variable PID TCM that accommodates hard-component shape evolution, describes comparisons of PID TCM results to \ppb\ spectrum and yield ratios and to ensemble \mmpt\ data in order to address competing interpretations of such data. As noted, the goal is to establish the fullest possible understanding of PID spectrum composition and its response to A-B centrality variation. 

%%%%%%%%%%%%%%%%
This article is arranged as follows:
Section~\ref{spectrumtcm} presents the basic structure of a TCM for 5 TeV \ppb\ PID spectra and introduces spectrum data from Ref.~\cite{aliceppbpid} that are the basis for the present study.
Section~\ref{prelim} resolves certain issues persisting from an earlier TCM analysis of the same \ppb\ PID spectrum data~\cite{ppbpid}.
Section~\ref{zxspectra} reports precise inference of centrality-dependent TCM parameters $z_{si}(n_s)$ and $z_{hi}(n_s)$ directly from PID spectra.
Section~\ref{zcentt} presents a comprehensive TCM description of the resulting centrality dependence of $z_{si}(n_s)$, $z_{hi}(n_s)$ and ratio parameter $\tilde z_i(n_s) \equiv z_{hi}(n_s)/z_{si}(n_s)$.
Section~\ref{newdetails} reports the results of a revised method to extract PID hard components from spectrum data. The new method gives access to subtle but essential aspects of jet-related hard-component evolution with \ppb\ centrality and hadron species.
Section~\ref{revisedspec} compares results from the present study with the TCM analysis reported in Ref.~\cite{ppbpid}.
Section~\ref{sys}  reviews systematic uncertainties.
Sections~\ref{disc} and~\ref{summ} present discussion and summary.

%%%%%%%%%%%%%%
\section{$\bf p$-$\bf Pb$ PID Spectrum $\bf TCM$} \label{spectrumtcm}

This section summarizes a previous TCM study of PID spectra from 5 TeV \ppb\ collisions reported in Ref.~\cite{ppbpid} and introduces PID spectrum data from Ref.~\cite{aliceppbpid} that are the subject of the present analysis. 

\subsection{p-Pb geometry inferred from non-PID data} \label{geom}

To best interpret \ppb\ PID \pt\ spectra collision centralities and geometry parameters should be estimated as accurately as possible. An analysis of \ppb\ centrality based on the Glauber model and including estimates of systematic biases for several methods was reported in Ref.~\cite{aliceglauber}. However, those results are strongly contradicted by a TCM study of \mmpt\ data for \ppb\ collisions reported in Ref.~\cite{tommpt}. The TCM geometry describes \mmpt\ data within their uncertainties whereas the Glauber geometry fails to do so. The large differences between Glauber and TCM estimates are explained in Refs.~\cite{tomglauber,tomexclude}.

Several aspects of the TCM require an estimate of quantity $\alpha$ in the relation $\bar \rho_{hNN} \approx \alpha(\sqrt{s_{NN}}) \bar \rho_{sNN}^2$ for the relevant collision energy. The value used in a previous study of 5 TeV \ppb\ collisions reported in Ref.~\cite{ppbpid} was 0.0113. However, recent results for spectrum data from 13 TeV \pp\ collisions reported in Ref.~\cite{tomnewppspec} suggest that the value at LHC energies should be increased by about 12\%. The value used for the present study is therefore 0.0127. The difference is within the stated uncertainty reported in Ref.~\cite{alicetomspec} in connection with its Fig.~16.

%%%%%%%%%%%%%%%%%%%%%%%%%%%%%%%%%%
\begin{table}[h]
	\caption{TCM fractional cross section $\sigma / \sigma_0$ (bin centers) and charge density $\bar \rho_0$, \nn\ soft component $\bar \rho_{sNN}$ and TCM hard/soft ratio $x(n_s)$ used for 5 TeV \ppb\ PID spectrum analysis~\cite{ppbpid}. $\sigma' / \sigma_0$ and $N_{bin}'$ values are from Table~2 of Ref.~\cite{aliceglauber}. Other parameter values are from Ref.~\cite{tomglauber} with $\alpha = 0.0113$.
	}
	\label{rppbdata}
	%%%%%%%%%%%%%%%%%%%%%%%%%%%%%%%
	\begin{center}
		\begin{tabular}{|c|c|c|c|c|c|c|c|c|} \hline
			%\multicolumn{8}{|c|}{\pp\ multiplicity classes} \\ \hline
			$n$ &   $\sigma' / \sigma_0$ & $N_{bin}'$  &  $\sigma / \sigma_0$     & $N_{bin}$  & $\nu$ & $\bar \rho_0$ & $\bar \rho_{sNN}$ & $x(n_s)$ \\ \hline
			1	   &      0.025  & 14.7  & 0.15   & 3.20   & 1.52 & 44.6 & 16.6  & 0.188 \\ \hline
			2	 &  0.075  & 13.0 & 0.24    & 2.59   & 1.43 & 35.9 &15.9  & 0.180 \\ \hline
			3	 &  0.15  & 11.7 & 0.37 & 2.16  &  1.37 & 30.0  & 15.2  & 0.172 \\ \hline
			4	 &  0.30 & 9.36 & 0.58  & 1.70   & 1.26  & 23.0  & 14.1  & 0.159  \\ \hline
			5	 &  0.50  & 6.42 &0.80    & 1.31   & 1.13 & 15.8 &   12.1 & 0.137  \\ \hline
			6	 &  0.70 & 3.81 & 0.95   & 1.07   & 1.03  & 9.7  &  8.7 & 0.098 \\ \hline
			7	 & 0.90  & 1.94 & 0.99  & 1.00  & 1.00  &  4.4  & 4.2 &0.047  \\ \hline
		\end{tabular}
	\end{center}
	%%%%%%%%%%%%%%%%%%%%%%%%%%%%%%%
\end{table}
%%%%%%%%%%%%%%%%%%%%%%%%%%%%%%

Table~\ref{rppbdata} presents TCM geometry parameters for 5 TeV \ppb\ collisions reported in Ref.~\cite{tomglauber}. Those geometry parameters, derived from \ppb\ \pt\ spectrum and \mmpt\ data for unidentified hadrons~\cite{tommpt}, are assumed to be valid also for each identified-hadron species and were used unchanged to process \ppb\ PID spectrum data in Ref.~\cite{ppbpid}. 

The $\bar \rho_0 = n_{ch} / \Delta \eta$ charge densities are inferred from $\eta$-density distributions in Fig.~16 of Ref.~\cite{aliceglauber} averaged over $|\eta_\text{lab}| < 0.5$. Those values agree with Table 1 of Ref.~\cite{aliceppbpid} within published uncertainties. Relations $N_{part} = N_{bin} + 1$ and $\nu = 2 N_{bin} / N_{part}$ involve the number of nucleon N participants and \nn\ binary collisions. $\bar \rho_{sNN}$ is the mean soft-component charge density per participant pair averaged over all pairs. $x = \bar \rho_{hNN}/\bar \rho_{sNN} \approx \alpha \bar \rho_{sNN}$ is the hard/soft density ratio. That approximation is based on TCM results from \pp\ \pt\ spectra~\cite{ppprd,ppquad} and is applied here also to \ppb\ collisions assuming they consist of linear superpositions of \pn\ collisions. Given that context the soft and hard components of \ppb\ charge density $\bar \rho_0$ are $\bar \rho_s = (N_{part}/2) \bar \rho_{sNN}$ and $\bar \rho_h = N_{bin} \bar \rho_{hNN}$.

Columns $\sigma' / \sigma_0$ and $N_{bin}'$ present the nominal centralities (bin centers) and \pn\ binary-collision numbers quoted by Ref.~\cite{aliceppbpid} in connection with measured charge densities $\bar \rho_0$ whereas column $\sigma / \sigma_0$ presents the values inferred in Ref.~\cite{tomglauber}. The remaining parameters in the table are a result of the latter analysis. The main finding of Ref.~\cite{tomglauber} is that \pn\ collisions within \pa\ collisions exhibit {\em exclusivity} -- the projectile proton interacts with only one nucleon {\em at a time}~\cite{tomexclude}. As a consequence, the number of participant nucleons is substantially less than what is predicted from a classical Glauber model: the \pa\ collisions are substantially less central and the actual mean \pn\ charge multiplicity is consequently greater to correspond with the measured $\bar \rho_0$ values. The result is a qualitative increase in predicted jet production based on relation $\bar \rho_{hNN} \approx \alpha \bar \rho_{sNN}^2$ inferred from \pp\ data.

The differences between centrality results from Ref.~\cite{aliceglauber} (from the Glauber model) and TCM results from Ref.~\cite{tomglauber} (from \mmpt\ data) are substantial. The difference can be illustrated in terms of impact parameters. The estimate $\sigma'/\sigma_0 \approx 0.025$ for most-central \ppb\ collisions quoted by Ref.~\cite{aliceppbpid} corresponds to impact-parameter ratio $b' / b_0 \approx  0.15$, where $b_0$ is approximately the Pb nuclear radius. For the TCM geometry reported in Ref.~\cite{tomglauber} the estimate $\sigma/\sigma_0 \approx 0.15$ translates to $b / b_0 \approx  0.40$. Those results suggest a factor 2.7 difference in mean path length of projectile protons through the target nucleus. But the presence of exclusivity in \pa\ collisions explains why $N_{bin}$ estimates in the two cases differ by factor 4.6.

The $N_{part}$ trend inferred via Glauber analysis implies that $\bar \rho_{0NN} \approx \bar \rho_{sNN} \approx 5$ (the NSD value at 5 TeV/c) for any \ppb\ centrality -- e.g.\ 44.6 / (15.7/2) = 5.68 for 0-5\% central \ppb. If that were the case then $\bar \rho_{hNN} \approx 0.0127 (5.5)^2 = 0.38$ and $\bar \rho_h = 14.7 \times 0.38 \approx 5.6$ is the predicted total hard component of $\bar \rho_0$ for central collisions. In contrast, the TCM geometry has $N_{part} \approx 4.2$ and $\bar \rho_{sNN} \approx 16.6$ for central collisions, implying $\bar \rho_{hNN} \approx 3.5$ and $\bar \rho_h = 3.2 \times 3.5 \approx 11.2$. Although $\bar \rho_h =$ 11.2 (TCM) is only twice 5.6 (Glauber) the difference in predicted \mmpt\ trends is dramatic. The key parameter for centrality evolution of $\bar p_t$ is hard/soft ratio $\bar \rho_h / \bar \rho_s = x \nu \approx \alpha \bar \rho_{sNN} \nu$~\cite{tommpt}. For \ppb\ collisions $\nu \in [1,2]$ for either model. For the Glauber model $\bar \rho_{sNN}$ is approximately constant whereas for the TMC $\bar \rho_{sNN}$ increases by factor 4 as in Table~\ref{rppbdata}. Thus, \mmpt\ trends are dramatically different as noted. In Fig.~11 (left) of Ref.~\cite{tommpt} the \mmpt\ trend for \ppb\ follows the trend for \pp\ within data uncertainties up to $\bar \rho_0 = 20$ implying that $N_{bin} \approx 1$ across that interval (the first four \ppb\ centrality bins), not increasing rapidly beyond 1.94 (for 80-100\% central) as in Ref.~\cite{aliceglauber}.

\subsection{$\bf p$-Pb PID  spectrum data} \label{alicedata}

The identified-hadron spectrum data adopted from Ref.~\cite{aliceppbpid} for the present analysis were produced by the ALICE collaboration at the LHC.  The event sample for charged hadrons is 12.5 million minimum-bias (MB) collisions and for neutral hadrons 25 million MB collisions. Collision events were divided into seven charge-multiplicity \nch\ or \ppb\ centrality classes based on yields in a VZERO-A (V0A) counter subtending $2.8 < \eta_{lab} < 5.1$ in the Pb direction.
Hadron species include charged pions $\pi^\pm$, charged kaons $K^\pm$, neutral kaons $K^0_\text{S}$, protons $p,~\bar p$ and Lambdas $\Lambda,~\bar \Lambda$. Spectra for charged vs neutral kaons and particles vs antiparticles are reported to be statistically equivalent.

\begin{figure}[h]
	\includegraphics[width=1.65in]{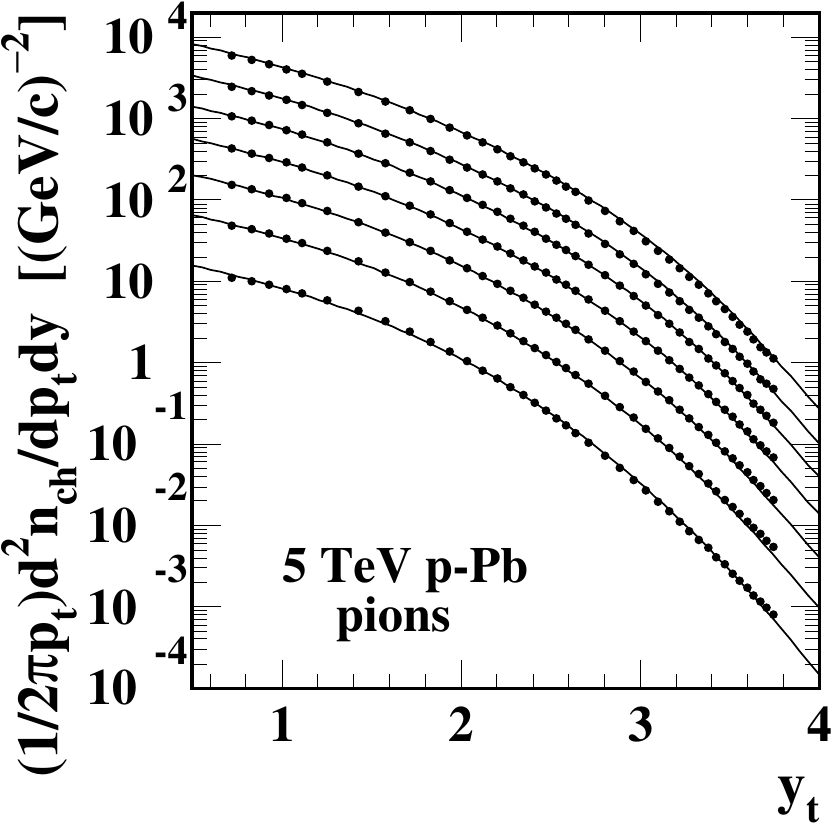}
	\includegraphics[width=1.65in]{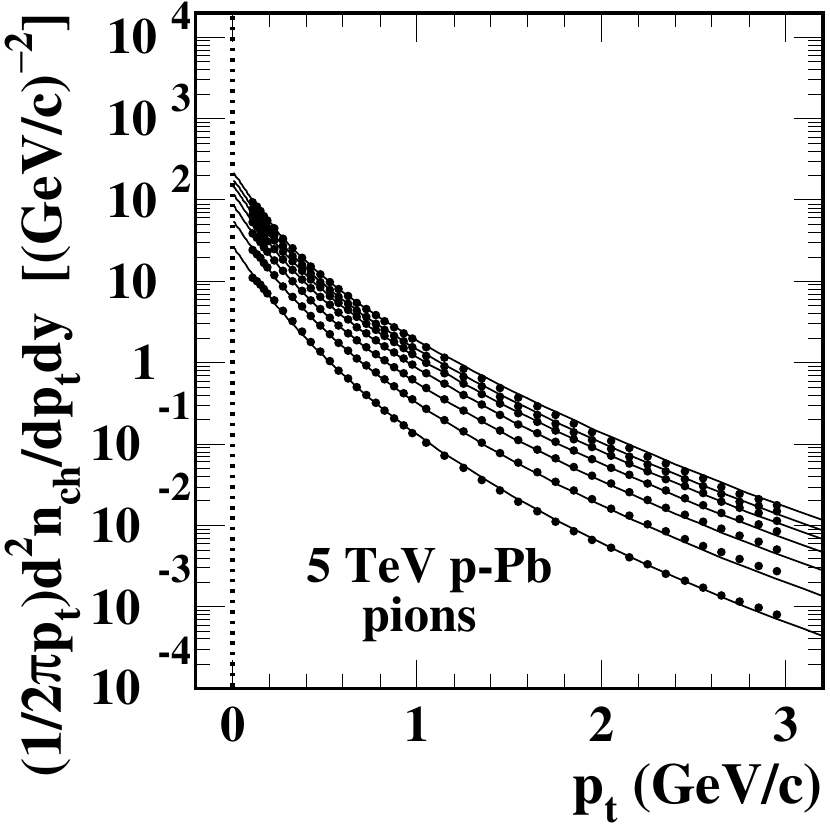}
	\put(-142,105) {\bf (a)}
	\put(-23,105) {\bf (b)}\\
	\includegraphics[width=1.65in]{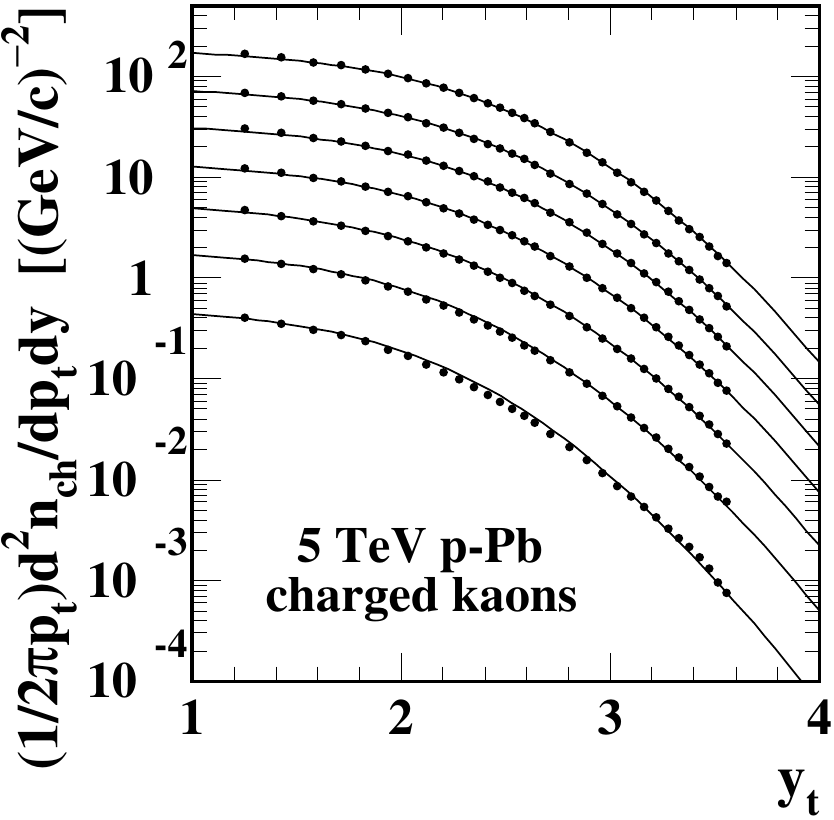}
	\includegraphics[width=1.65in]{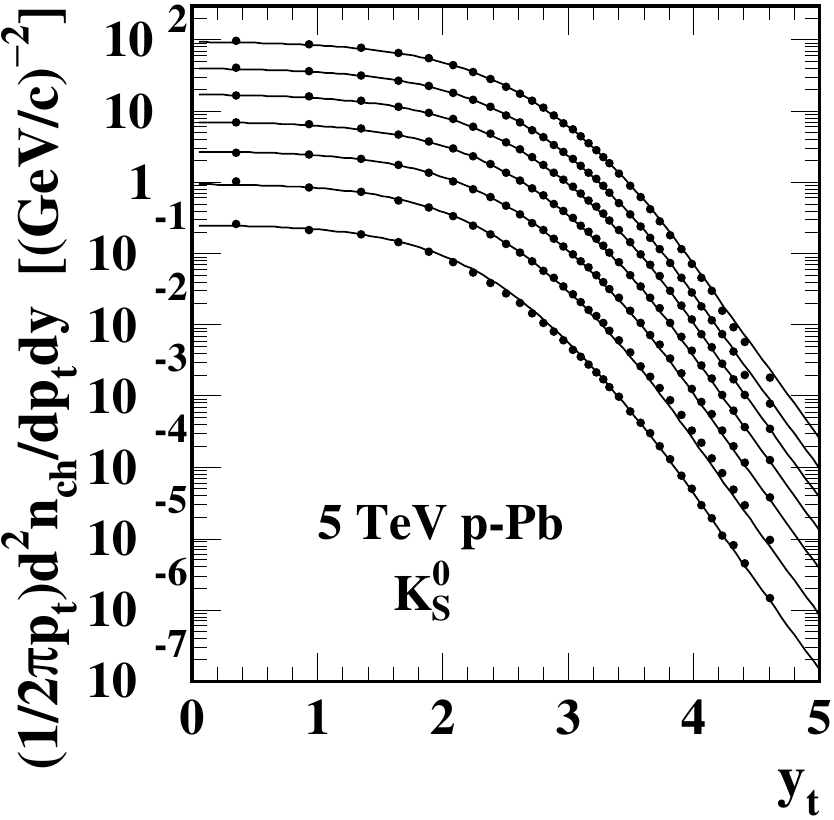}
	\put(-142,105) {\bf (c)}
	\put(-23,105) {\bf (d)}\\
	\includegraphics[width=1.65in]{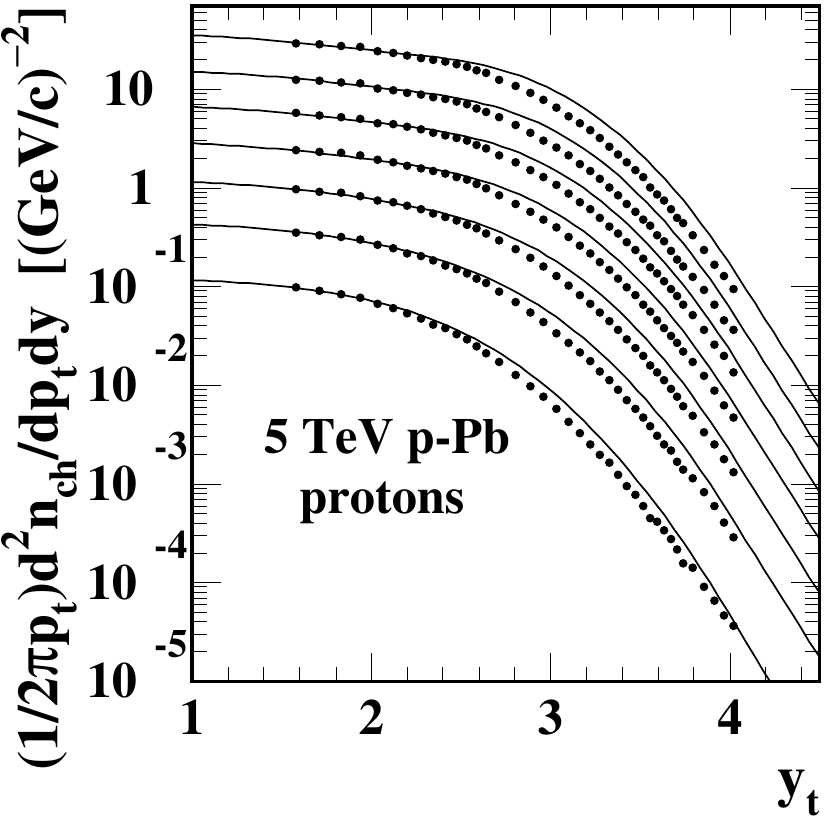}
	\includegraphics[width=1.65in]{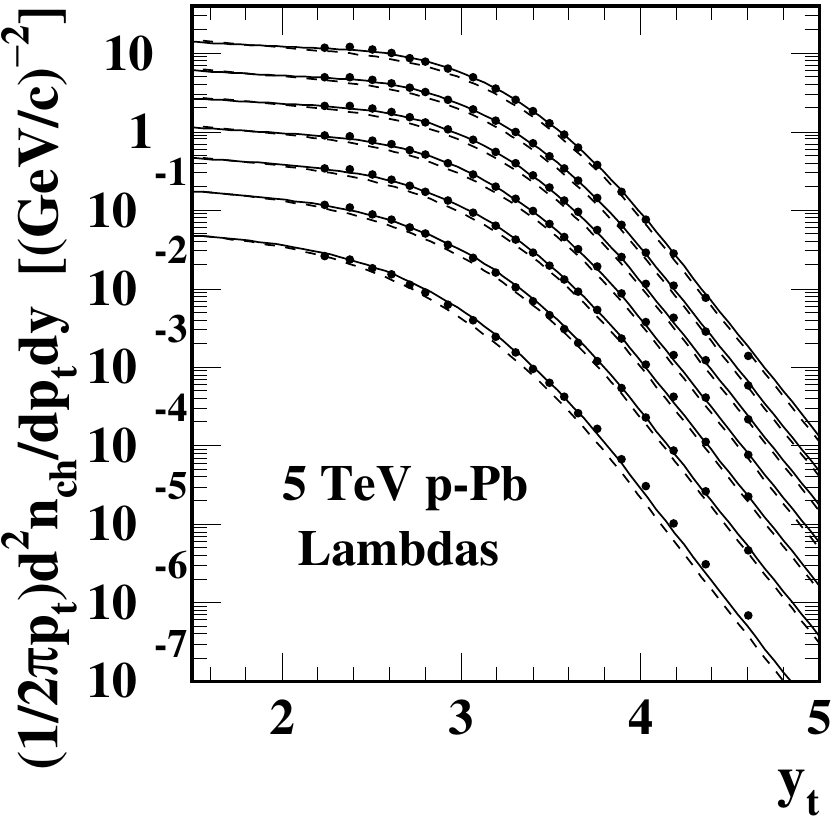}
	\put(-142,105) {\bf (e)}
	\put(-23,105) {\bf (f)}\\
	\caption{\label{piddata}
		\pt\ spectra for identified hadrons from 5 TeV \ppb~\cite{aliceppbpid} plotted vs transverse rapidity \yt\ for pions:
		(a) pions,
		(b) pions without multiplicative factors,
		(c) charged kaons,
		(d) neutral kaons,
		(e) uncorrected protons,
		(f) Lambdas.
		Solid curves represent the PID spectrum TCM from Sec.~\ref{pidspecc}. The dashed curves in (f) repeat the proton TCM curves in (e) divided by 2 for direct comparison. Pion \yt\ is the default for all hadrons.
	}%   alice600aay, aax
%	\\alice630aay, 640aay
%	\\ alice610aayno, 620aa
\end{figure}

Figure~\ref{piddata} shows PID spectrum data from Ref.~\cite{aliceppbpid} (points) in a plotting format vs logarithmic variable \yt. That format (effectively log-log with respect to \pt) provides detailed access to low-\pt\ structure (where most jet fragments appear) and clearly shows power-law trends at higher \pt. The curves are TCM parametrizations described in Sec.~\ref{newoldspec}.  The spectra for panels (a) and (c-f) have been scaled up by powers of 2 according to $2^{n-1}$ where $n \in [1,7]$ is the centrality class index and $n = 1$ is {\em least} central (following the usage in Ref.~\cite{aliceppbpid}). In all other instances in this paper $n=1$ denotes the {\em most}-central data as in Table~\ref{rppbdata}. Panel (b) shows pion spectra with no such scaling, the variation then due solely to the different \ppb\ centrality classes. It is notable that the largest charge-density increases correspond to the most peripheral \nch\ classes $n \in [1,3]$ where the actual \ppb\ centrality varies most slowly (see Table~\ref{rppbdata}). That trend is consistent with the \ppb\ centrality determination in Ref.~\cite{tomglauber}. Panel (b) also demonstrates that spectra plotted on linear \pt\ tend to obscure both spectrum details at lower \pt\ and approximate power-law trends at higher \pt.

$K_\text{S}^0$ data are approximately a factor 2 below $K^\pm$ data as expected, and the present study confirms that there is no statistically significant difference between spectrum shapes where they overlap. The baryon spectra in panels (e) and (f) are expected to correspond closely in shape, as reflected by the TCM trends. However, the proton data  (points) fall substantially ($\sim 40$\%) below the proton TCM expectation (curves) above $p_t \approx 0.6$ GeV/c ($y_t \approx 2.15$) (see Sec.~\ref{ineff}). The dashed curves in (f) are the proton TCM in (e) divided by 2 to illustrate the close agreement in TCM spectrum shapes. 

%%%%%%%%%%
%%%%%%%%%%%%

\subsection{p-Pb spectrum TCM for identified hadrons} \label{pidspecc}

What follows is a summary of the TCM analysis of \ppb\ PID spectra reported in Ref.~\cite{ppbpid}. To establish a TCM for A-B PID \pt\ spectra it is assumed that (a) \nn\ parameters $\alpha$, $\bar \rho_{sNN}$ and $\bar \rho_{hNN}$ have been inferred from unidentified-hadron (nonPID) data and (b) geometry parameters $N_{part}$, $N_{bin}$, $\nu$ and $x$ are a common property (relating to centrality) of a specific A-B collision system independent of identified-hadron species. Quantity $n_s = \Delta \eta \bar \rho_s$ (referring to the soft component of total charge density $\bar \rho_0$) is employed as a centrality index. Transverse rapidity  $y_{ti} \equiv \ln[(p_t + m_{ti})/m_i]$ is define for hadron species $i$. For pion rapidity $y_{t\pi}$ the correspondence with \pt\ is 1 vs 0.16 GeV/c, 2 vs 0.5 GeV/c, 2.67 vs 1 GeV/c, 4 vs 3.8 GeV/c and 5 vs 10.4 GeV/c. $y_{t\pi}$ is the default plotting rapidity for all hadron species in this study.

Given the A-B spectrum TCM for nonPID spectra, reviewed for example in Ref.~\cite{ppbpid} Sec.~4, a corresponding TCM for identified hadrons can be formulated by assuming that each hadron species $i$ comprises certain {\em fractions} of soft and hard TCM components denoted by $z_{si}(n_s)$ and $z_{hi}(n_s)$  (both $\leq 1$) and assumed {\em independent of \yt}. That assumed factorization is reconsidered in Sec.~\ref{newdetails}. The PID spectrum TCM is then expressed as
\bea \label{pidspectcm}
\bar \rho_{0i}(y_t) &=& S_i(y_t)+ H_i(y_t)
\\ \nonumber
& \approx& \bar \rho_{si} \hat S_{0i}(y_t) + \bar \rho_{hi} \hat H_{0i}(y_t) 
\\ \nonumber
&\approx& z_{si}(n_s)  \bar \rho_{s} \hat S_{0i}(y_t) +  z_{hi}(n_s)  \bar \rho_{h} \hat H_{0i}(y_t) 
\nonumber \\ \label{eq4}
\frac{\bar \rho_{0i}(y_t)}{ \bar \rho_{si}} &\equiv&   X_i(y_t)
\\ \nonumber
&\approx & \hat S_{0i}(y_t) +  \tilde z_{i}(n_s)x(n_s)\nu(n_s) \hat H_{0i}(y_t),
\eea
where ratio $ z_{hi}(n_s) / z_{si}(n_s) \equiv \tilde z_{i}(n_s)$ and unit-integral model functions $\hat S_{0i}(y_t)$ and $\hat H_{0i}(y_t)$ depend on hadron species $i$. Integrating over \yt\ and rearranging gives
$\bar \rho_{si}$  in terms of $ \bar \rho_{s}$ for unidentified hadrons previously determined
\bea \label{rhosi}
{\bar \rho_{si}(n_s)} &\approx&  \left[\frac{1 + x(n_s) \nu(n_s)}{1 + \tilde z_{i}(n_s) x(n_s) \nu(n_s)}  \cdot {z_{0i}} \right] {\bar \rho_{s}} 
%\\ \nonumber
%&\equiv& q_{si}(n_s) {z_{0i}} \cdot {\bar \rho_{s}}
%\equiv z_{si}(n_s) \bar \rho_s
\\ \nonumber
{\bar \rho_{hi}(n_s)} &=&  z_{hi}(n_s) \bar \rho_h
\\ \nonumber
\bar \rho_{hi} / \bar \rho_{si} &=&   \tilde z_{i}(n_s) x(n_s) \nu(n_s),
\eea
and $\bar \rho_{0i} \equiv z_{0i} \bar \rho_0$ defines $z_{0i}$. The expression in square brackets in the first line defines $z_{si}(n_s)$. Model function $\hat S_{0i}(y_t)$ is defined on proper $m_{ti}$ for a given hadron species $i$ and then transformed to $y_{t\pi}$. $\hat H_{0i}(y_t)$ is always defined on $y_{t\pi}$. The $\hat S_{0i}(y_t)$ model is defined as the limit of normalized data spectra $\bar \rho_{0i}(y_t) / \bar \rho_{si}$ as $n_{ch} \rightarrow 0$. Soft-component models are thus derived from data spectra.

Normalized data spectra $X_i(y_t)$ can be combined with soft-component model function $\hat S_{0i}(y_t)$ per Eq.~(\ref{eq4}) to extract data spectrum hard components in the form
\bea \label{pidhard}
Y_i(y_t) &\equiv& \frac{1}{\tilde z_{i}(n_s) x(n_s) \nu(n_s) } \left[ X_i(y_t) -  \hat S_{0i}(y_t) \right]~~~
\eea
that may then be compared directly with hard-component model functions $\hat H_{0i}(y_t)$.

 \subsection{$\bf p$-$\bf Pb$ PID TCM spectrum parameters} \label{pidfracdata}
 
This subsection reviews parameter values reported in Ref.~\cite{ppbpid} based on best descriptions of inferred PID spectrum hard components and on assuming fixed values for hard/soft ratios $z_h / z_s = \tilde z_{i}$ and total fractions $z_{0i}$. In Sec.~\ref{zxspectra} a more-detailed procedure is used to infer $z_{si}(n_s)$ and $z_{hi}(n_s)$ parameters directly from spectrum data. In Sec.~\ref{newdetails} the approach is further generalized to obtain $z_{hi}(y_t,n_s)$ distributions on \yt\ for the most differential description of \ppb\ PID hard components.
  
Table~\ref{pidparams} shows TCM model parameters for hard component $\hat H_{0i}(y_t)$ (first three) and soft component $\hat S_{0i}(y_t)$ (last two). Hard-component model parameters vary slowly but significantly with hadron species. Centroids $\bar y_t$ shift to greater \yt\ with increasing hadron mass. Widths $\sigma_{y_t}$ are significantly larger for mesons than for baryons. Only $K_\text{S}^0$ and $\Lambda$ data extend to sufficiently high \pt\ to determine exponent $q$ from those data. $q$ is substantially greater for baryons than for mesons.  The combined centroid, width and exponent trends result in near coincidence among the several species for hard components at larger \yt. Evolution of TCM hard-component parameters with hadron species is consistent with measured PID FFs (refer to Fig.~7 of Ref.~\cite{ppbpid}) and with a common underlying parton (jet) spectrum for all TCM hard components.
  
 %%%%%%%%%%%%%%%%%%%%%%%%%%%%%%%%%%
 \begin{table}[h]
 	\caption{TCM model parameters for unidentified hadrons $h$ from Ref.~\cite{alicetomspec} and for identified hadrons from 5 TeV \ppb\ collisions from Ref.~\cite{ppbpid}: hard-component parameters $(\bar y_t,\sigma_{y_t},q)$ and soft-component parameters $(T,n)$. Numbers without uncertainties are adopted from a comparable hadron species with greater accuracy (e.g.\ $p$ vs $\Lambda$). 
 	}
  	\label{pidparams}
 	%%%%%%%%%%%%%%%%%%%%%%%%%%%%%%%
 	\begin{center}
 		\begin{tabular}{|c|c|c|c|c|c|} \hline
 			%\multicolumn{8}{|c|}{\pp\ multiplicity classes} \\ \hline
 			& $\bar y_t$ & $\sigma_{y_t}$ & $q$ & $T$ (MeV) &  $n$  \\ \hline
 			$ h $     &  $2.64\pm0.03$ & $0.57\pm0.03$ & $3.9\pm0.2$ & $145\pm3$ & $8.3\pm0.3$ \\ \hline
 			$ \pi^\pm $     &  $2.52\pm0.03$ & $0.56\pm0.03$ & $4.0\pm1$ & $145\pm3$ & $8.5\pm0.5$ \\ \hline
 			$K^\pm$    & $2.65$  & $0.58$ & $4.0$ & $200$ & $14$ \\ \hline
 			$K_\text{S}^0$          &  $2.65\pm0.03$ & $0.58\pm0.02$ & $4.0\pm0.2$ & $200\pm5$ & $14\pm2$ \\ \hline
 			$p$        & $2.92\pm0.02$  & $0.47$ & $4.8$ & $210\pm10$ & $14\pm4$ \\ \hline
 			$\Lambda$       & $2.96\pm0.02$  & $0.47\pm0.03$ & $4.8\pm0.5$  & $210$ & $14$ \\ \hline	
 		\end{tabular}
 	\end{center}
 	%%%%%%%%%%%%%%%%%%%%%%%%%%%%%%%
 \end{table}
 %%%%%%%%%%%%%%%%%%%%%%%%%%%%%%%
  
 Soft-component model parameter $T \approx 145$ MeV for pions ($\pi^\pm$) is the same as that for unidentified hadrons $h$ found to be universal over all A-B collision systems and collision energies~\cite{alicetomspec}.  The values for higher-mass hadrons are substantially larger. L\'evy exponent $n \approx 8.5$ for pions is also consistent with that for unidentified hadrons at 5 TeV and has a  $\log(\sqrt{s}/\text{10 GeV})$ energy dependence~\cite{alicetomspec}. Soft-component exponent $n$ values for more-massive hadrons are not well-defined because the hard-component contribution is much larger than for pions. Varying $n$ then has little impact on model spectra.
 
 Table~\ref{otherparams} shows PID parameters $z_{0i}$ and $\tilde z_{i} = z_{hi} / z_{si}$ for five hadron species that are determined from \ppb\ PID spectrum data as fixed values independent of event centrality or multiplicity \nch.  The choice to hold $z_0$ and $\tilde z_{i}$ fixed rather than $z_{si}$ and $z_{hi}$ separately in Ref.~\cite{ppbpid} was based on spectrum trends at low \pt\ ($p_t < 0.5$ GeV/c) such that properly normalized spectra in the form of Eq.~(\ref{eq4}) should coincide with model function $\hat S_0(y_t)$ independent of \nch\ at low \pt\ where spectrum hard components are {observed} to make a negligible contribution to spectra~\cite{ppbpid}. 
 While Table~\ref{otherparams} is intended to duplicate the corresponding Table 4 in Ref.~\cite{ppbpid} the value $z_{0i} = 0.70$ for pions in that table was incorrectly entered. It should be 0.80. 
   
 %%%%%%%%%%%%%%%%%%%%%%%%%%%%%%%%%%
 \begin{table}[h]
	\caption{TCM model parameters for identified hadrons from 5 TeV \ppb\ collisions~\cite{ppbpid}. Numbers without uncertainties are adopted from a comparable hadron species with greater accuracy. Parameters $ \bar p_{tsi}$ and $\bar p_{th0i}$ are determined by model functions $\hat S_{0i}(y_t)$ and $\hat H_{0i}(y_t)$ with parameters from Table~\ref{pidparams}.  $h$ represents results for nonPID hadrons. The value 0.70 for pion $z_0$ in Ref.~\cite{ppbpid} was an incorrect entry. The intended value from that analysis was 0.80. Values of $z_0$ for charged particles should sum to 1 within uncertainties. See Eq.~(\ref{sums}).
 	}
 	\label{otherparams}
 	%%%%%%%%%%%%%%%%%%%%%%%%%%%%%%%
 	\begin{center}
 		\begin{tabular}{|c|c|c|c|c|} \hline
 			%\multicolumn{8}{|c|}{\pp\ multiplicity classes} \\ \hline
 			&   $z_0$    &  $\tilde z = z_h / z_s$ &   $ \bar p_{ts}$ (GeV/c)  & $ \bar p_{th0}$ (GeV/c)  \\ \hline
 			$ h$        &  $\equiv 1$  & $\equiv 1$  & $ 0.40\pm0.02$ &    $1.30\pm0.03$  \\ \hline
 			$ \pi^\pm$        &   $0.80\pm0.02$  & $0.80\pm0.05$  & $0.40\pm0.02$ &    $1.15\pm0.03$  \\ \hline
 			$K^\pm $   &  $ 0.125\pm0.01$   &  $2.8\pm0.2$ &  $0.60$&  $1.34$   \\ \hline
 			$K_\text{S}^0$        &  $0.062\pm0.005$ &  $3.2\pm0.2$ &  $0.60\pm0.02$ &   $1.34\pm0.03$  \\ \hline
 			$p $        & $ 0.07\pm0.005$    &  $7.0\pm1$ &  $0.73\pm0.02$&   $1.57\pm0.03$   \\ \hline
 			$\Lambda $        &  $0.037\pm0.005$    & $7.0$ &   $0.76\pm0.02$ &    $1.65\pm0.03$ \\ \hline	
 		\end{tabular}
 	\end{center}
 	%%%%%%%%%%%%%%%%%%%%%%%%%%%%%%%
 \end{table}
 %%%%%%%%%%%%%%%%%%%%%%%%%%%%%%
 
 It is notable that while parameters $z_{0i}$ and $\tilde z_{i}(n_s)$ in Eq.~(\ref{rhosi}) (first line) describe the soft component of a PID charge density at low \pt, the ratio $\tilde z_{i}(n_s)$ in Eq.~(\ref{pidhard}) in effect determines the hard component of the PID charge density. When  $z_{0i}$ and $\tilde z_{i}(n_s)$ are adjusted so as to describe the centrality evolution of the soft component the hard-component trend is effectively {\em predicted}. That fortuitous result arises because the mean event charge $\bar n_{ch}$ or density $\bar \rho_0$ that defines a \pp\ event class acts as a constraint on events in that class. If the hard-component fraction of a particular hadron species increases, the soft component must decrease to compensate. The two components are thus directly linked via the constraint $\bar \rho_0$. 
 
 The nonPID yield $\bar n_{ch}$ includes a proton hard component with {\em no significant inefficiency}. Equation~(\ref{rhosi}) (first line) then implicitly relates $\bar \rho_0$ to $\bar \rho_{si}$ at low \pt\ where there is also no significant proton inefficiency. Ratio $\tilde z_{i}(n_s)$ estimated at low \yt\ must then correspond to the full proton hard component even though only a fraction may be detected in a full PID spectrum. Those relationships explain why the TCM can determine proton inefficiencies accurately by the method described below.

 %%%%%%%%%%%%%%%%%%
\section{Preliminary TCM updates} \label{prelim}

Prior to considering new material from the present analysis three preliminary issues are addressed: 
(a) A contribution to pion spectra in addition to TCM soft and hard components is quantified and incorporated into TCM analysis. 
(b) An apparent proton detection inefficiency is assessed and a method is developed to correct published proton spectra.
(c) Two distinct applications of the TCM to spectrum data are distinguished. 

\subsection{Pion soft component and resonances} \label{pionsoft}

For non-pion hadron species the L\'evy distribution form
\bea \label{s0model}
\hat S_{0i}(m_{ti}) &=& \frac{A_i}{[1+(m_{ti} - m_i)/n_iT_i]^{n_i}},
\eea
with $A_i$ determined so that $\hat S_{0i}(m_{t_i})$ is unit normal, provides a good description of spectrum data. A notable example is $K_\text{S}^0$ data in panel (d) of Fig.~\ref{piddata} that are described within data uncertainties down to $p_t = 0$ (\yt\ = 0). In the limit $1/n \rightarrow 0$ the L\'evy distribution goes to a Maxwell-Boltzmann distribution on \mt. Model functions $\hat S_{0i}(y_t)$ are defined as the limits of normalized data spectra $X_i(y_t) = \bar \rho_{0i}(y_t) / \bar \rho_{si}$ as $n_{ch} \rightarrow 0$, and the L\'evy functional form generally describes those limits well.
The exception is pion spectra where model function $\hat S_{0i}(m_{ti})$ defined by Eq.~(\ref{s0model}) must be modified to accommodate spectrum data. Additional factor $f(y_t)$ must be applied to $\hat S_{0i}(y_t) = (m_{ti} p_t / y_t) \hat S_{0i}(m_{ti})$.

\begin{figure}[h]
	\includegraphics[width=1.65in]{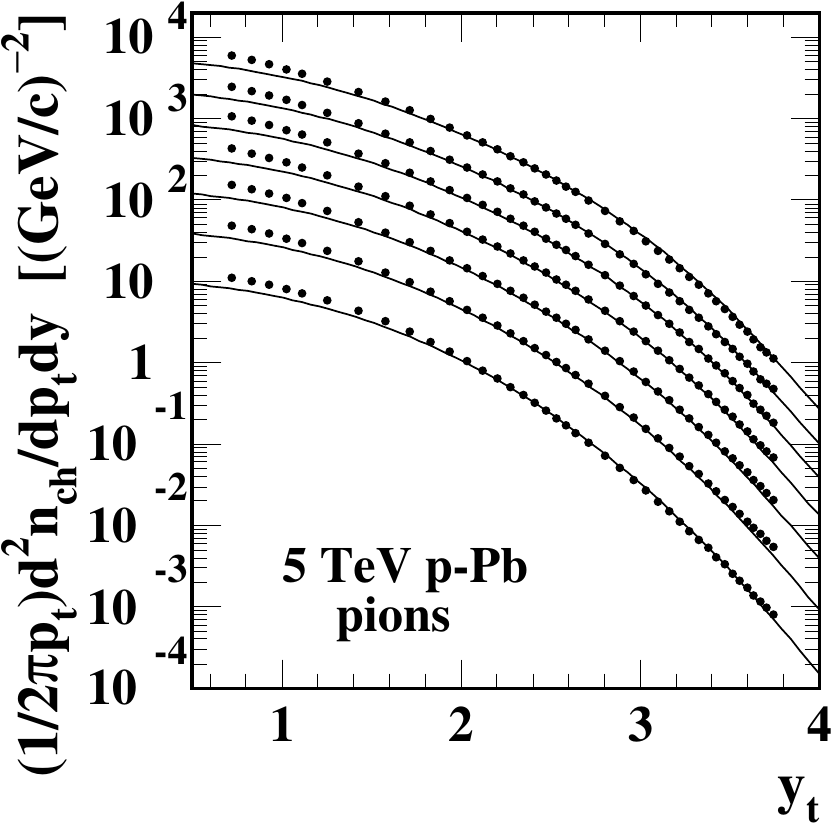}
	\includegraphics[width=1.65in]{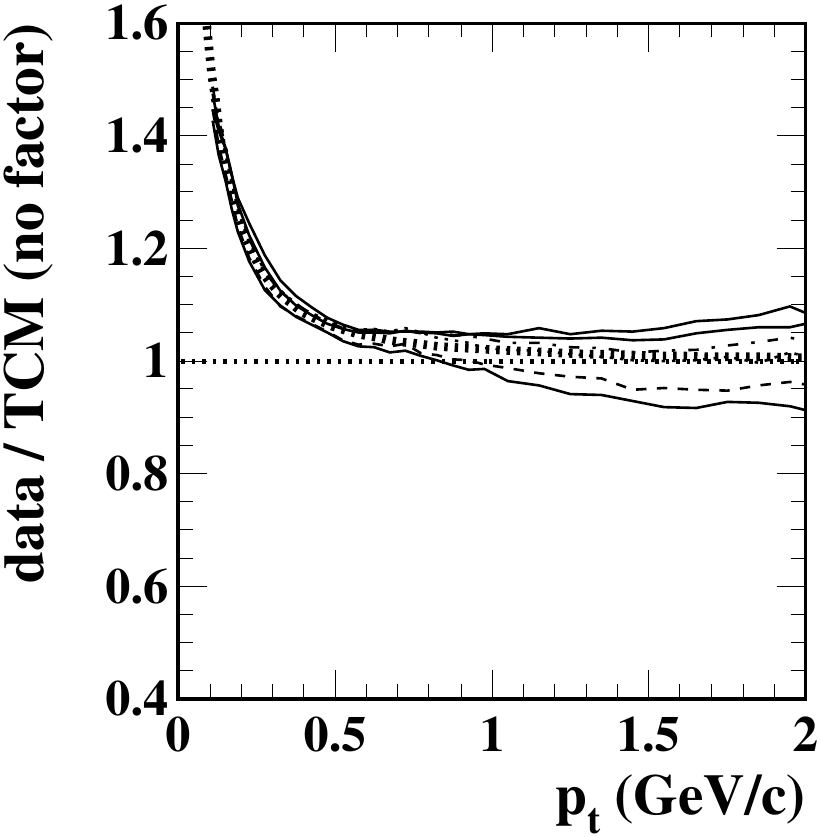}
	\caption{\label{fac1}
		Left: Data from Fig.~\ref{piddata} (a) (points) and modified TCM (solid) for seven centralities of 5 TeV \ppb\ collisions. The TCM soft component in this case does not include factor $f(y_t)$ as described in the text below.
		Right: Ratios of \ppb\ data to TCM without factor $f(y_t)$ applied to the soft component. Line styles vary with centrality from most-central collisions as solid, dashed, dotted, dash-dotted and returning to solid. The bold dotted curve following data is defined by Eq.~(\ref{bolddotted}).
		}	%    alice600aacal, 600aa2cal
\end{figure}

Fig.~\ref{fac1} (left) shows pion data spectra from panel (a) of Fig.~\ref{piddata} (points) and TCM curves with required factor $f(y_t)$ omitted. Soft-component model $\hat S_{0i}(y_t)$ is derived from  Eq.~(\ref{s0model}) with pion model parameters from Table~\ref{pidparams}. Those parameter values are required to describe the spectra above \yt\ = 2 ($p_t \approx 0.5$ GeV/c), but spectrum data rise increasingly above the TCM $\hat S_{0i}(y_t)$ below \yt\ = 2.

%%%%%%%%%%
%%%%%%%%%%%%

Fig.~\ref{fac1} (right) shows the ratio of spectrum data to TCM [with factor $f(y_t)$ omitted] as they appear in the left panel for six centrality classes of 5 TeV \ppb\ collisions (the data hard component is strongly biased for $n = 7$). Deviations from 1 at higher \pt\ correspond to evolution of spectrum hard components in Fig.~\ref{pioncomp} (left) taken from Ref.~\cite{ppbpid} where the hard-component model was adjusted to reflect an intermediate form relative to data. The most-central data then fall below 1 whereas for the treatment in Sec.~\ref{newdetails} the most-central data should coincide with 1 at higher \pt. In Fig.~\ref{fac1} (right) deviations of data hard components from $\hat H_{0i}(y_t)$ below the mode in Fig.~\ref{pioncomp} are overwhelmed by the soft component at lower \pt.

The bold dotted curve defined by the expression
\bea \label{bolddotted}
f(y_t) &=& 2/\left\{1+\tanh[(y_t - 0.3)/1.2]\right\}
\eea
optimizes data-model agreement at lower \yt\ as in Fig.~\ref{piddata} (a). In effect, unit-normal model function $\hat S_0(y_t)$ in the TMC is replaced by $S_0'(y_t)$ (not normalized) defined by
\bea \label{s0prime}
S_{0i}'(y_t) &=& f(y_t) \hat S_{0i}(y_t),
\eea
where $\hat S_{0i}(y_t)$ for pions has soft-component ensemble-mean $\bar p_{tsi} \approx 0.44$ GeV/c while modified model function $ S_{0i}'(y_t) $ has soft-component mean $\bar p_{tsi} \approx 0.40$ GeV/c due to the additional contribution at lower \pt. Quantity $[f(y_t) - 1]\hat S_{0i}(y_t)$ represents a {\em third} component not described by the TCM. One may conjecture that this third component is the pion contribution from $\rho$ and $\omega$ resonance decays predicted to appear below 0.5 GeV/c~\cite{resonances}. Figure~\ref{fac1} (right) indicates that the resonance yield scales exactly with the soft component within data uncertainties. Neither the resonance contribution nor the TCM soft component changes shape significantly with varying \ppb\ centrality. Factor $f(y_t)$ as defined by Eq.~(\ref{bolddotted}) was already adopted for the analysis reported in Ref.~\cite{ppbpid}.

\subsection{Proton inefficiency correction} \label{ineff}

In Ref.~\cite{ppbpid} a strong ($\approx 40$\%) suppression of the proton spectrum hard component in comparison to TCM predictions was noted. Whether the suppression was an instrumental effect or novel physics was not determined. The apparent proton suppression is one of several motivations for the present study. Just as for Ref.~\cite{pbpbpid} a comparison of proton spectrum data with a PID TCM is used to derived an inefficiency correction. The basis for expecting a correction based on the present revised TCM to be accurate is summarized at the end of Sec.~\ref{pidfracdata}.

\begin{figure}[h]
	\includegraphics[width=3.3in]{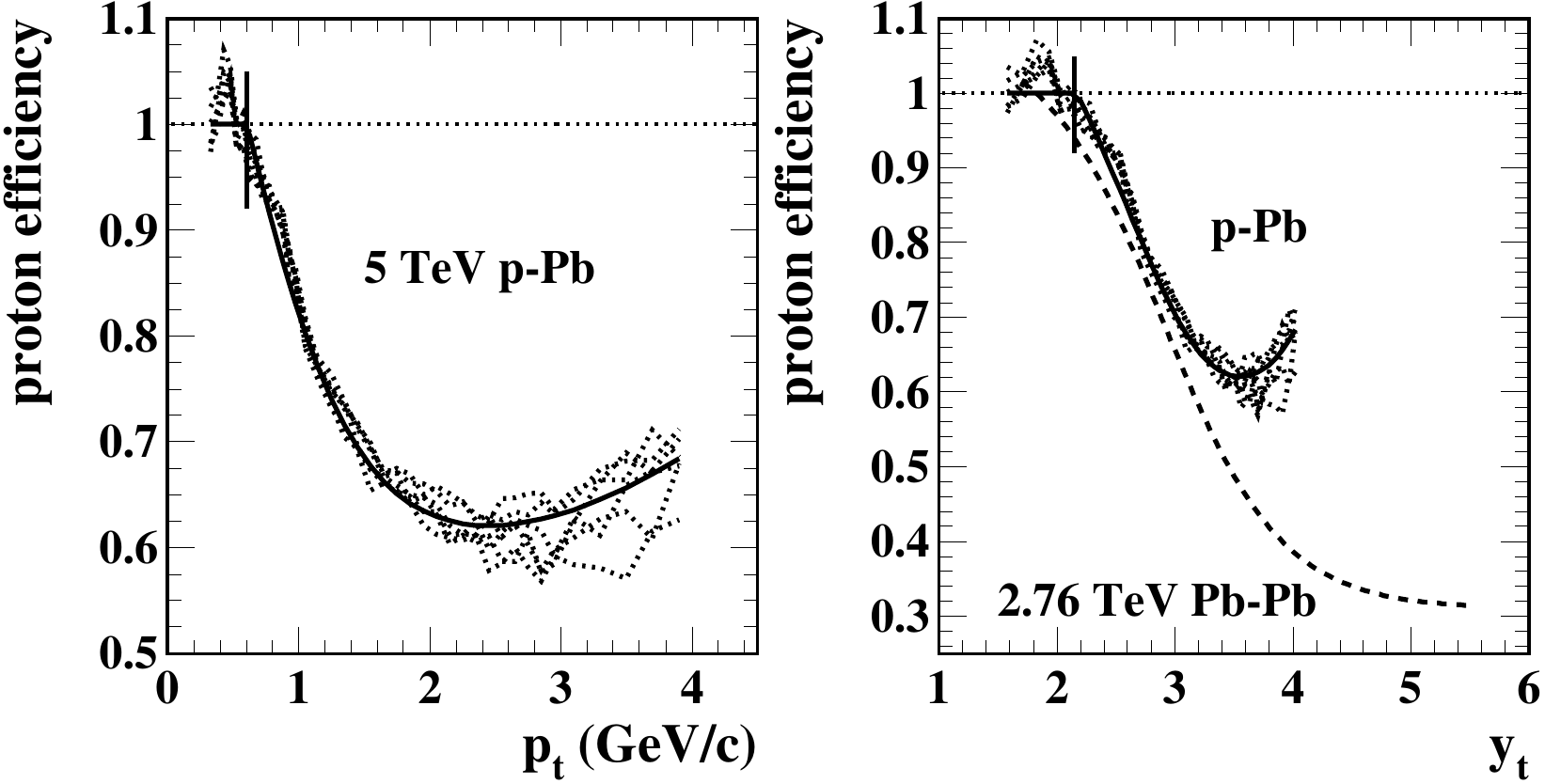}
	\caption{\label{eppsproton}
		Left: Ratios of uncorrected identified-proton \pt\ spectra as reported in Ref.~\cite{aliceppbpid} to TCM spectra defined in the present study (dotted curves) for seven centralities of 5 TeV \ppb\ collisions. The solid curve is a proton efficiency model defined by Eq.~(\ref{effppb}).
		Right: Curves at left plotted vs transverse rapidity \yt. The dashed curve is a proton efficiency trend defined as in Eq.~(\ref{effpbpb}) for \pp\ and \pbpb\ PID spectra in Ref.~\cite{pbpbpid}.
	}  % alice610c
\end{figure}

Figure~\ref{eppsproton} (left) shows ratios (dotted curves) of uncorrected proton spectrum data from Ref.~\cite{aliceppbpid} to TCM spectra derived for seven centrality classes from the present study. It is important to note that the TCM does not result from model fits to individual spectra. It is a comprehensive data model that must self-consistently describe a broad array of A-B collision systems and must be consistent with all jet measurements. The proton TCM (solid) in Fig.~\ref{piddata} (e) is based on Eqs.~(\ref{rhosi}) (first line) and (\ref{pidhard}) with fixed $\tilde z_{i}(n_s) \rightarrow \tilde z_{i} = 5.8 \pm 0.1$ derived by fitting Eq.~(\ref{zsinew}) to $z_{si}(n_s)$ values from Table~\ref{zsixx} obtained at low \yt\ where proton inefficiency is not an issue. Further details relating to proton inefficiencies are discussed in Sec.~\ref{resolve}

%%%%%%%%%%
%%%%%%%%%%%%

The data-model comparison in Fig.~\ref{eppsproton} (left) reveals systematic suppression of protons not confined to the data hard component alone as was conjectured in Ref.~\cite{ppbpid}. The solid curve defined by
\bea \label{effppb}
\epsilon_p(\ppb) \hspace{-.05in} &=& \hspace{-.05in} \left\{0.58+(1\hspace{-.02in} - \hspace{-.02in}\tanh[(p_t \hspace{-.02in}- \hspace{-.02in}0.45)/0.95])/2\right\} 
\\ \nonumber &\times& (1+0.003 p_t^3)
\eea
is assumed to describe, {\em independent of \ppb\ centrality}, an instrumental inefficiency affecting $dE/dx$ PID measurements reported in Ref.~\cite{aliceppbpid}. The inefficiency appears significant only above 0.6 GeV/c (vertical lines in both panels of Fig.~\ref{eppsproton}). In the present case the same correction (bold solid) is applied consistently to all event classes.

The source of inefficiency is conjectured here to result from systematic bias in the $dE/dx$ analysis where the proton contribution is vulnerable to losses during a subtraction procedure. ``In the regions where signals from several [hadron] species overlap [e.g.\ $> 0.6$ GeV/c for protons] [the] $dE/dx$ [distribution] is fit with two Gaussian distributions.... The [Gaussian] fit of the overlapping species [i.e.\ kaons] is then integrated in the signal region [encompassing the signal species, i.e.\ protons] and subtracted from the [total integrated] signal.''~\cite{alicepppid}. The sensitivity of that procedure to systematic bias can be appreciated from Fig.~5 of Ref.~\cite{alicepbpbpidspec} where the $dE/dx$ peaks for protons and charged kaons overlap strongly.

Figure~\ref{eppsproton} (right) shows the curves in the left panel now plotted on transverse rapidity \yt. The dashed curve is the proton detection efficiency
\bea \label{effpbpb}
\epsilon_p(\pp) &=& 0.31 + 0.77 \left[ 1 - \tanh(y_t - 2.9)   \right]/2 
\eea
inferred in Ref.~\cite{pbpbpid} from 2.76 TeV \pbpb\ PID spectrum data. These several indicators suggest that an instrumental proton inefficiency results from the $dE/dx$ procedure utilized in Refs.~\cite{aliceppbpid,alicepbpbpidspec,alicepppid} (also see Ref.~\cite{alicepppidx}), and that the efficiency trend in Eq.~(\ref{effppb}) may be used to correct \ppb\ proton spectra. 

%%%%%%%%%%
\begin{figure}[h]
	\includegraphics[width=1.65in]{alice610aayno}
	\includegraphics[width=1.65in]{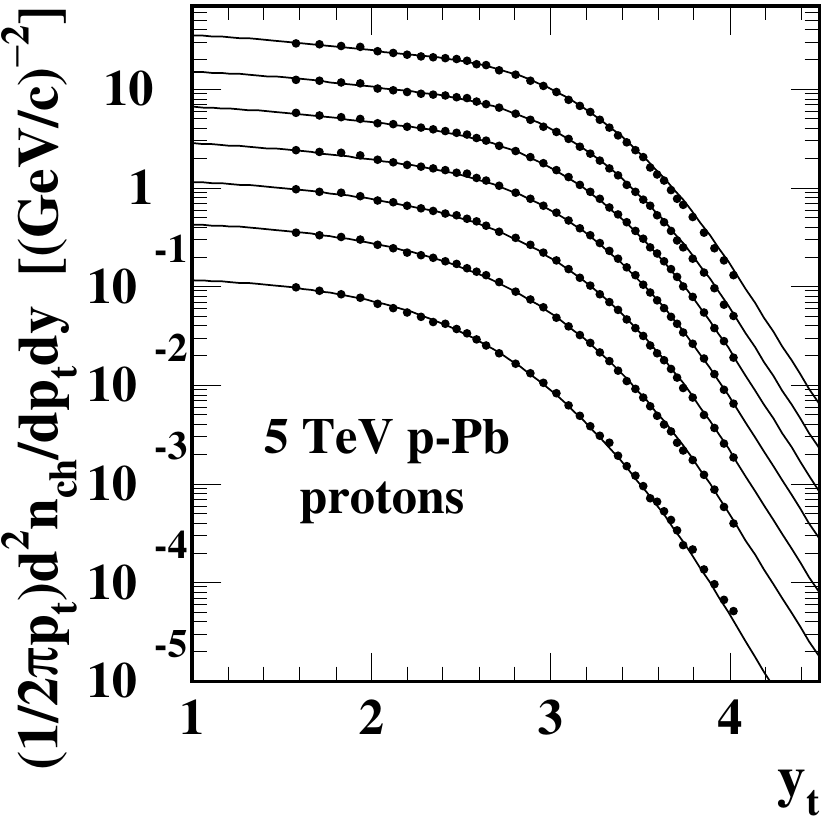}
	\caption{\label{eppsprotonx}
		Left: Panel (e) of Fig.~\ref{piddata} repeated.
		Right: The left panel but with proton spectra corrected via Eq.~(\ref{effppb}). The correction is applied only above 0.6 GeV/c or $y_t \approx 2.15$.
	}  %  alice610aano, aax
\end{figure}
%%%%%%%%%%%%

Figure~\ref{eppsprotonx} (left) repeats Fig.~\ref{piddata} (e) showing the systematic disagreement between proton data and TCM. The proton spectra appearing in Fig.~\ref{eppsprotonx} (right) have been corrected with the function in Fig.~\ref{eppsproton} (left) and are utilized as such in the remainder of this study. Correction of proton spectra is necessary to proceed with determination of individual $z_{si}(n_s)$ and $z_{hi}(n_s)$ trends in what follows.

Results from this study strongly suggest that the proton suppression is an instrumental effect not confined to the spectrum hard component: (a) The PID TCM provides accurate descriptions of hadron species other than protons, (b) the newly-measured centrality dependence of soft-component fractions $z_{si}(n_s)$ at low \pt\ is able to generate TCM-predicted proton hard-component yields, (c) ensemble-mean \mmpt\ values for nonPID \ppb\ data compared to those constructed from PID spectra in a related study also favor the proton yield predicted by the TCM. Another source of information is TCM analysis of 2.76 TeV \pp\ and \pbpb\ PID spectra in Ref.~\cite{pbpbpid} that strongly suggests a large proton detection inefficiency above 1 GeV/c for both systems. Reference~\cite{aliceppbpid} (ALICE) notes that ``The [PID] separation power achieved in p-Pb collisions is identical to that in pp collisions.'' 

There are also indications from theory: In connection with statistical-model fits to PID yield data Ref.~\cite{stoeckerstatmodel} (referring to \pp, \ppb\ and \pbpb\ systems) notes that ``The [statistical] model overpredicts systematically the $p/\pi$ ratio, roughly  on  a  2$\sigma$ level.   Separately,  the  proton  yields are  overpredicted  on  a  1$\sigma$ level,  while  the  yields  of  pions  are  underpredicted  on  a  1$\sigma$ level.'' Reference~\cite{thermalprotons} refers to the ``Thermal proton yield anomaly in \pbpb\ collisions....'' and continues with ``The statistical hadronization model predicts about 25\% more protons and antiprotons [integrated yields] than measured by the ALICE collaboration in central Pb-Pb collisions at the LHC. This constitutes the much debated `proton-yield anomaly' in heavy-ion collisions at the LHC.'' The report suggests that the proton anomaly may be resolved as follows: ``By implementing the essential features of the empirical $\pi N$ scattering -- the effects of broad resonances and the presence of nonresonant contributions -- and using LQCD results on the baryon-charge susceptibility in the statistical model, we find a reduction of the proton yield relative to the HRG [statistical model] result.  It is then in excellent agreement with experiment.'' That result is intriguing but deals only with integrated yields. It fails to address the observation that proton spectra at lower \pt, where a {\em thermal or statistical} model should be most applicable, are consistent with other hadron species. Quite similar inefficiencies are observed for \pp, \ppb\ and \pbpb\ collisions mainly in the higher-\pt\ regions of spectra where jet fragments should dominate (at least in \pp\ and \ppb).

\subsection{TCM description of full PID spectra} \label{newoldspec}

The TCM appearing in Fig.~\ref{piddata} (curves) follows Eqs.~(\ref{eq4}) and (\ref{rhosi}) as in Ref.~\cite{ppbpid} with model-function parameters from Table~\ref{pidparams} except $\bar y_t \rightarrow 2.96$ for protons and $q \rightarrow 3.8$ for pions and kaons. The proton and Lambda hard-component centroids are shifted step-wise with centrality as described in Ref.~\cite{ppbpid} and as shown in Fig.~\ref{baryoncomp} (left). However, the TCM is altered from Ref.~\cite{ppbpid} via parameters $\tilde z_i$ and $z_{0i}$. Based on the present analysis parameters $\tilde z_i(n_s)$ now vary with centrality according to the linear trends in Fig.~\ref{zrat}, and resulting updated values for $z_{0i}$ used for Fig.~\ref{piddata} are 0.82, 0.128, 0.064, 0.065 and 0.034. 

It is important to distinguish two different applications of the TCM as a model: (a) a fixed reference model against which data evolution is assessed differentially and quantitatively with the goal of accessing new physics. 
(b) a variable TCM that accommodates data evolution to the extent permitted by its basic structure and assumptions. 
Examples of the former are Ref.~\cite{tomnewppspec} and the results in Sec.~\ref{newdetails}. Examples of the latter are Ref.~\cite{tommodeltests} and the variable hard-component models in Fig.~\ref{baryoncomp} (left) for baryons. The two types of application are compared in Sec.~\ref{revisedspec}.

%%%%%%%%%%%
\section{$\bf z_{si}$ and $\bf z_{hi}$ Inferred from spectra} \label{zxspectra}

In Ref.~\cite{ppbpid} fixed ratio $\tilde z_{i}$ was adjusted for each hadron species $i$ to achieve coincidence of all seven normalized spectra as $y_t \rightarrow 0$. Parameter $z_{0i}$ was then adjusted to match the rescaled spectra to unit-normal $\hat S_{0i}(y_t)$, also as $y_t \rightarrow 0$. That procedure was based on the assumption  that $\tilde z_{i}$ and $z_{0i}$ are approximately independent of \ppb\ centrality. The present analysis determines separate multiplicity evolution trends for $z_{si}(n_s)$ and $z_{hi}(n_s)$ for each hadron species and reviews those previous assumptions.

This section introduces a method to obtain estimates of $z_{si}(n_s)$ and $z_{hi}(n_s)$ directly from PID spectrum data . In what follows  relations $\bar \rho_s = (N_{part}/2) \bar \rho_{sNN}$ (soft charge density) and $\bar \rho_h = N_{bin} \bar \rho_{hNN}$ (hard charge density) relate soft and hard charge densities to 5 TeV \ppb\ geometry parameters reported in Table~\ref{rppbdata}. No estimates for hard-component coefficient $z_{hi}(n_s)$ are provided for the most-peripheral $n = 7$ class per Ref.~\cite{aliceppbpid}: The most-peripheral centrality class ``...is well below the corresponding multiplicity in pp minimum-bias collisions and therefore likely to be subject to a strong selection bias.''

\subsection{Estimating $\bf z_{si}(n_s)$ and $\bf z_{hi}(n_s)$ from spectra}

Based on the structure of Eq.~(\ref{pidspectcm}) nearly model-independent estimates for soft-component coefficients $z_{si}(n_s)$ can be obtained from PID spectra $\bar \rho_{0i}(p_t,n_s)$ via
\bea \label{zsi}
\lim_{p_t \to 0} \bar \rho_{0i}(p_t,n_s) / \bar \rho_s \hat S_{0i}(p_t) &\approx &  
\bar \rho_{si} / \bar \rho_s = z_{si}(n_s),
\eea
where $\hat S_{0i}(p_t)$ is a soft-component model inferred in Ref.~\cite{ppbpid} per parameters in Table~\ref{pidparams} and $\bar \rho_s$ is obtained from parameters in Table~\ref{rppbdata} derived in Ref.~\cite{tommpt}.

Given measured values $z_{si}(n_s)$ corresponding hard-component coefficients $z_{hi}(n_s)$ can be inferred via
\bea \label{zhi}
\lim_{y_t \to \bar y_t}  \left[\bar \rho_{0i}(y_t,n_s) \hspace{-.03in} - \hspace{-.03in} z_{si}\bar \rho_s \hat S_{0i}(y_t)\right]\hspace{-.03in} / \bar \rho_h \hat H_{0i}(\bar y_t) \hspace{-.07in} &\approx &  \hspace{-.07in} z_{hi}(n_s),~~~~
\eea
where $\bar y_t$ denotes the mode of the {\em data} spectrum hard component, $\bar \rho_h(n_s) =N_{bin} \alpha \bar \rho_{sNN}^2(n_s)$ is derived from Table~\ref{rppbdata} (from corresponding \mmpt\ data in Ref.~\cite{tommpt}), and unit-normal hard-component models $\hat H_{0i}(y_t)$ are determined by parameters in Table~\ref{pidparams}.
%(but see updates in Table~\ref{pidparamsx}). 
Accurate results for this procedure require that the value $\hat H_{0i}(\bar y_t)$ be inferred from a correctly-normalized model function. The above assumes hard components factorized as
\bea \label{factorize}
H_i(y_t,n_s) &\rightarrow & z_{hi}(\bar y_t,n_s) \bar \rho_h(n_s) \hat H_{0i}(y_t)
\eea
so that
\bea
H_i(\bar y_t,n_s)/ \bar \rho_h(n_s) \hat H_{0i}(\bar y_t) &\approx& z_{hi}(\bar y_t,n_s) .
\eea
An alternative approach in which $z_{hi}(y_t,n_s)$ is obtained from data without assumptions is described in Sec.~\ref{newdetails}.

\subsection{Accommodating incomplete PID spectrum data} \label{resolve}

The procedure outlined above is challenged by incomplete PID spectrum data from Ref.~\cite{aliceppbpid}, either because of acceptance limitations (e.g.\ no Lambda data below 0.65 GeV/c) or because of systematic bias (i.e.\ substantial proton inefficiency above 0.6 GeV/c). As a result $z_{hi}(n_s)$ data for protons and $z_{si}(n_s)$ data for Lambdas cannot be obtained directly from spectrum data. However, an iterative approach leads to a self-consistent system.

For spectrum data from Ref.~\cite{aliceppbpid} the absolute yield of protons above 0.6 GeV/c is uncertain by as much as 40\%. The published proton data thus establish only a lower bound, but as noted in Sec.~\ref{pidfracdata} the centrality-{\em averaged} hard/soft ratio $\tilde z_{i}$ for protons can be inferred from spectrum evolution at low \yt\  as in Ref.~\cite{ppbpid}.  The $\tilde z_{i}$ value $5.8 \pm 0.1$ thus inferred establishes an estimate of the true proton hard component but does not provide a detailed $z_{hi}(n_s)$ centrality trend. The initial $\tilde z_{i}$ value is used via Eq.~(\ref{rhosi}) to predict a corrected proton spectrum from which a fixed proton efficiency correction as in Sec.~\ref{ineff} is applied uniformly to all \ppb\ event classes. A detailed $z_{hi}(n_s)$ centrality dependence is then inferred from corrected proton spectra as demonstrated below.

The lower-\pt\ cutoff for Lambda spectra at 0.65 GeV/c in Fig.~\ref{protonxxx} (c) precludes direct inference of $z_{si}(n_s)$. However, it is expected that the Lambda trend may be closely related to the proton trend. Proton and Lambda spectra are related at 0.85 GeV/c (vertical lines in Fig.~\ref{protonxxx} left) by factor 1.8 which was initially applied to proton $z_{si}(n_s)$ to estimate Lambda $z_{s}(n_s)$ values. Comparison of the resulting TCM to measured Lambda spectra in Fig.~\ref{piddata} (f) suggested that the $z_{si}(n_s)$ values should be reduced from that estimate by 10\%. Lambda $z_{si}(n_s)$ values were thereafter determined as proton values divided by 2. The Lambda spectra at higher \pt\ appear to be unbiased.

\subsection{Differential spectrum analysis} \label{diffanal}

\begin{figure}[h]
	\includegraphics[width=3.3in]{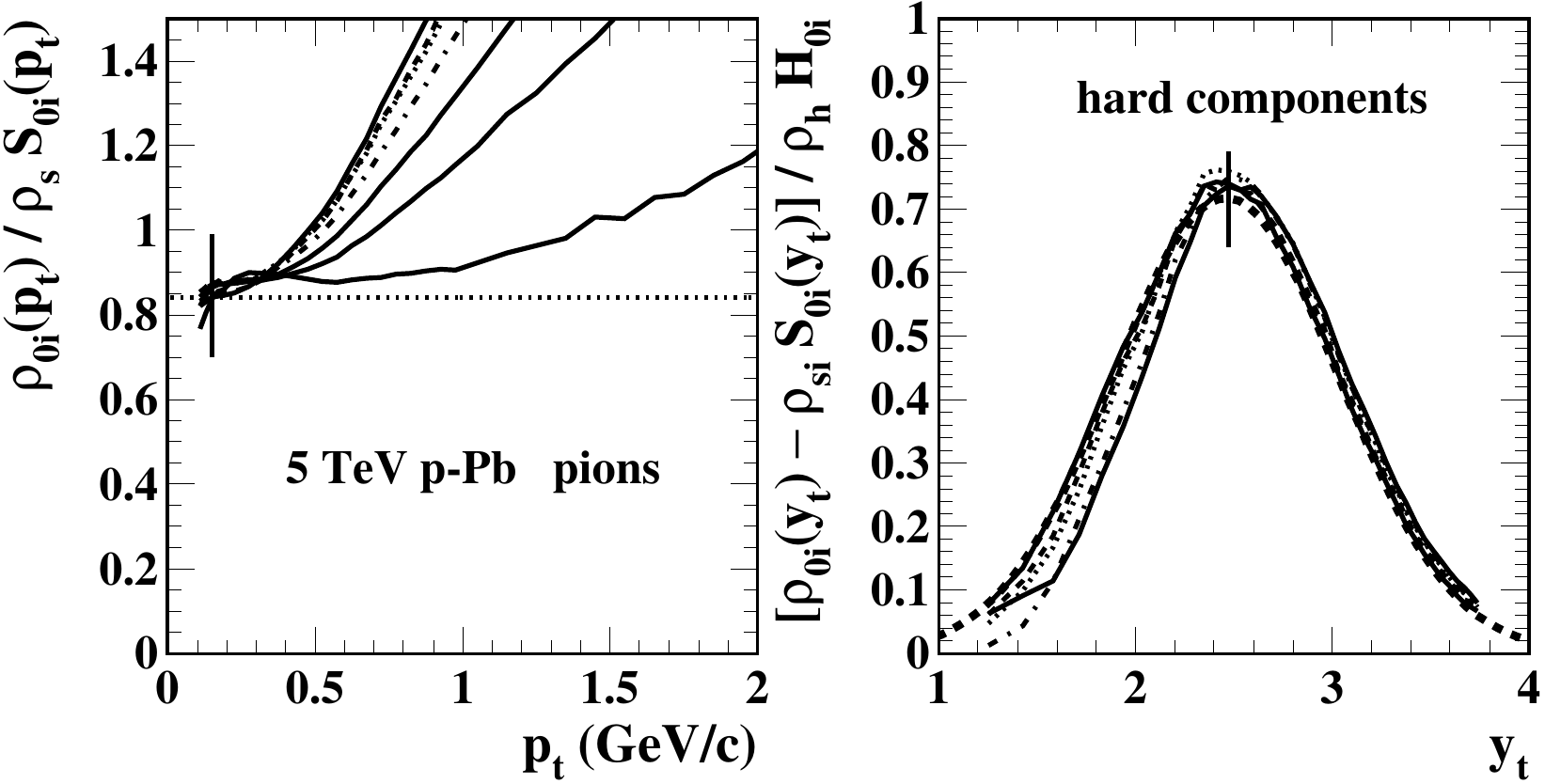}
	\caption{\label{pionx}
Left: Pion spectrum ratios for seven centralities of 5 TeV \ppb\ collisions (curves) processed as in Eq.~(\ref{zsi}), with values of $z_{si}$ inferred at 0.15 GeV/c as denoted by the vertical line. The bold dotted line indicates the average value of $z_{si}(n_s)$
Right: Pion spectrum hard components processed as in Eq.~(\ref{zhi}), with $z_{hi}$ inferred at the distribution mode (line).
	}   % alice600aa2
\end{figure}

Figure~\ref{pionx} (left) shows the expression on the left of Eq.~(\ref{zsi}) from pion spectrum data for seven centrality classes of 5 TeV \ppb\ collisions. For pion spectra special treatment is required corresponding to the procedure described in Sec.~\ref{pionsoft}. In  Eq.~(\ref{zsi}) model function $\hat S_{0i}(p_t)$ is replaced by $S_{0i}'(p_t)$ derived from Eq.~(\ref{s0prime}). In effect, that replacement removes the resonance contribution from the data/model ratio [factor $f(y_t)$ cancels the resonance contribution to data]. The remaining ``nonresonance'' pion soft component is then treated the same as for other hadron species. The vertical bar indicates the point at 0.15 GeV/c ($y_t \approx 1$) where hard-component contributions are small and values for $z_{si}(n_s)$ are inferred. In contrast to other hadron species variation of $z_{si}(n_s)$ for pions is small because the hard/soft ratio $\tilde z_{i}(n_s)$ is close to 1 (see Fig.~\ref{zrat}, left).  With the effective elimination of the resonance contribution the trend of $z_{si}(n_s)$ for pions is consistent with coefficient $z_{0i} \approx 0.82$. 

%%%%%%%%%%
%%%%%%%%%%%%

Figure~\ref{pionx} (right) shows spectrum data corresponding to the expression on the left of Eq.~(\ref{zhi}). The bold dashed curve is the TCM model function determined to best  coincide with data for the {\em most-central} ($n = 1$) event class. The same criterion is applied to all hadron species. The vertical bar indicates the corresponding model mode $\bar y_t = 2.46$ included in Table~\ref{pidparamsx}. Coefficients $z_{hi}(n_s)$ are inferred from data values at the mode.  

\begin{figure}[h]
	\includegraphics[width=3.3in]{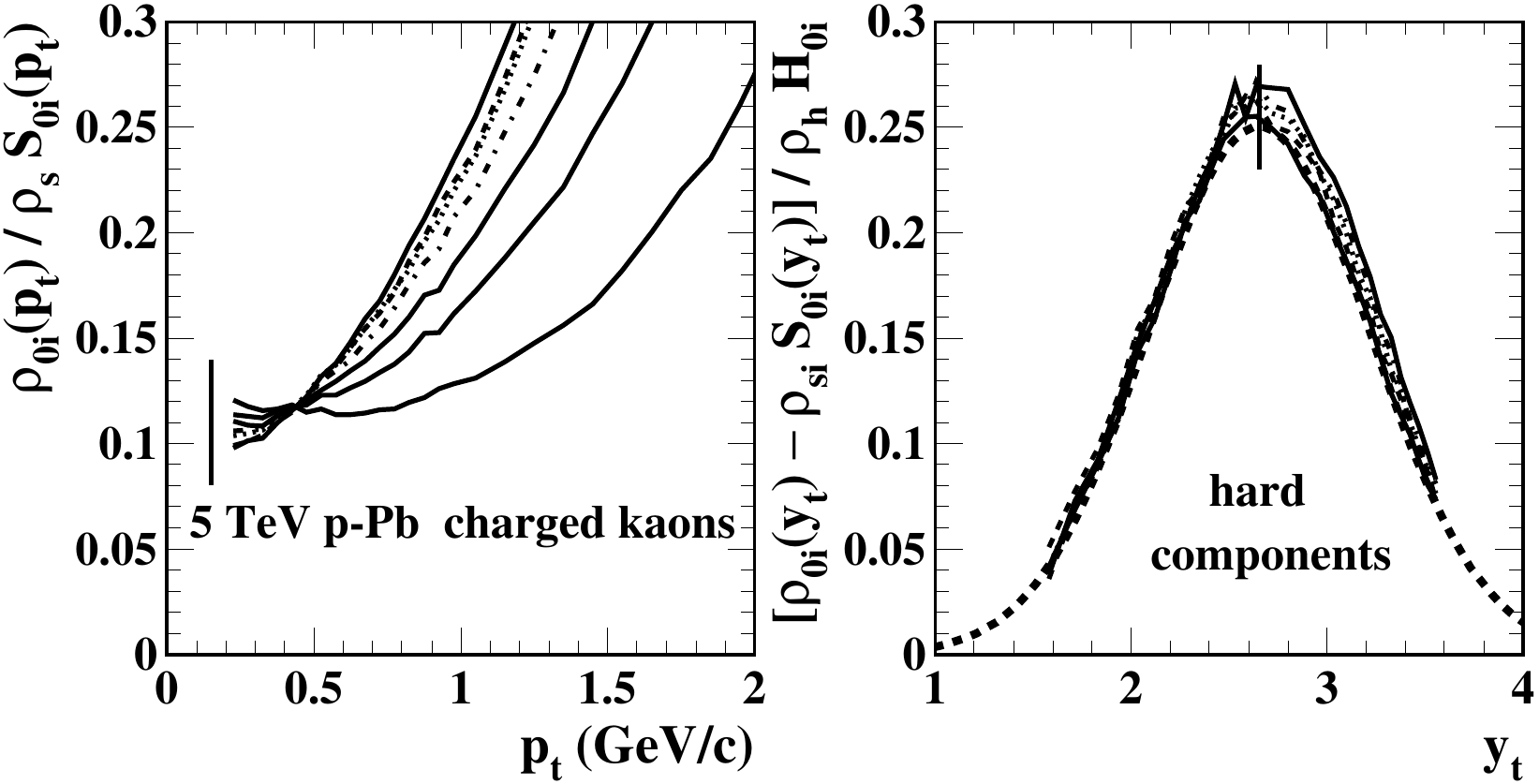}
\put(-202,105) {\bf (a)}
\put(-23,105) {\bf (b)}\\
	\includegraphics[width=3.3in]{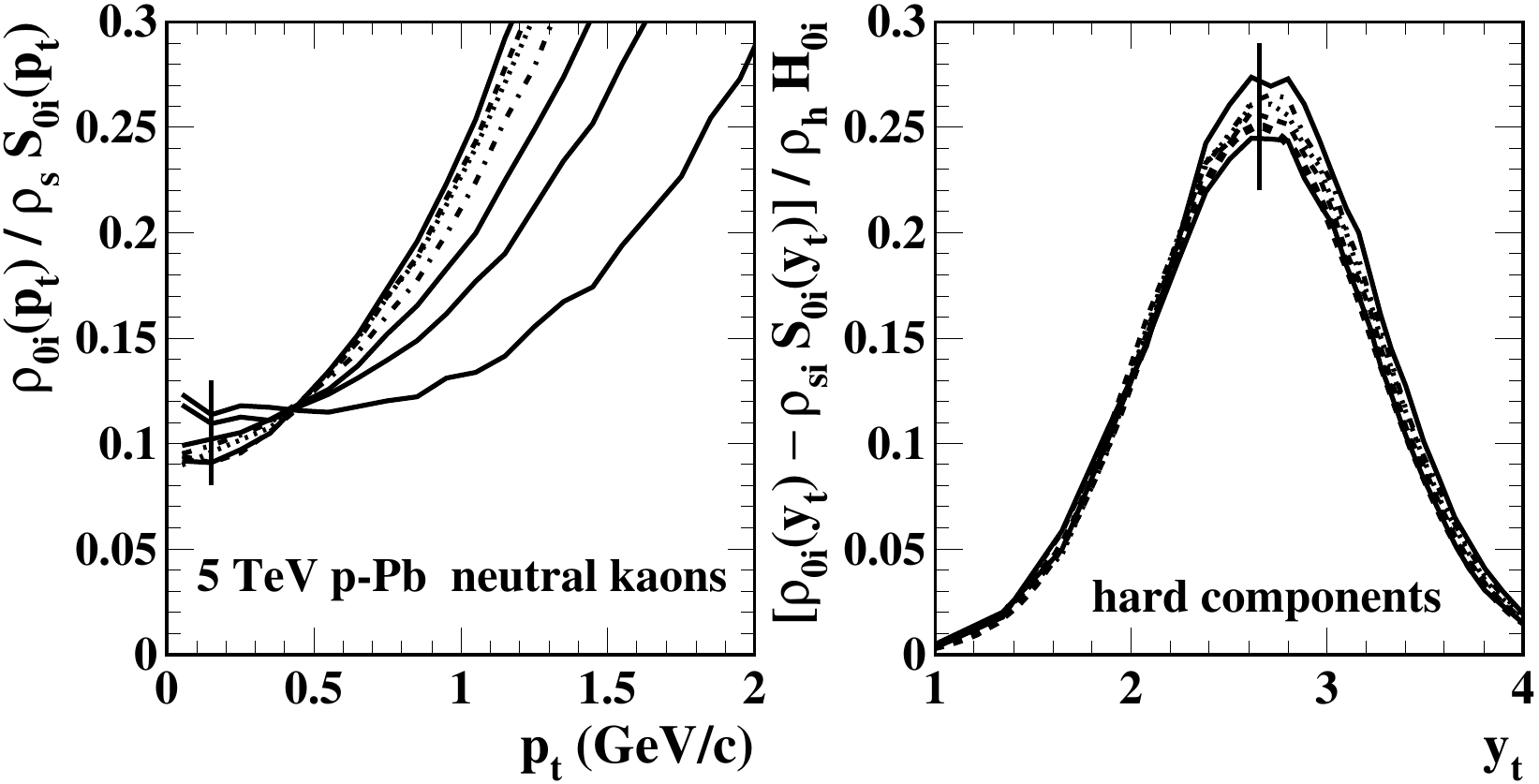}
\put(-202,105) {\bf (c)}
\put(-23,105) {\bf (d)}\\
	\caption{\label{kchx}
Spectrum data for $K^\pm$ (a), (b) and $K^0_\text{S}$ (c), (d) processed as for pions in Fig.~\ref{pionx}. The $K^0_\text{S}$ data are multiplied by 2 for direct comparison with $K^\pm$ data.
	}  %  alice630aa2, 640aa2
\end{figure}

Figure~\ref{kchx} shows results for charged and neutral kaons carried out with the same procedure. The two kaon species are treated identically after a factor-two scale-up of the neutral-kaon spectra. Because of the reduced low-\pt\ acceptance for charged kaons in panel (a) $z_{s}(n_s)$ values estimated from $K_\text{S}^0$ data are also applied to charged kaons. The excellent consistency of extracted hard components at low \pt\ (b,d) arises from accurate determination of $z_{si}(n_s)$ for each event class in (c). Evaluation of $z_{si}(n_s)$ for $K_\text{S}^0$ at 0.15 GeV/c ($y_t \approx 1$) insures a negligible contribution from the hard component in (d).

%%%%%%%%%%
%%%%%%%%%%%%

\begin{figure}[h]
	\includegraphics[width=3.3in]{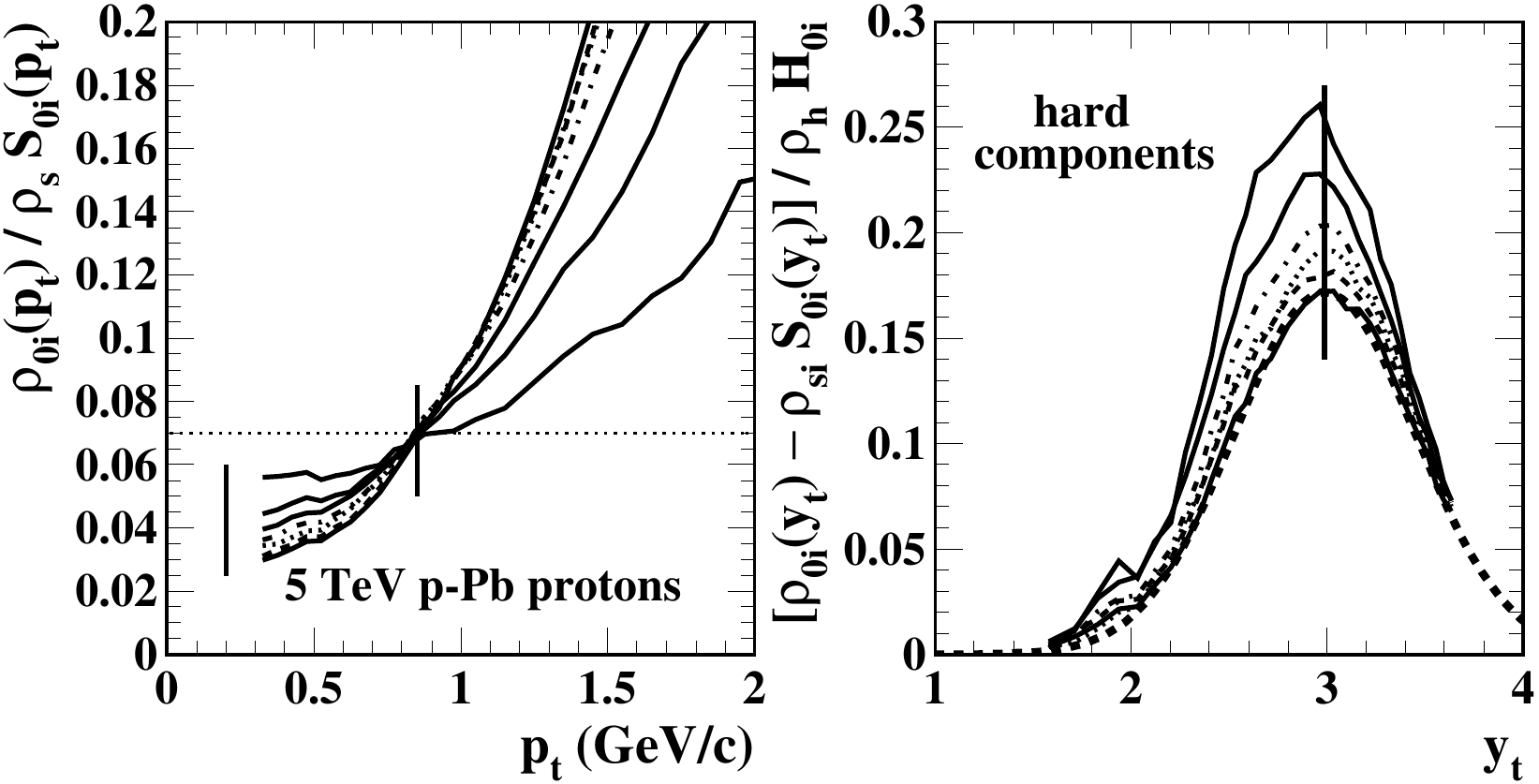}
	\put(-202,105) {\bf (a)}
	\put(-23,105) {\bf (b)}\\
	\includegraphics[width=3.3in]{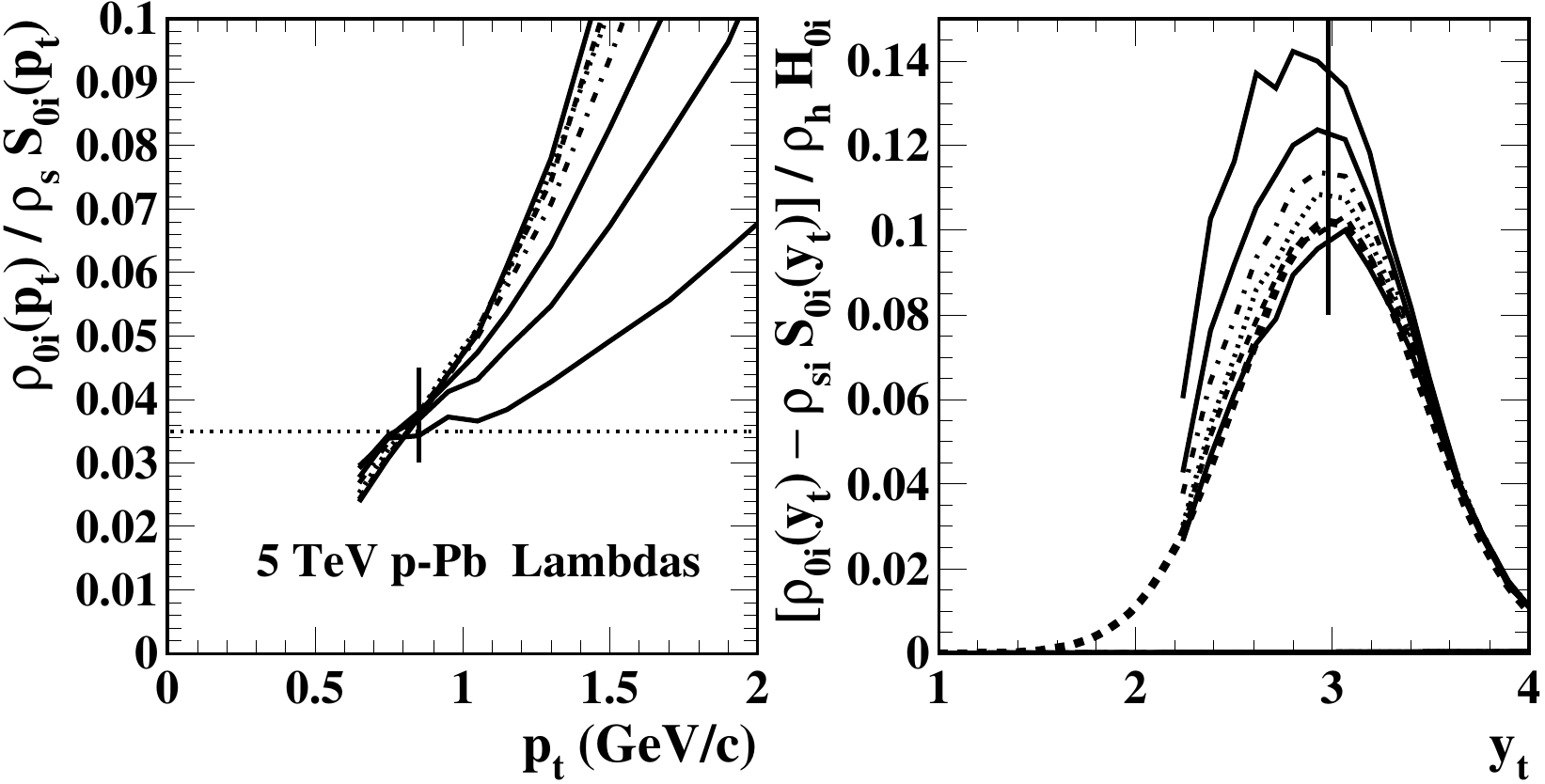}
	\put(-202,105) {\bf (c)}
	\put(-23,105) {\bf (d)}\\
	\caption{\label{protonxxx}
		Spectrum data for corrected protons (a), (b) and Lambdas (c), (d) processed as for pions in Fig.~\ref{pionx}. The vertical scales differ by factor 2 to better determine the relation between proton and Lambda $z_{xi}(n_s)$ coefficients.
	}  %  alice610aa2, 620aa2
\end{figure}

Figure~\ref{protonxxx} (a,c) shows proton and Lambda spectra from seven centrality classes of 5 TeV \ppb\ collisions in relation to the expression on the left of Eq.~(\ref{zsi}). The left vertical bar for protons indicates the point at 0.2 GeV/c where hard-component contribution is negligible and where values for $z_{si}(n_s)$ are inferred (proton data trends are extrapolated down to 0.2 GeV/c). Proton data have been corrected as described in Sec.~\ref{ineff}, but the correction does not affect $z_{si}$ inferred {\em below} 0.6 GeV/c. 
Lambda spectra do not extend low enough to obtain meaningful estimates for $z_{si}$. Based on comparison of proton and Lambda data in panels (a) and (c) at $p_t = 0.85$ GeV/c (vertical lines) relative to horizontal reference lines (dotted) at 0.035 and 0.07 proton values for $z_{si}$ are divided by 2 to obtain values for Lambdas.

%%%%%%%%%%
%%%%%%%%%%%%

Figure~\ref{protonxxx} (b,d) shows proton and Lambda spectrum hard components in relation to the expression on the left of Eq.~(\ref{zhi}). The vertical bars indicate the average model mode $\bar y_t \approx 2.99$. The detailed procedure for estimation of $z_{hi}(n_s)$ from data is as described above. In these panels hard components for six event classes are presented because the hard/soft ratio is large for baryons. For other hadron species only five event classes are plotted.

It is notable that baryon hard components fall rapidly below their modes and are negligible below \yt\ = 1.5 ($p_t \approx 0.30$ GeV/c) whereas meson hard components are non-negligible at \yt = 1. However, baryon hard components comprise a much larger fraction of their full spectra than mesons. Estimation of $z_{si}(n_s)$ is still nontrivial given the need to extrapolate proton spectra down to 0.2 GeV/c.

Tables~\ref{zsixx} and \ref{zhixx} show the resulting $z_{si}(n_s)$ and $z_{hi}(n_s)$ values respectively from differential spectrum analysis.

%%%%%%%%%%%%%%%%%%%%%%%%%%%%%%%%%%
\begin{table}[h]
	\caption{	\label{zsixx}
		Soft-component fractions $z_{si}$ for five hadron species from seven centralities of 5 TeV \ppb\ collisions. The proton $z_{si}$ are extrapolated down to 0.2 GeV/c.
		$^*$Note: $z_{si}$ estimates for charged kaons and Lambdas are uncertain due to limited \pt\ acceptance. Charged-kaon values are obtained from neutral-kaon data. Lambda values are obtained from proton values per low-\pt\ spectrum data in Fig.~\ref{protonxxx} (a,c).
	}

	%%%%%%%%%%%%%%%%%%%%%%%%%%%%%%%
	\begin{center}
		\begin{tabular}{|c|c|c|c|c|c|} \hline
			%\multicolumn{8}{|c|}{\pp\ multiplicity classes} \\ \hline
			$n$	& $ \pi^\pm $  & 	$K^{\pm*}$  & 	$p$  & 	$2K_\text{S}^0$ & 	$\Lambda^*$  \\ \hline
			1  & $0.837$  & $0.091$ & $0.028$ & $0.091$ & $0.014$  \\ \hline
			2   & $0.845$  & $0.091$ & $0.029$ & $0.091$ & $0.014$  \\ \hline
			3   & $0.866$  & $0.096$ & $0.033$ & $0.096$ & $0.016$  \\ \hline
			4   & $0.874$  & $0.099$ & $0.035$ & $0.099$ & $0.017$  \\ \hline
			5   & $0.868$  & $0.102$ & $0.038$ & $0.102$ & $0.019$  \\ \hline
			6  & $0.861$  & $0.110$ & $0.044$ & $0.110$ & $0.022$  \\ \hline
			7    & $0.832$  & $0.113$ & $0.053$ & $0.113$ & $0.027$  \\ \hline
		\end{tabular}
	\end{center}
	%%%%%%%%%%%%%%%%%%%%%%%%%%%%%%%
\end{table}
%%%%%%%%%%%%%%%%%%%%%%%%%%%%%%%

%%%%%%%%%%%%%%%%%%%%%%%%%%%%%%%%%%
\begin{table}[h]
	\caption{Hard-component fractions $z_{hi}$ for five hadron species from seven centralities of 5 TeV \ppb\ collisions. The most-peripheral \ppb\ centrality class ($n$ = 7) has a mean charge density less than that for non-single-diffractive \pp\ collisions. The spectrum hard components are  strongly biased. 
	}
	\label{zhixx}
	%%%%%%%%%%%%%%%%%%%%%%%%%%%%%%%
	\begin{center}
		\begin{tabular}{|c|c|c|c|c|c|} \hline
			%\multicolumn{8}{|c|}{\pp\ multiplicity classes} \\ \hline
			$n$	& $ \pi^\pm $  & 	$K^\pm$  & 	$p$  & 	$2K_\text{S}^0$ & 	$\Lambda$  \\ \hline
			1  & $ 0.73$  & $0.255 $ & $ 0.17 $ & $0.244 $ & $ 0.097 $  \\ \hline
			2  & $ 0.73$  & $ 0.262$ & $ 0.18 $ & $ 0.254$ & $ 0.100 $  \\ \hline
			3  & $ 0.76 $  & $ 0.265 $ & $ 0.19 $ & $ 0.261$ & $ 0.108 $  \\ \hline
			4  & $0.75 $  & $ 0.271$ & $ 0.20 $ & $ 0.266$ & $ 0.113 $  \\ \hline
			5  & $ 0.74$  & $0.270 $ & $0.23  $ & $0.269 $ & $ 0.124 $  \\ \hline
			6  & $ 0.70$  & $ 0.27 $ & $ 0.26  $ & $0.27 $ & $ 0.142 $  \\ \hline
			7  & $- $  & $ - $ & $ 0.29  $ & $- $ & $ 0.158 $  \\ \hline
			%7  & $$  & $$ & $$ & $$ & $$  \\ \hline
		\end{tabular}
	\end{center}
	%%%%%%%%%%%%%%%%%%%%%%%%%%%%%%%
\end{table}
%%%%%%%%%%%%%%%%%%%%%%%%%%%%%%%

%%%%%%%%%%%%%%%%
\section{$\bf z_{si}(n_s)$ and $\bf z_{hi}(n_s)$ $\bf vs$ $\bf p$-$\bf Pb$ centrality} \label{zcentt}

In this section the relation between TCM predictions and inferred $z_{si}(n_s)$ and $z_{hi}(n_s)$ data trends is examined. 
Given measured values for $z_{si}(n_s)$ and $z_{hi}(n_s)$ inferred above from differential spectrum analysis a comprehensive TCM description can be developed as follows: The centrality dependence of ratios $\tilde z_{i}(n_s)$ is inferred from $z_{si}(n_s)$ and $z_{hi}(n_s)$ data. A simple parametrization of $\tilde z_{i}(n_s)$ centrality trends for each hadron species is obtained. Those parametrizations are then applied to Eqs.~(\ref{zsinew}) and (\ref{zhinew}) to provide self-consistent TCM expressions for coefficients $z_{si}(n_s)$ and $z_{hi}(n_s)$ in Eqs.~(\ref{rhosi}).

The spectrum TCM defined by Eq.~(\ref{pidspectcm}) assumes that soft and hard components are approximately factorizable on variables $(y_t,n_s)$. In general, the $z_{xi}(n_s)$ are then independent of \yt\ assuming model functions $\hat X_{0i}(y_t)$ are approximately independent of centrality. That assumption is relaxed in Sec.~\ref{newdetails} wherein $z_{hi}(n_s) \rightarrow z_{hi}(y_t,n_s)$.  An explicit expression for $z_{si}$ appearing in Eq.~(\ref{rhosi}) is
\bea \label{zsinew}
z_{si}(n_s) &=& \left[\frac{1 + x(n_s) \nu(n_s)}{1 + \tilde z_{i}(n_s) x(n_s) \nu(n_s)} \right]  {z_{0i}}.
\eea
The corresponding expression for $z_{hi}$ is
\bea \label{zhinew}
z_{hi}(n_s) &=& \tilde z_{i}(n_s) z_{si}(n_s).
\eea
Given the structure of Eq.~(\ref{pidspectcm}) and definitions in Eq.~(\ref{rhosi}) the two PID coefficients should satisfy the following sum rules for sums over charged-hadron species $i$:
\bea \label{sums}
\sum_i z_{si}(n_s) &\approx & 1~\text{and}~ \sum_i z_{hi}(n_s) \approx  1, ~
\eea
and for each species $\bar \rho_{si} + \bar \rho_{hi} = \bar \rho_{0i} = z_{0i} \bar \rho_0$. Given Eq.~(\ref{zsinew})  $\sum_i z_{0i} \approx  1$ is an implicit consequence of the first relation above. Those sum rules then provide a test of TCM self consistency. 

\subsection{$\bf z_{si}(n_s)$ and $\bf z_{hi}(n_s)$ centrality trends} \label{zcent}

\begin{figure}[h]
	\includegraphics[width=3.3in]{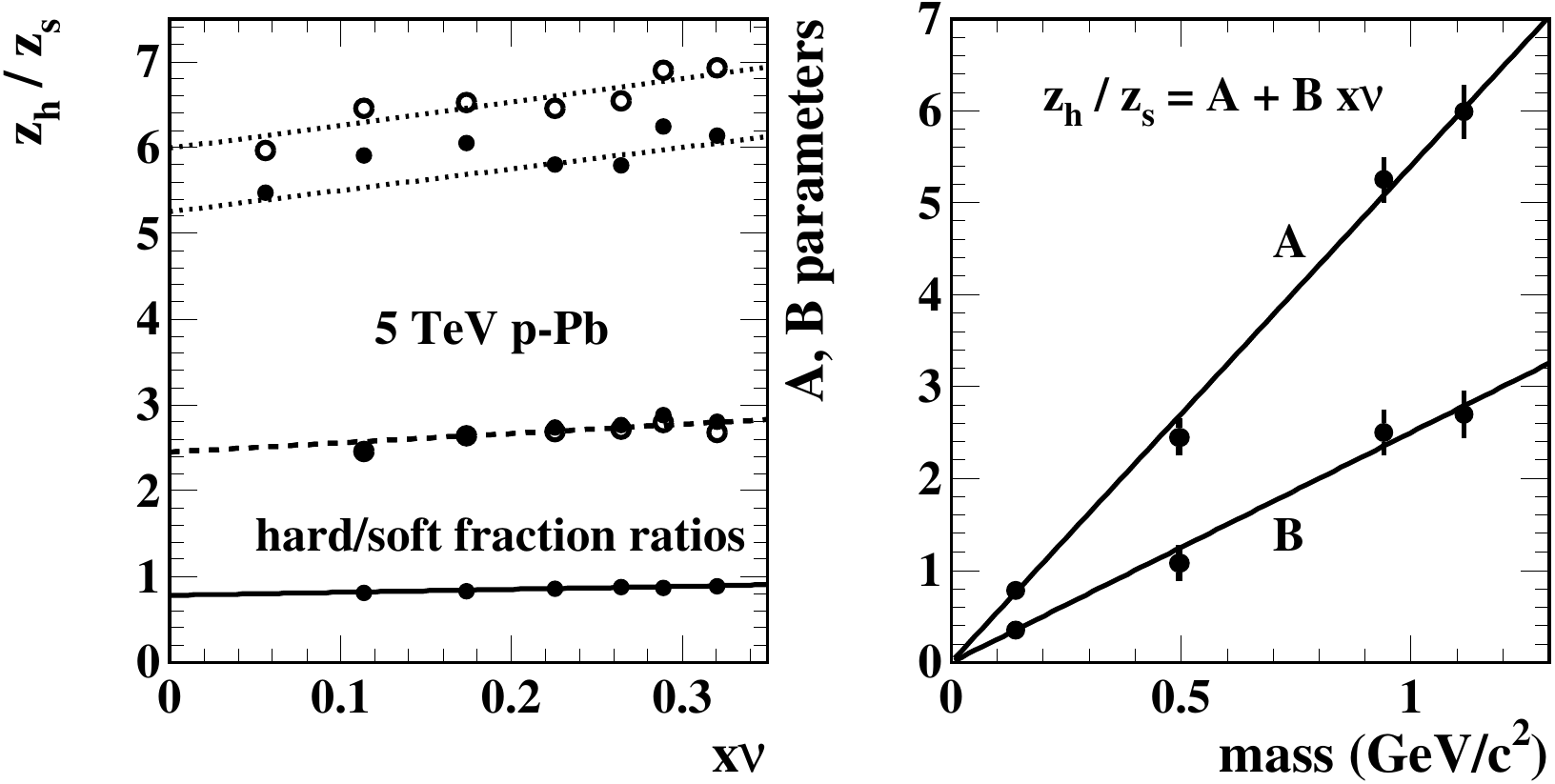}
	\caption{\label{zrat}
		Left: Ratios $z_{hi}/z_{si} = \tilde z_{i}(n_s)$ inferred from $z_{si}$ and $z_{hi}$ entries in Tables~\ref{zsixx} and \ref{zhixx} for charged (solid dots) and neutral (open circles) hadrons. The lines are linear parametrizations that describe the ratio data: solid, dashed and dotted for pions, kaons and baryons respectively.
		Right: Coefficients $A$ ($y$ intercept) and $B$ (slopes) for linear descriptions of ratio data in the left panel plotted vs hadron mass. The lines provide a visual reference.
	} %   alice680e
\end{figure}

Figure~\ref{zrat} (left) shows ratios $\tilde z_{i}(n_s) = z_{hi}(n_s)/z_{si}(n_s)$ (points) inferred from $z_{si}(n_s)$ and $z_{hi}(n_s)$ data in Tables~\ref{zsixx} and \ref{zhixx} for charged hadrons (solid dots) and neutral hadrons (open circles) obtained from spectrum data in the previous section, ranging from pions at the bottom to Lambdas at the top. Significant ratio variation with \ppb\ centrality, increasingly for more-massive hadrons, contradicts an important assumption in Ref.~\cite{ppbpid}. The solid, dashed and dotted lines corresponding to pions, kaons and baryons  are linear fits by eye to ratio data of the form $\tilde z_{i}(n_s) = A + B \,x\nu$ with $A = 0.78, 2.45, 5.25$ and 6.00 and $B = 0.35, 1.08, 2.5$ and 2.7 for pions, kaons, protons and Lambdas respectively. 

%%%%%%%%%%
%%%%%%%%%%%%

Figure~\ref{zrat} (right) shows values $A$  and $B$ (points) plotted vs mass for the lines in the left panel. Parameters $A$ and $B$ approximately follow linear trends on hadron mass. The solid lines with slopes 5.4 and 2.5 provide a linear reference. Those parametrizations of the $\tilde z_{i}(n_s)$ are then applied to  Eqs.~(\ref{zsinew}) and (\ref{zhinew}) to generate TCM curves describing $z_{si}(n_s)$ and $z_{hi}(n_s)$ data trends as explained below. The linear trends in the left panel are those referred to in Sec.~\ref{newoldspec}.
 
\begin{figure}[h]
	\includegraphics[width=3.3in]{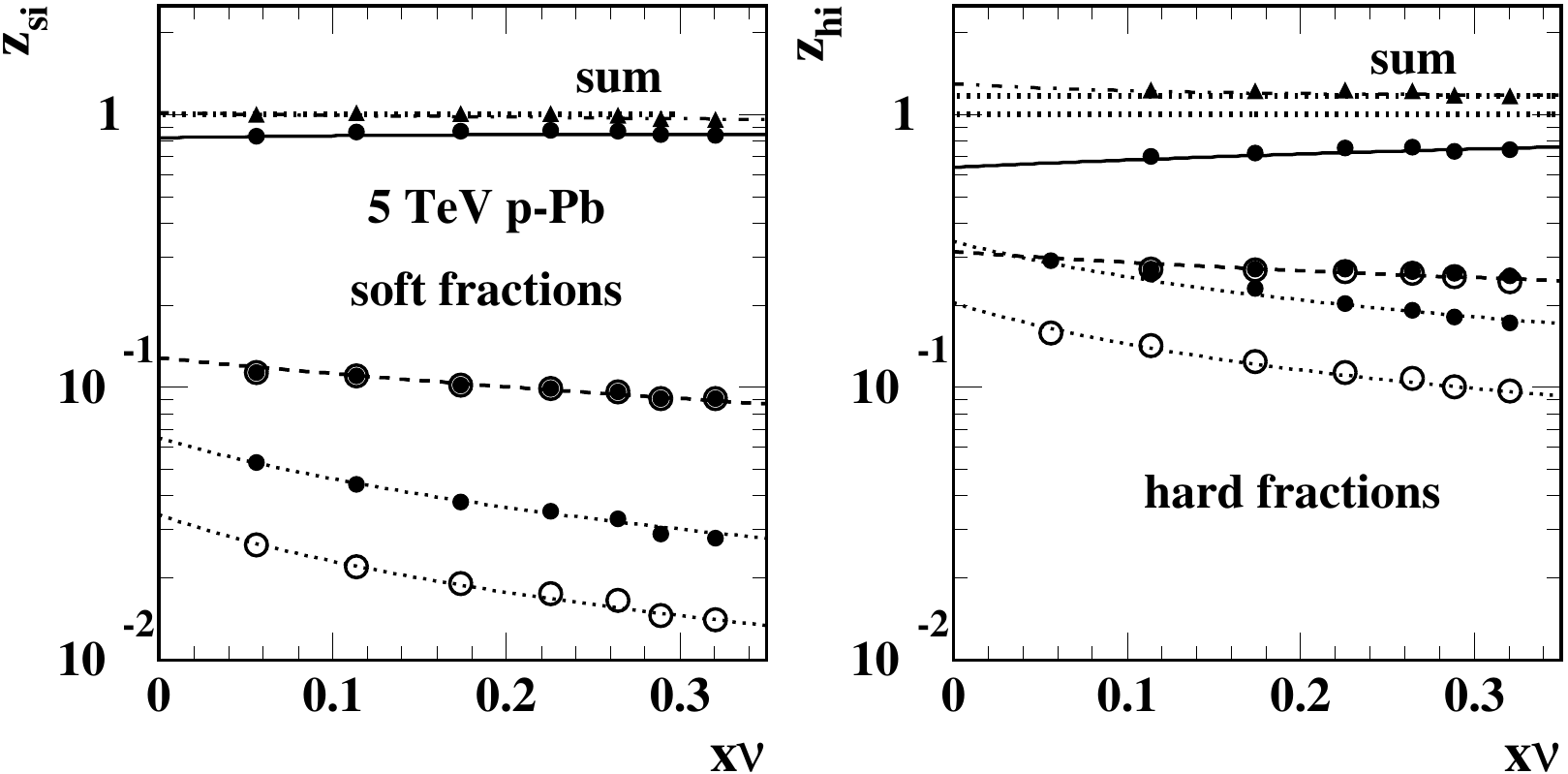}
	\caption{\label{zshx}
		Left: Soft-component PID fractions $z_{si}(n_s)$ (points) from Table~\ref{zsixx} for pions, kaons, protons and Lambdas from top to bottom. Charged particles are solid dots, neutral particles are open circles. The curves are Eq.~(\ref{zsinew}) with values for $\tilde z_i(n_s)$ from Fig.~\ref{zrat} (left) and values for $z_{0i}$ from Table~\ref{pidparamsx}. Triangles and dash-dotted curve represent the first sum rule in Eq.~(\ref{sums}).
	Right:  Hard-component PID fractions $z_{hi}(n_s)$ (points) from Table~\ref{zhixx} as in the left panel. The curves are Eq.~(\ref{zhinew}).  Triangles and dash-dotted curve represent the second sum rule in Eq.~(\ref{sums}). The Lambda values at left are simply scaled from the proton values. The proton values at right rely on the correction procedure in Sec.~\ref{ineff}.
	}  %  alice680c
\end{figure}

Figure~\ref{zshx} (left) shows soft-component PID fractions $z_{si}(n_s)$ (points) from Table~\ref{zsixx} for charged hadrons (solid dots) and neutral hadrons (open circles). As noted, the charged-kaon values are neutral-kaon values multiplied by 2, and the Lambda values are proton values divided by  2 (the latter factor obtained by comparing proton and Lambda spectra  at low \pt). The curves represent Eq.~(\ref{zsinew})  with $\tilde z_{i}(n_s)$ ratio trends corresponding to the lines in Fig.~\ref{zrat} (left). The solid triangles and dash-dotted curve at the top are the sums of charged-hadron points and curves showing good agreement with the expected sum rule. Asymptotic values of $z_{si}(n_s)$ for $x\nu \rightarrow 0$ are $z_{0i} = 0.82, 0.128, 0.064, 0.065$, and 0.034 for pions, charged kaons, neutral kaons, protons and Lambdas respectively (see Table~\ref{pidparamsx}). Those values agree within uncertainties with the entries in Table~\ref{otherparams} (from Ref.~\cite{ppbpid}).

%%%%%%%%%%
%%%%%%%%%%%%

Figure~\ref{zshx} (right) shows hard-component PID fractions $z_{hi}(n_s)$ (points) from Table~\ref{zhixx}. The curves are the curves in the left panel multiplied by linear trends for ratios $\tilde z_{i}(n_s)$ shown in Fig.~\ref{zrat} (left).  The neutral-kaon data and curve are multiplied by 2 for precise comparison. The upper solid triangles and dash-dotted curve represent sums of points and curves for charged hadrons. The $z_{hi}(n_s)$ data derived from differential analysis of (corrected) PID spectra are generally well described by the revised TCM.

The overall scale of the $z_{si}(n_s)$ TCM (and hence consistency with the expected sum rule) is determined by $\bar \rho_s$ in Eq.~(\ref{zsi}) which in turn depends on values in Table~\ref{rppbdata} derived from \pp\ spectrum data as in Ref.~\cite{alicetomspec} and \mmpt\ data as in Ref.~\cite{tommpt}. The overall scale of the measured $z_{hi}(n_s)$ depends on the same parameters but in addition depends on the value of $\alpha(\sqrt{s})$ used to determine $\bar \rho_h = N_{bin} \alpha \bar \rho_{sNN}^2$ in Eq.~(\ref{zhi}). The nominal value for 5 TeV \nn\ collisions is 0.0113 per Eq.~(15) of Ref.~\cite{alicetomspec} $\rightarrow 0.0127$ for the present analysis. The $\alpha(\sqrt{s})$ trend reported in Ref.~\cite{alicetomspec}, derived from nonPID \pp\ spectra over a broad range of collision energies, has an estimated 10\% uncertainty. The sums in Fig.~\ref{zshx} (right) are about 10\% high (dotted line above 1).

\subsection{Are $\bf z_{si}$ and $\bf z_{hi}$ coefficients self-consistent?}

Based on the results above the complete system of $z_{si}(n_s)$ and $z_{hi}(n_s)$ coefficients for 5 TeV \ppb\ collisions is effectively determined by two parameters: the slopes of mass trends in Fig.~\ref{zrat} (right). Those trends determine the linear trends for $\tilde z_i(n_s)$ in Fig.~\ref{zrat} (left) which in turn determine $z_{si}(n_s)$ and $z_{hi}(n_s)$ per Eqs.~(\ref{zsinew}) and (\ref{zhinew}). Also required are quantities $z_{0i}$ that determine the overall hadron species abundances and can presumably be compared with statistical-model predictions. Initially, $\tilde z_i(n_s)$ and $z_{0i}$ had been assumed independent of \ppb\ centrality, but that assumption can now be tested by assuming some degree of centrality dependence $z_{0i}(n_s)$ and attempting to estimate it via PID spectrum data.

The relation $\bar \rho_{0i} \equiv z_{0i} \bar \rho_0$ was introduced just below Eq.~(\ref{rhosi}). On that basis $z_{0i}$ appears as a fixed coefficient in Eq.~(\ref{zsinew}), but integrating Eq.~(\ref{pidspectcm}) over \yt\ results in
\bea \label{}
 z_{0i}(n_s) (\bar \rho_{s} +\bar \rho_{h} )
&=& z_{si}(n_s)  \bar \rho_{s} +  z_{hi}(n_s)  \bar \rho_{h}.
\eea
Dividing through by nonPID soft-component density $\bar \rho_s$ and some rearrangement leads to
\bea
 z_{0i}(n_s) -z_{si}(n_s) &=& [z_{hi}(n_s) - z_{0i}(n_s)] x \nu.
\eea
That relation makes clear why measurement of $z_{si}(n_s)$ at lower \yt\ leads to quantitative predictions of $z_{hi}(n_s)$ at higher \yt\ {\em assuming that $z_{0i}$ is constant}: Variation of $z_{si}(n_s)$ relative to $z_{0i}$ depends on details of jet production as measured by basic hard/soft ratio $x \nu$ and $z_{hi}(n_s)$. That is just the argument presented in Sec.~\ref{pidfracdata} to justify estimation of a proton detection inefficiency at higher \pt\ by inferring ratio $\tilde z_i = z_{hi}(n_s) / z_{si}(n_s)$ at low \pt\ where there appears to be no inefficiency (also see Sec.~\ref{ineff}).

The separate coefficients $z_{si}(n_s)$ and $z_{hi}(n_s)$ are measured with minimal assumptions as reported in Sec.~\ref{zxspectra}. $z_{0i}(n_s)$ is then determined from measured coefficients by
\bea \label{z0ns}
 z_{0i}(n_s) &=& \frac{z_{si}(n_s)  +  z_{hi}(n_s)  x\nu}{1 + x\nu}.
\eea
The inferred $ z_{0i}(n_s)$ trends can in turn be compared with values $z_{0i}(0)$ evaluated as the asymptotic limits of measured $z_{si}(n_s)$ trends as $n_s \rightarrow 0$.

\begin{figure}[h]
	\includegraphics[width=1.62in]{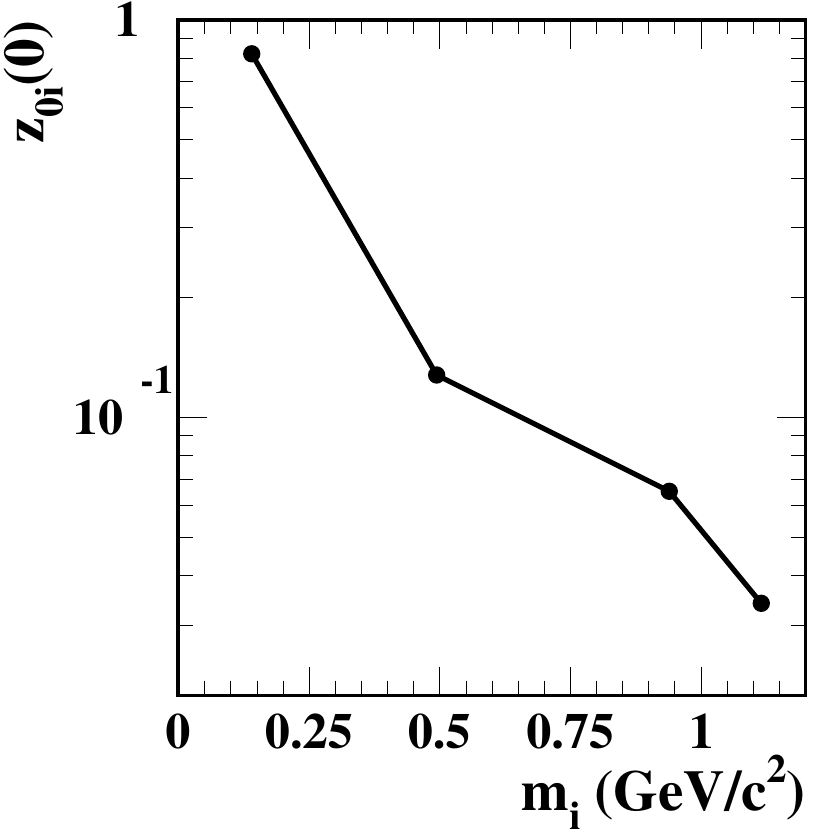}
	\includegraphics[width=1.67in]{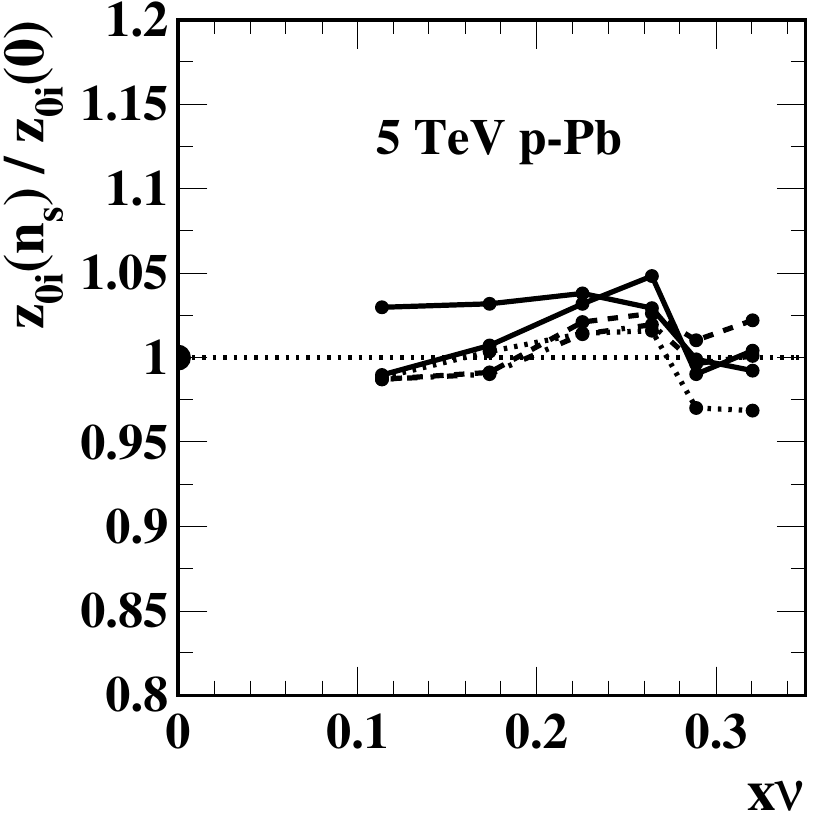}
	\caption{\label{vvv}
		Left: Asymptotic $z_{0i}$ values inferred from $z_{si}(n_s)$ trends in Fig.~\ref{zshx} (left) per Eq.~(\ref{zsinew}).
		Right: Ratio $z_{0i}(n_s) / z_{0i}(0)$ inferred from $z_{si}(n_s)$ and $z_{hi}(n_s)$ values as presented in Fig.~\ref{zshx} per Eq.~(\ref{z0ns}).
		}  %  alice680i
\end{figure}

Figure~\ref{vvv} (left) shows $z_{0i}(0)$ values inferred from $z_{si}(n_s)$ trends in Fig.~\ref{zshx} (left) (see text describing that panel) vs hadron mass $m_i$. Those asymptotic $z_{si}(n_s)$ (for $y_t \rightarrow 0$) limits are independent of jet production.

%%%%%%%%%%
%%%%%%%%%%%%

Figure~\ref{vvv} (right) shows ratios $z_{0i}(n_s) / z_{0i}(0)$. The line styles proceed through solid, dashed, dotted, dash-dotted, solid -- corresponding to pions, charged kaons, neutral kaons, protons, Lambdas. At the percent level those results indicate that the assumption $z_{0i}(n_s) \approx$ constant is consistent with PID spectrum data.

%%%%%%%%%%%%%%%
\section{$\bf H_i(y_t,n_s)$ shape evolution} \label{newdetails}

The structure of spectrum hard component $H_i(y_t,n_s)$ revealed by the format of Eq.~(\ref{zhi}) in Sec.~\ref{zxspectra} demonstrates that assumed hard-component factorization in Eq.~(\ref{factorize}) is strongly violated by all hadron species. Details can be investigated by initially assuming no factorization. Data are scaled only by nonPID charge densities $\bar \rho_s$ and $\bar \rho_h$ common to all hadron species. The results from Sec.~\ref{zxspectra} demonstrate that soft-component factorization remains consistent with data:
\bea \label{zsiz}
\lim_{y_t \to 0}\bar \rho_{0i}(y_t,n_s)  &\approx &  
 z_{si}(n_s) \bar \rho_s \hat S_{0i}(y_t).
\eea
However, accurately representing hard-component structure requires (as one possible strategy) modifying the definition of $z_{hi}$ to describe strong \yt\ dependence:
\bea \label{zhiz}
 \bar \rho_{0i}(y_t) -  z_{si}(n_s)\bar \rho_s \hat S_{0i}(y_t) &\approx& H_i(y_t,n_s)
  \\ \nonumber
  &\approx& z_{hi}(y_t,n_s)  \bar \rho_h  \hat H_{0i}( y_t).~~~~
\eea
The first line follows from the first line of Eq.~(\ref{pidspectcm}). However, the second line presents an alternative factorization of $H_i(y_t,n_s)$ such that modified coefficient $z_{hi}(y_t,n_s)$ retains all information carried by the data hard component beyond basic fixed TCM reference $\hat H_{0i}( y_t)$. The data hard component on the left is then divided by nonPID density $\bar \rho_h$ derived from Table~\ref{rppbdata} and a TCM model function (that must be defined) to obtain $z_{hi}(y_t,n_s)$ on the right. In what follows the centrality dependence of PID hard-component shapes, relative to fixed TCM model functions, is represented by revised quantity $z_{hi}(y_t,n_s)$.

\begin{figure}[h]
	\includegraphics[width=3.3in]{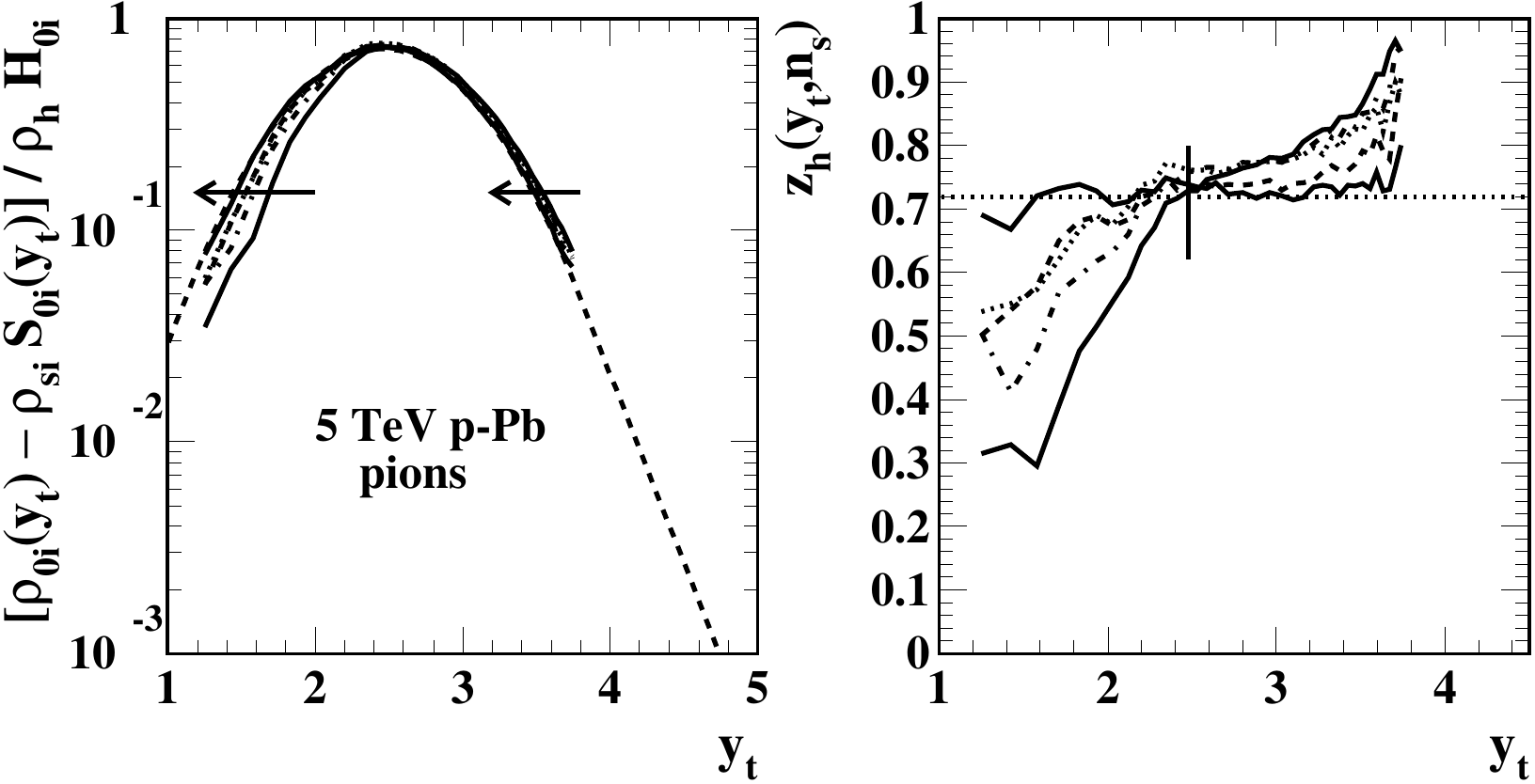}
	\caption{\label{pionaa4}
		Left: Pion spectrum hard components from Fig.~\ref{pionx} (right) plotted in a log-log format (log density vs \yt). The dashed curve corresponds to TCM model $\hat H_{0i}(y_t)$ adjusted to describe the most-central ($n = 1$) data.
		Right: Data in the left panel divided by model function $\hat H_{0i}(y_t)$ to reveal $z_{hi}(y_t,n_s)$ per Eq.~(\ref{zhiz}). Line styles in this and related figures proceed through solid, dashed, dotted, dash-dotted and then solid from most central to least central. The horizontal dotted line denotes the reference value $z_{hi}(\bar y_t,n_s)$ corresponding to most-central ($n = 1$) data.
	}  %  alice600aa4
\end{figure}

Figure~\ref{pionaa4} (left) shows pion hard components from Fig.~\ref{pionx} (right) plotted in a semilog format to better reveal structure in the tails. The arrows (in this and similar plots) indicate the direction of significant variation from peripheral to central \ppb\ collisions. Model function $\hat H_{0i}(y_t)$ (bold dashed) is not varied to match data evolution as in previous studies~\cite{alicetomspec,tomnewppspec}. Instead it is adjusted (for each hadron species) to best describe only the {\em most-central} data that are symmetric about the mode.

%%%%%%%%%%
%%%%%%%%%%%%

Figure~\ref{pionaa4} (right) shows coefficients $z_{hi}(y_t,n_s)$ vs \yt\ inferred from spectrum data and defined model $\hat H_{0i}(y_t)$ per Eq.~(\ref{zhiz}). The line styles proceed from most-central as solid, dashed, dotted and dash-dotted before returning to solid. The horizontal dotted line denotes the reference value $z_{hi}(\bar y_t,n_s)$ corresponding to $n = 1$ (most-central) data. The vertical line denotes the model mode. Values of $z_{hi}(y_t,n_s)$ at that point  are consistent with Table~\ref{zhixx} pion entries which indicate very little variation. However, it is apparent from Fig.~\ref{pionaa4} that there is strong {\em overall} variation of the pion hard component with \ppb\ centrality, with a tendency to shift the effective distribution centroid to {\em lower} \yt\ with increasing centrality.

\begin{figure}[h]
	\includegraphics[width=3.3in]{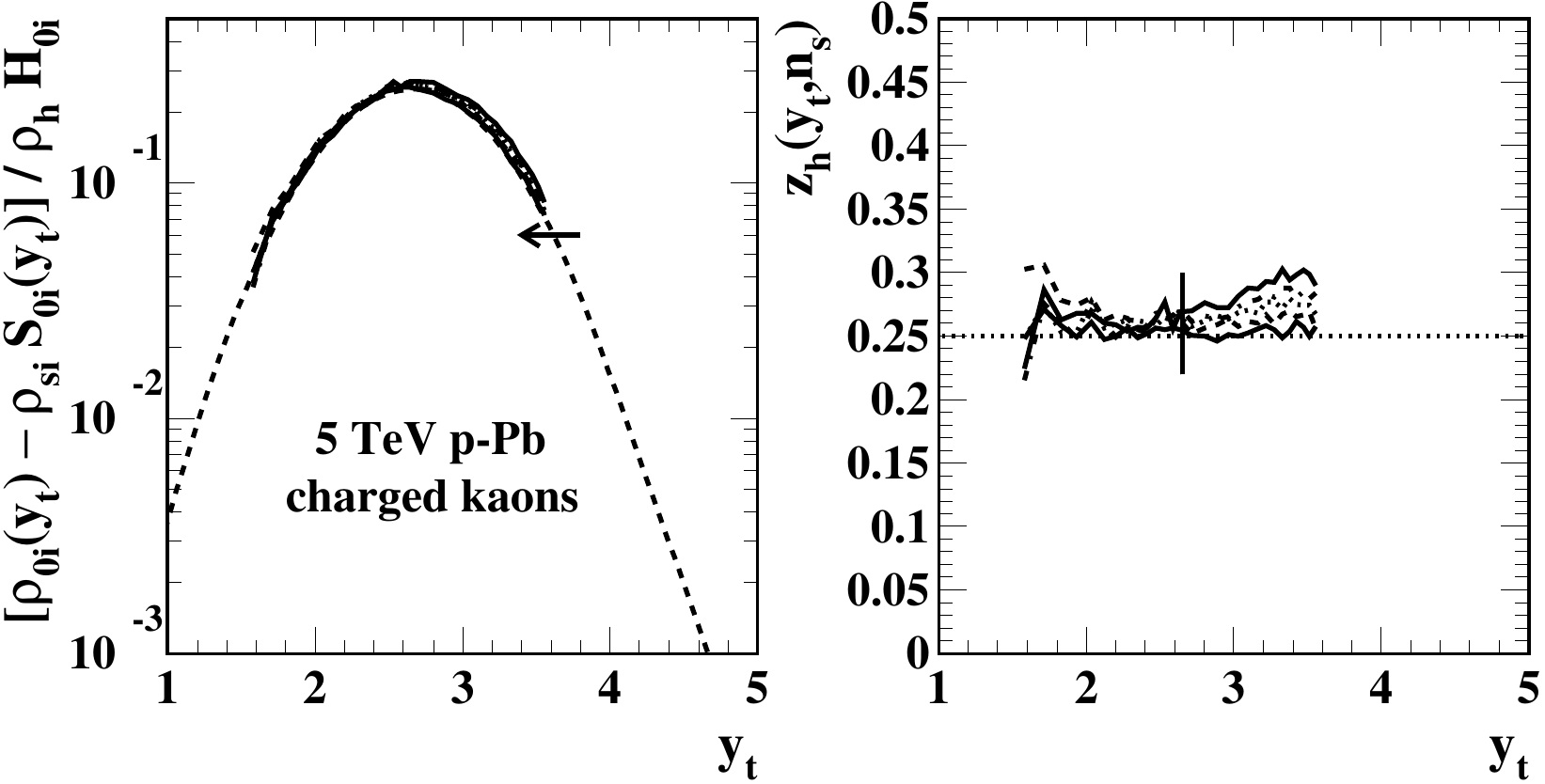}
	\put(-140,105) {\bf (a)}
\put(-23,105) {\bf (b)}\\
	\includegraphics[width=3.3in]{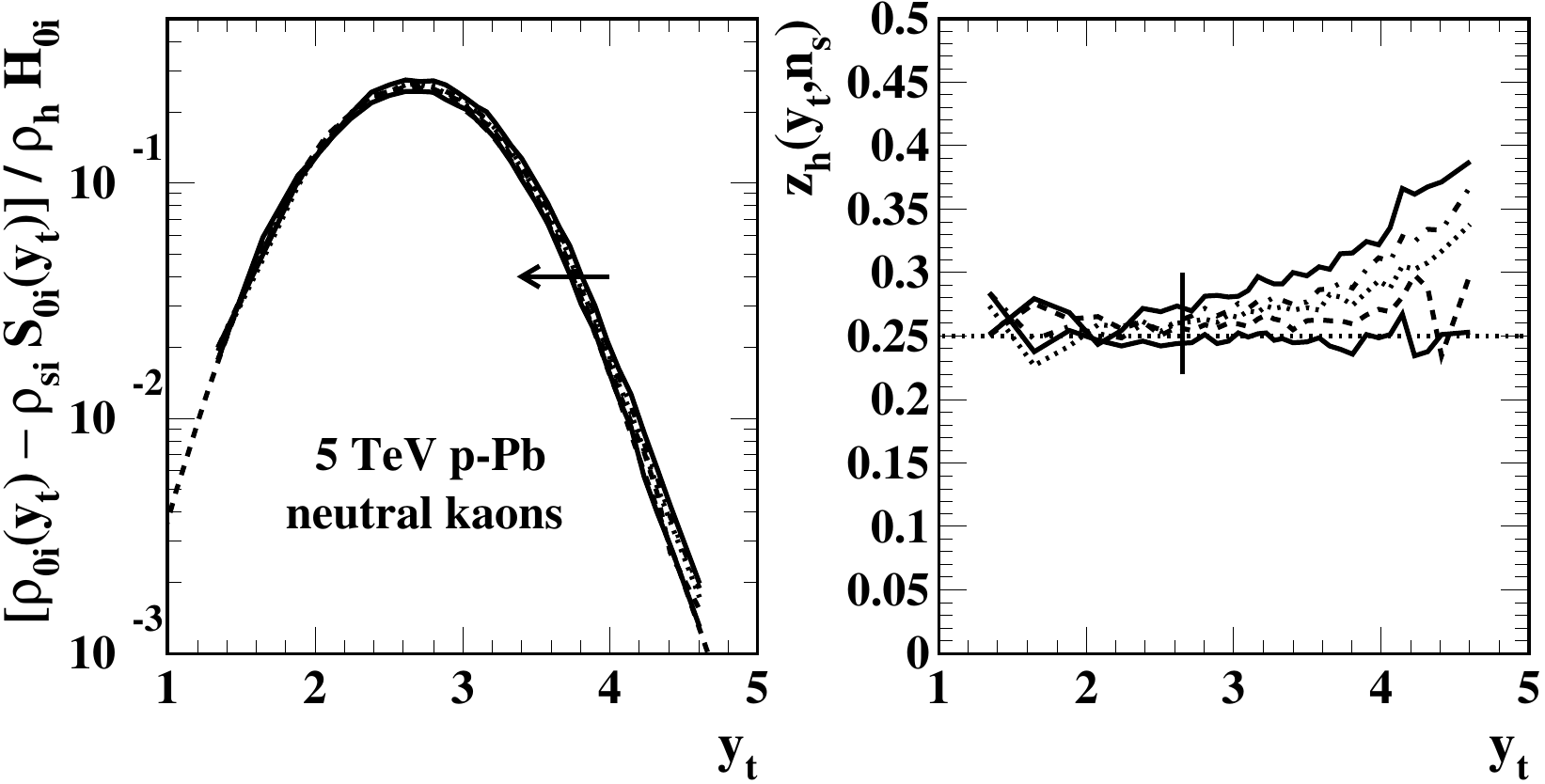}
	\put(-140,105) {\bf (c)}
	\put(-23,105) {\bf (d)}
	\caption{\label{kaonaa4}
		Left: Kaon spectrum hard components from Fig.~\ref{kchx}  plotted as log density vs \yt. The dashed curves correspond to TCM model $\hat H_{0i}(y_t)$ adjusted to describe the most-central data.	
	 Right: Data in the left panels divided by model functions $\hat H_{0i}(y_t)$ to reveal $z_{hi}(y_t,n_s)$ per Eq.~(\ref{zhiz}). 
 }  %  alice630aa4, 640aa4
\end{figure}

Figure~\ref{kaonaa4} (left) shows spectrum hard components for charged kaons (a) and neutral kaons (c) from Fig.~\ref{kchx} (right) plotted in a semilog format to better reveal structure in the tails. The distribution below the mode shows no significant evolution with centrality whereas the distribution above the mode does show significant variation.

%%%%%%%%%%
%%%%%%%%%%%%

Figure~\ref{kaonaa4} (right) shows coefficient $z_{hi}(y_t,n_s)$ vs \yt\ inferred from spectrum data per Eq.~(\ref{zhiz}). These results confirm that there is no significant difference between charged and neutral kaon spectra down to the statistical limits of data. Panel (c) especially demonstrates that the jet fragment distribution can be isolated precisely down to low \yt\ $\approx$ 1.3 ($p_t \approx 0.25$ GeV/c).  The spectrum hard component does in fact turn over below its mode and descend over more than a decade.

\begin{figure}[h]
	\includegraphics[width=3.3in]{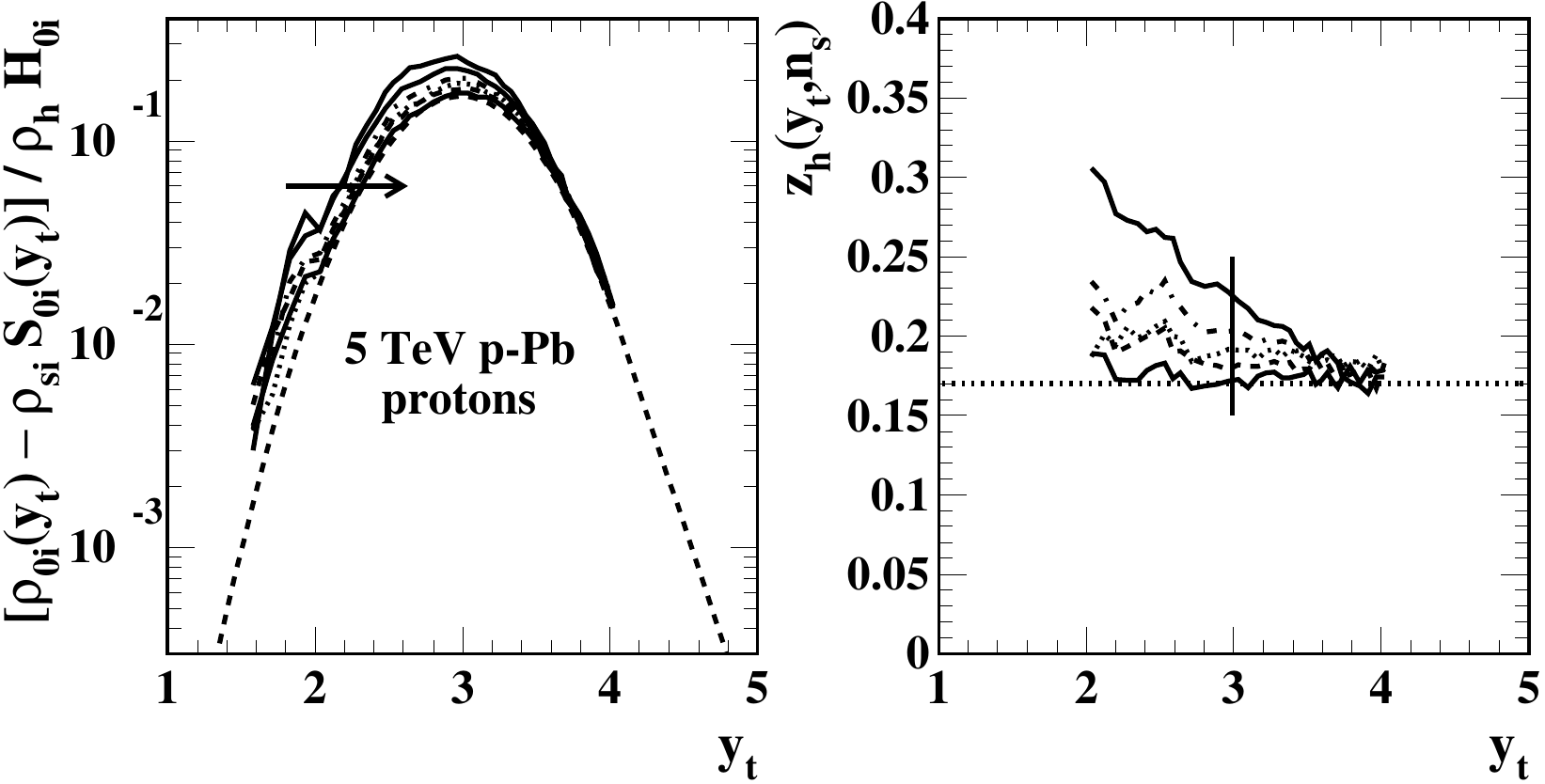}
	\put(-140,105) {\bf (a)}
\put(-23,105) {\bf (b)}\\
\includegraphics[width=3.3in]{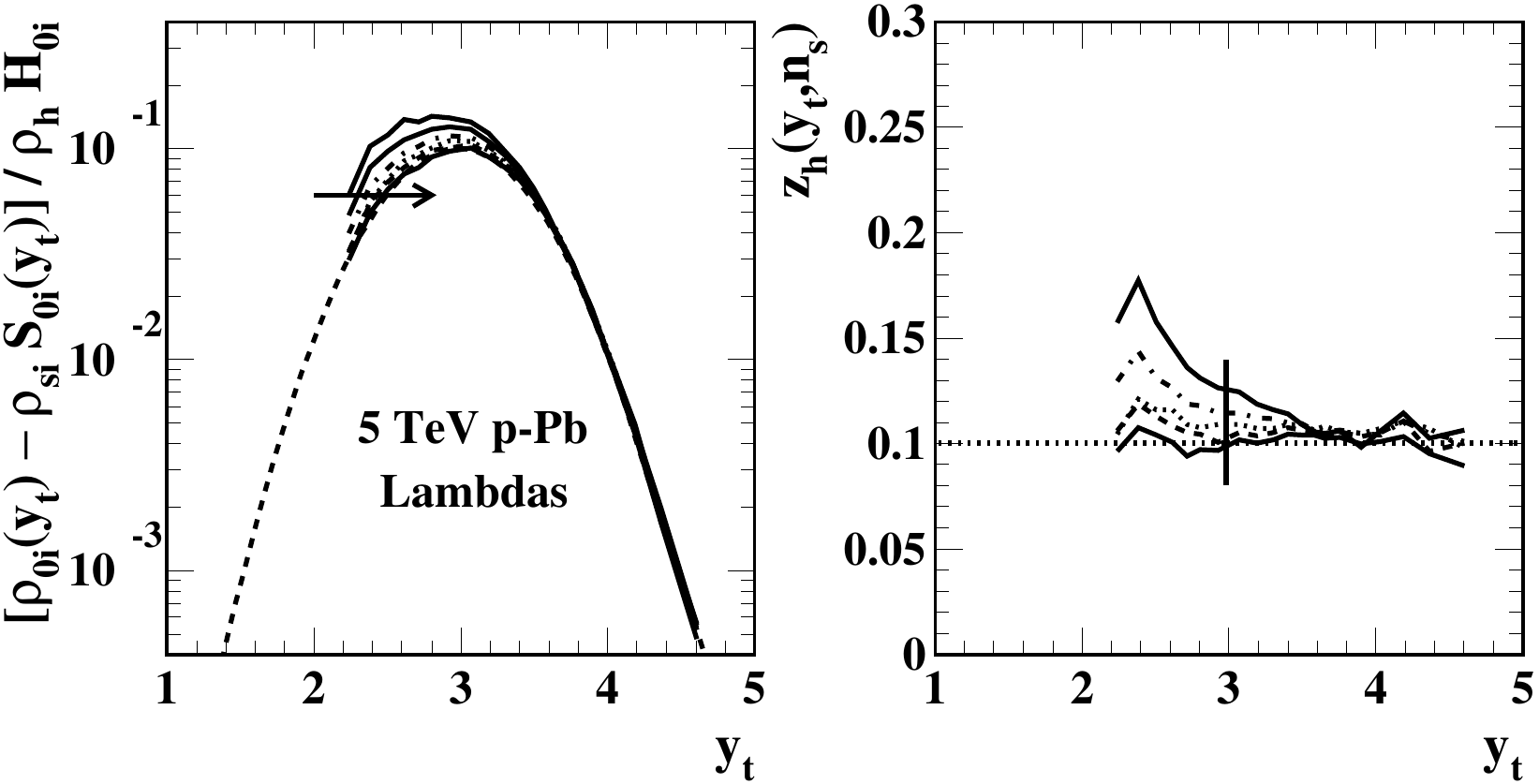}
	\put(-140,105) {\bf (c)}
	\put(-23,105) {\bf (d)}
	\caption{\label{baryonaa4}
		Left: Proton and Lambda spectrum hard components from Fig.~\ref{protonxxx}. The dashed curves correspond to TCM models $\hat H_{0i}(y_t)$ adjusted to describe the most-central data.	
		Right: Data in the left panels divided by model functions $\hat H_{0i}(y_t)$ to reveal $z_{hi}(y_t,n_s)$ per Eq.~(\ref{zhiz}). 	
}  %  alice610aa4, 620aa4
\end{figure}

Figure~\ref{baryonaa4} (left) shows spectrum hard components for protons (a) and Lambdas (c) from Fig.~\ref{protonxxx} (right) plotted in a semilog format to better reveal structure in the tails. In contrast to kaons the baryon hard components show strong variation with centrality below and near the mode, moving to {\em higher} \yt\ with increasing centrality, but the exponential tail at high \yt\ (for Lambdas at least) shows no significant variation with centrality.

%%%%%%%%%%
%%%%%%%%%%%%

Figure~\ref{baryonaa4} (right) shows  coefficient $z_{hi}(y_t,n_s)$ vs \yt\ inferred from spectrum data per Eq.~(\ref{zhiz}). It is clear from those details that strong variation at the mode decreases to negligible at or above $y_t = 3.75$ ($p_t \approx 3$ GeV/c). The Lambda data extend from 0.65 GeV/c up to 7 GeV/c.

It is important to note that these data deviations from fixed TCM model functions $\hat H_{0i}(y_t)$ are the {\em only deviations} of spectrum data from the TCM applied here as a {\em fixed reference}. The data-model deviations do not reflect a deficiency of the model. Instead they reveal important new aspects of hadron production in \ppb\ collisions. The observed trends admit no possibility of ``collective'' radial flow: some hadron species (baryons) are ``boosted'' to higher \yt\ while others (mesons) are shifted to lower \yt. Those changes relate to jet production in \pn\ collisions.

Table~\ref{pidparamsx} presents updated hard-component model parameters based on optimized descriptions of spectrum hard components for the {\em most-central} ($n = 1$) \ppb\ event class as described above. That condition plus the more-differential format in the right panels above leads to improved precision of inferred parameter values. Also included are PID fractions $z_{0i}$ for five hadron species.

%%%%%%%%%%%%%%%%%%%%%%%%%%%%%%%%%%
\begin{table}[h]
	\caption{Revised PID TCM hard-component model parameters $(\bar y_t,\sigma_{y_t},q)$ for identified hadrons from 5 TeV \ppb\ collisions derived from the differential study in this section.  Parameter $z_{0i}$ values inferred from $z_{si}(n_s)$ trends in Fig.~\ref{zshx} (left) are also included. Uncertainties are determined as one half the change that would produce an obvious variation in right-panel ratios above. Values with no uncertainties are duplicated from a related particle type.
	}
	\label{pidparamsx}
	%%%%%%%%%%%%%%%%%%%%%%%%%%%%%%%
	\begin{center}
		\begin{tabular}{|c|c|c|c|c|} \hline
			%\multicolumn{8}{|c|}{\pp\ multiplicity classes} \\ \hline
			&  $\bar y_t$ & $\sigma_{y_t}$ & $q$ & 	$z_{0i}$  \\ \hline
%			$ h $     &  $2.64\pm0.03$ & $0.57\pm0.03$ & $3.9\pm0.2$ & $145\pm3$ & $8.3\pm0.3$ \\ \hline
			$ \pi^\pm $     
			&	   $2.46\pm0.005$ & $0.575\pm0.005$ & $4.1\pm0.5$  &  0.82 $\pm$0.01 \\ \hline
			$K^\pm$    
			&	  $2.655$  & $0.568$ & $4.1$  & 0.128 $\pm$0.002 \\ \hline
			$K_\text{S}^0$          
			& 	   $2.655\pm0.005$ & $0.568\pm0.003$ & $4.1\pm0.1$ & 0.064 $\pm$0.002  \\ \hline
			$p$        
			& 	  $2.99\pm0.005$  & $0.47\pm0.005$ & $5.0$ & 0.065 $\pm$0.002  \\ \hline
			$\Lambda$       
			& 	  $2.99\pm0.005$  & $0.47\pm0.005$ & $5.0\pm0.05$  & 0.034 $\pm$0.002 \\ \hline	
		\end{tabular}
	\end{center}
	%%%%%%%%%%%%%%%%%%%%%%%%%%%%%%%
\end{table}
%%%%%%%%%%%%%%%%%%%%%%%%%%%%%%%

One may question why the detailed hard-component shape evolutions presented in this section were not reported in Sec.~6.2 of Ref.~\cite{ppbpid} based on the same PID spectra. That question is addressed in Sec.~\ref{revisedspec}.

%%%%%%%%%%%%%%%
\section{Revised $\bf p$-$\bf Pb$\ PID spectrum TCM} \label{revisedspec}

The main emphasis in Ref.~\cite{ppbpid} was to define a PID TCM for \ppb\ spectrum data, extract PID hard components and identify their average features. In the present study the extended PID spectrum TCM with newly-parametrized $z_{si}(n_s)$ and $z_{hi}(n_s)$ in Eqs.~(\ref{zsinew}) and (\ref{zhinew}) is applied to PID spectra from 5 TeV \ppb\ collisions reported in Ref.~\cite{aliceppbpid}. The present emphasis is to isolate hard components with as little systematic distortion as possible and examine details of their centrality evolution.

\subsection{PID TCM from Ref.~\cite{ppbpid}}
 
For the analysis of PID \pt\ spectra from 5 TeV \ppb\ collisions presented in Sec.~6 of Ref.~\cite{ppbpid} fixed values of $z_{0i}$ and $\tilde z_{i}$ were determined to describe low-\pt\ parts of spectra, in effect fitting implicit $z_{si}(n_s)$ trends with two-parameter Eq.~(15) from that study. 
Given the fitted $z_{si}(n_s)$ model, full spectra $\bar \rho_{0i}(y_t,n_s)$ were normalized by
\bea
X_i(y_t,n_s) &\equiv& \frac{\bar \rho_{0i}(y_t,n_s)}{z_{si}(n_s)  \bar \rho_{s}}
\eea
comparable with $\hat S_{0i}(y_t)$. Using the fitted $\tilde z_{i}$ values normalized spectrum hard components were obtained by
\bea \label{yold}
Y_i(y_t,n_s) &\equiv& \frac{X_i(y_t,n_s) -  \hat S_{0i}(y_t)}{\tilde z_{i}\, x(n_s) \nu(n_s) } \approx  \frac{H_i(y_t,n_s)}{z_{hi}(n_s)\bar \rho_{h}}
\eea
which can be compared with $\hat H_{0i}(y_t)$. Results of that procedure are shown below in Figs.~\ref{pioncomp}, \ref{kaoncomp} and \ref{baryoncomp} (left). The present analysis reveals that that approach obscures information in the data hard components in  two ways: (a) The structure below the hard-component mode may be strongly biased and (b) subtleties in hard-component amplitude variations may be distorted or concealed.

\subsection{PID TCM from the present study}

A major goal of the present study is to obtain $z_{si}(n_s)$ and $z_{hi}(n_s)$ values directly from published spectra as accurately as possible and thereby to reexamine the assumptions about $z_{0i}$ and $\tilde z_{i}$ made in Ref.~\cite{ppbpid}. Experience from the previous analysis further suggests modification of the TCM formulation to maximize information obtainable from extracted spectrum hard components. Given the structure of Eq.~(\ref{zhiz}) (second line) a revised expression for PID spectrum hard components is
\bea \label{zhizz}
Y'_i(y_t,n_s)&=&\frac{\bar \rho_{0i}(y_t,n_s) -  z_{si}(n_s)\bar \rho_s \hat S_{0i}(y_t)}{\bar \rho_h  \hat H_{0i}(\bar  y_t) }
\\ \nonumber 
 &\approx& \frac{\tilde z_i(n_s) z_{si}(n_s)}{\hat H_{0i}(\bar  y_t)}Y_i(y_t,n_s) 
 \\ \nonumber
 &\equiv& z_{hi}(y_t,n_s) \frac{\hat H_{0i}(y_t)}{\hat H_{0i}(\bar  y_t)},
\eea
where $\bar  y_t$ is the mode of the $\hat H_{0i}(y_t)$ model function so that $Y'_i(\bar y_t,n_s) \approx z_{hi}(\bar y_t,n_s)$ as reported in Table~\ref{zhixx}.	

That formulation has two improved characteristics: (a) Accurate individual estimates of $z_{si}(n_s)$ [actually, {\em product} $z_{si}(n_s) \bar \rho_s$] are derived directly from differential analysis of spectra as in Sec.~\ref{diffanal}, greatly reducing systematic bias of hard components below the mode. (b) Parameters $z_{hi}(n_s)$ are not used to further normalize the inferred hard components because there is a problem of definition regarding hard components $H_{i}(y_t,n_s)$. Inferred data hard components are in effect scaled only by nonPID density $\bar \rho_h = \bar \rho_s x\nu$ and maxima of models $\hat H_{0i}(y_t)$.  Results of that procedure are shown in Figs.~\ref{pioncomp}, \ref{kaoncomp} and \ref{baryoncomp} (right).

\subsection{Method comparisons}

\begin{figure}[h]
	\includegraphics[width=1.65in,height=1.65in]{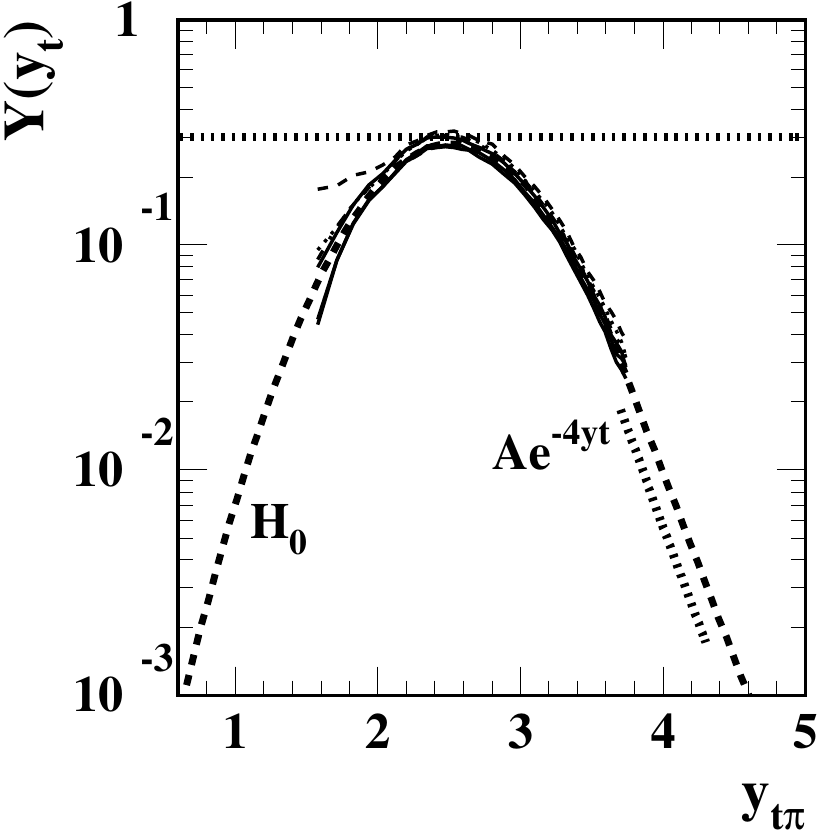}
	\includegraphics[width=1.65in,height=1.65in]{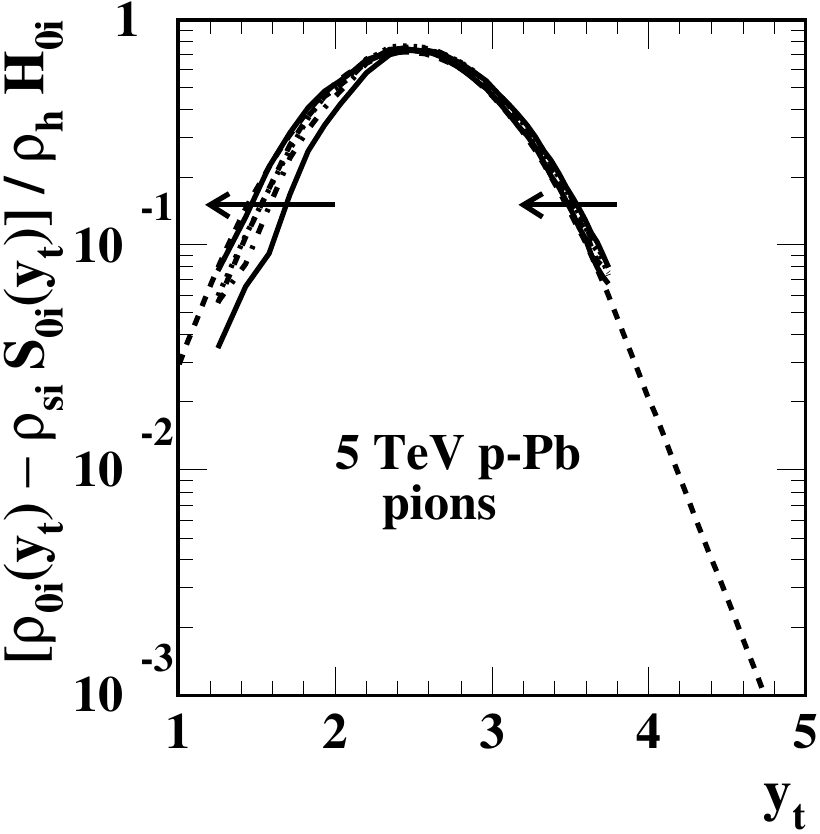}
	\caption{\label{pioncomp}
		Left: Pion hard components as reported in Ref.~\cite{ppbpid} and obtained via Eq.~(\ref{yold}). An incorrect entry $z_{0i} = 0.70$ in Table~4 of Ref.~\cite{ppbpid} has been corrected to 0.80 bringing the data into general agreement with TCM model $\hat H_0(y_t)$ (bold dashed).
		Right:	Pion hard components obtained via Eq.~(\ref{zhizz}).
	} % alice 600b, 600aa4s
\end{figure}

Figure~\ref{pioncomp} (left) shows pion spectrum hard components reported in Ref.~\cite{ppbpid}. There is significant bias of hard components below the mode because of imprecise estimation of parameter $z_{si}(n_s)$. Likewise, there is a potential overall scale error because of systematic bias of fixed parameters $\tilde z_{i}$ (e.g.\ they actually vary with centrality).

%%%%%%%%%%
%%%%%%%%%%%%

Figure~\ref{pioncomp} (right) shows spectrum hard components obtained with Eq.~(\ref{zhizz}) (first line). The structure below the mode is now meaningful down to $y_t \approx 1.2$ ($p_t \approx 0.2$ GeV/c) and clearly shows a substantial shift to lower \yt\ with increasing \ppb\ centrality. A similar significant shift (albeit less pronounced) is indicated above the mode.

\begin{figure}[h]
\includegraphics[width=1.65in,height=1.61in]{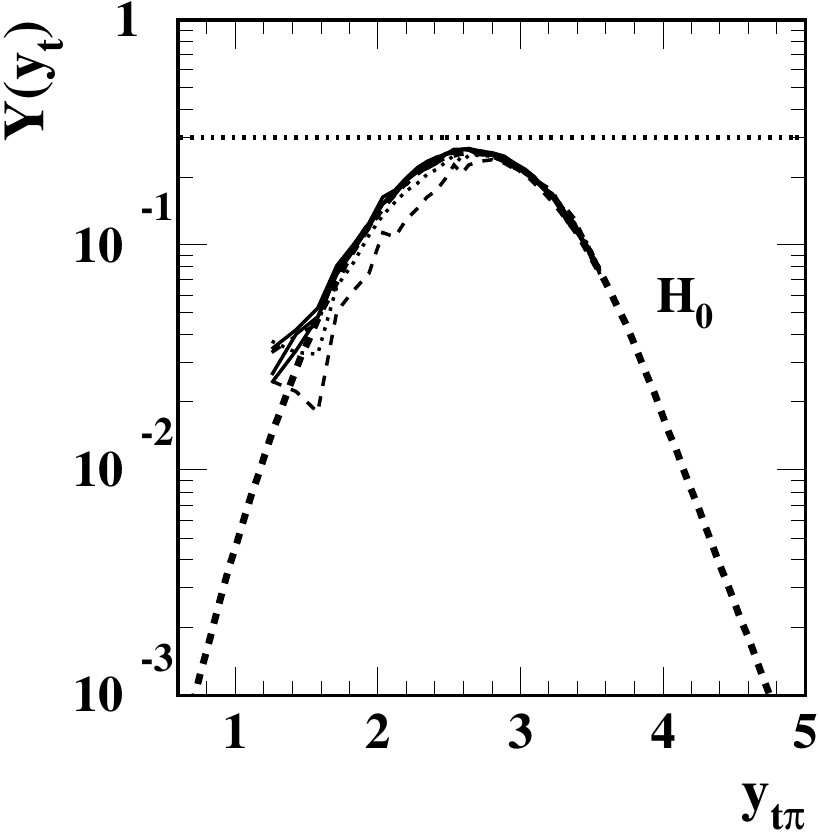}
	\includegraphics[width=1.65in,height=1.58in]{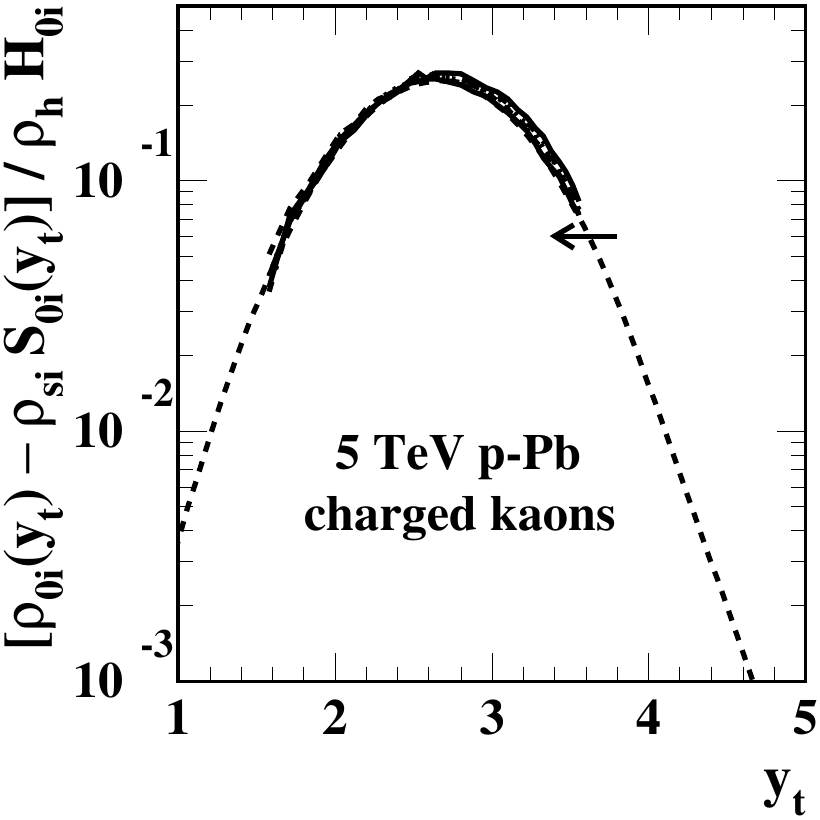}
	\put(-140,102) {\bf (a)}
	\put(-23,102) {\bf (b)}\\
\includegraphics[width=1.65in,height=1.61in]{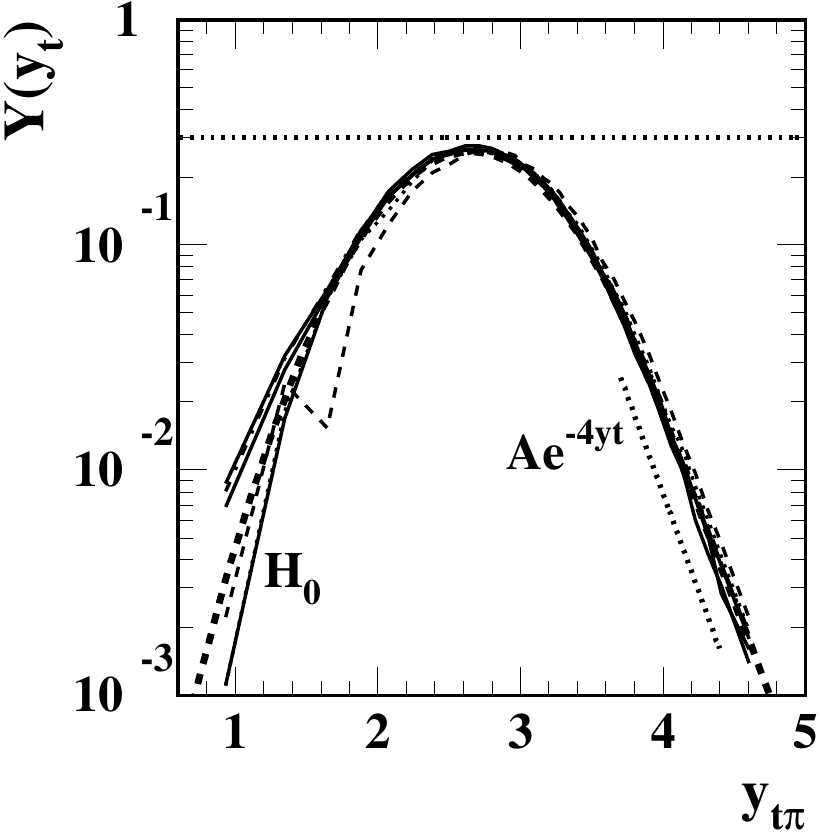}
	\includegraphics[width=1.65in,height=1.58in]{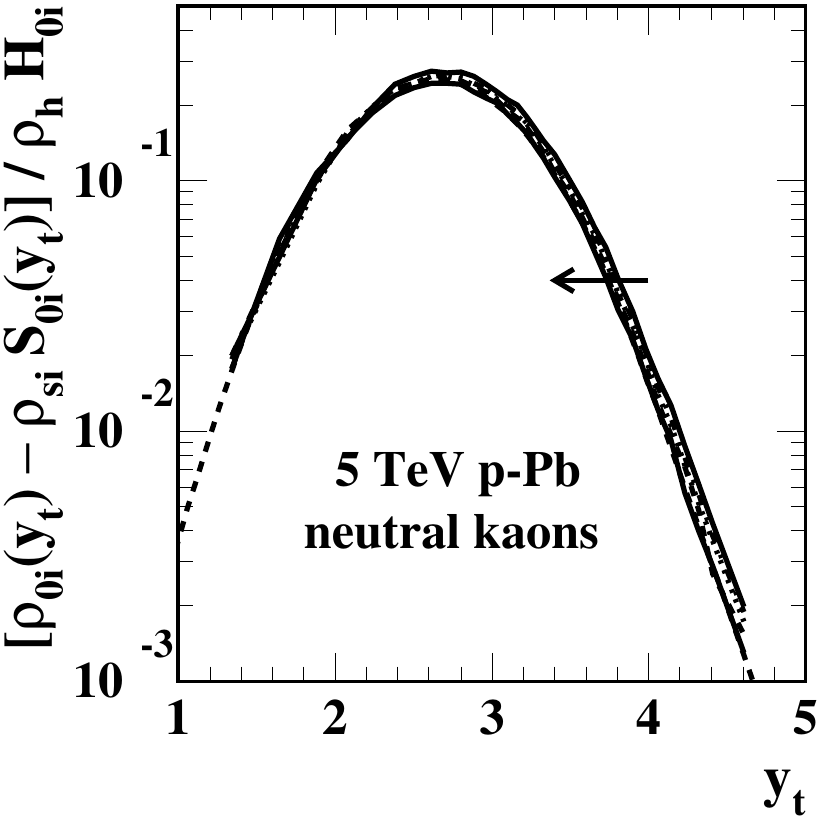}
	\put(-140,102) {\bf (c)}
	\put(-23,102) {\bf (d)}
	\caption{\label{kaoncomp}
Comparison between charged and neutral kaon PID hard components for 5 TeV \ppb\ collisions derived in Ref.~\cite{ppbpid} via Eq.~(\ref{yold}) (a,c) and revised results from the present study via Eq.~(\ref{zhizz}) (b,d).
	}  %  alice630b, 630aa4s, 640b, 640aa4s
\end{figure}

Figure~\ref{kaoncomp} (left) shows charged-kaon (a) and neutral-kaon (c) spectrum hard components reported in Ref.~\cite{ppbpid}. Aside from reasonable agreement with the overall shape of model function $\hat H_0(y_t)$ (common to both hadron species) there is no further information accessible.

%%%%%%%%%%
%%%%%%%%%%%%

Figure~\ref{kaoncomp} (right) shows  kaon spectrum hard components obtained via Eq.~(\ref{zhizz}) (first line). The charged-kaon (b) acceptance is limited, but within that acceptance results are consistent with neutral kaons (d). Below its mode the neutral-kaon hard component shows no significant centrality dependence down to \yt\ = 1.3 ($p_t \approx 0.25$ GeV/c), whereas above the mode the significant shifts to lower \yt\ with increasing centrality are similar in structure and magnitude to those for pions.

\begin{figure}[h]
	\includegraphics[width=1.65in,height=1.61in]{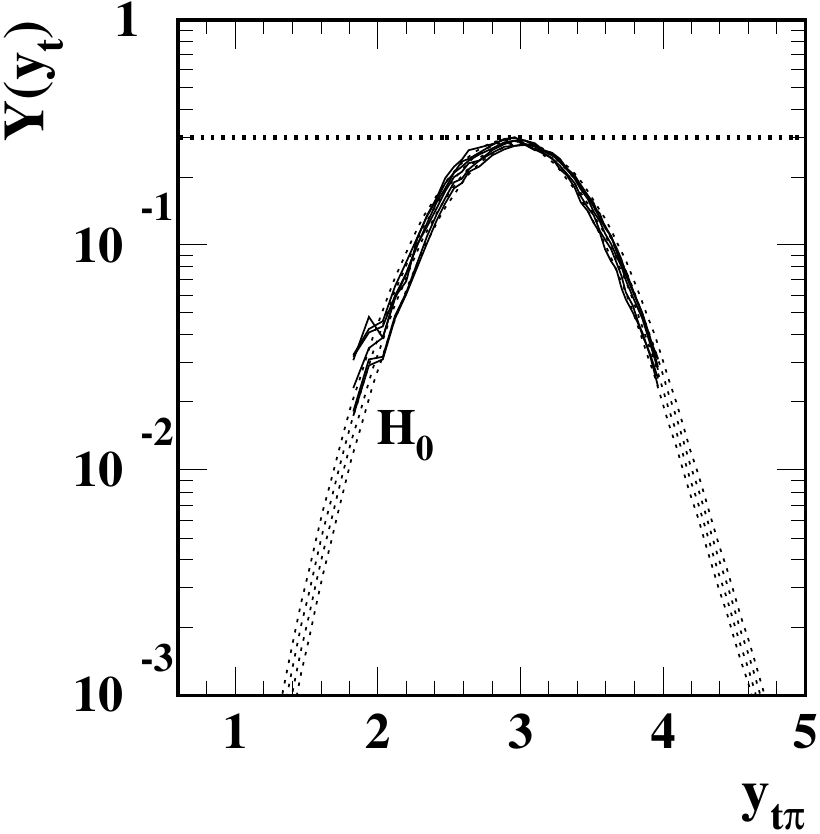}
	\includegraphics[width=1.65in,height=1.58in]{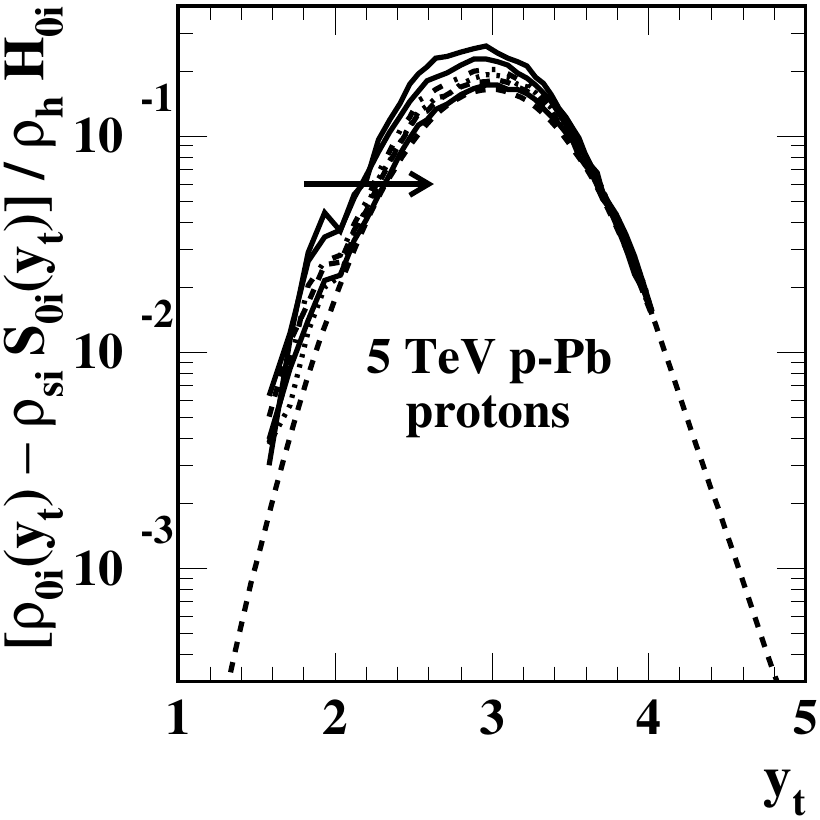}
	\put(-140,102) {\bf (a)}
	\put(-23,102) {\bf (b)}\\
	\includegraphics[width=1.65in,height=1.61in]{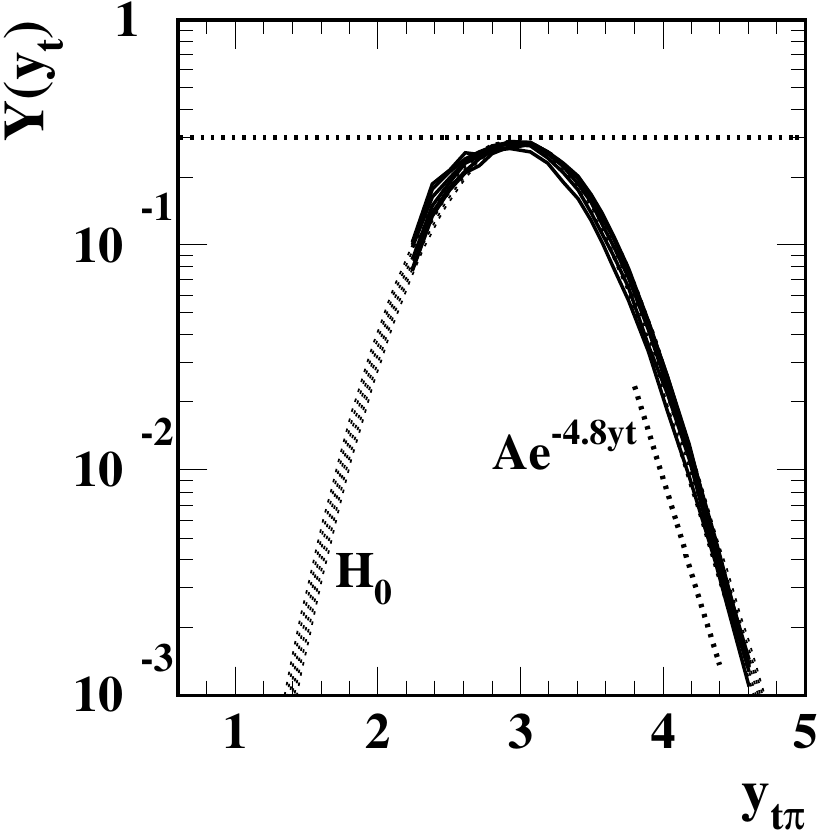}
	\includegraphics[width=1.65in,height=1.58in]{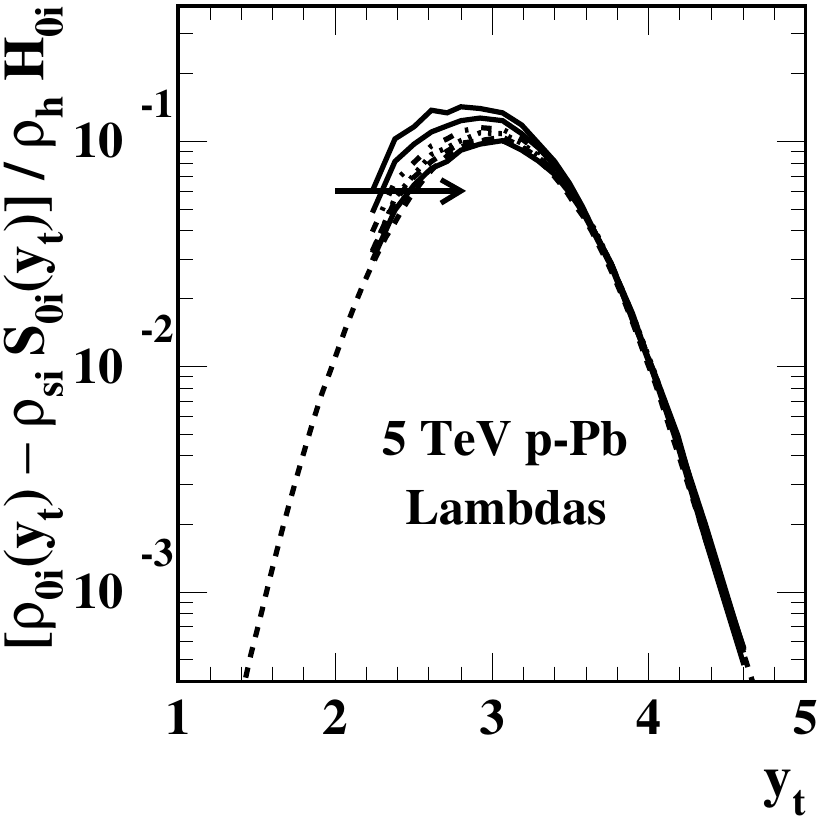}
	\put(-140,102) {\bf (c)}
	\put(-23,102) {\bf (d)}
	\caption{\label{baryoncomp}
Comparison between proton and Lambda PID hard components for 5 TeV \ppb\ collisions derived in Ref.~\cite{ppbpid} via Eq.~(\ref{yold}) (a,c) and revised results from the present study via Eq.~(\ref{zhizz}) (b,d).
	}  %  alice610b, 610aa4s, 620b, 620aa4s
\end{figure}

Figure~\ref{baryoncomp} (left) shows proton (a) and Lambda (c) spectrum hard components as inferred via Eq.~(\ref{yold}). The Lambda results are as reported in Ref.~\cite{ppbpid} but the proton results include inefficiency corrections as described in Sec.~\ref{ineff}. Results for different centralities have the same amplitudes {\em by construction} and exhibit significant shifts to {\em higher} \yt\ with increasing centrality. Those shifts are emulated by shifted $\hat H_0(y_t,n_s)$ models (dotted curves).

%%%%%%%%%%
%%%%%%%%%%%%

Figure~\ref{baryoncomp} (right) shows corrected-proton (b) and Lambda (d) spectrum hard components obtained with Eq.~(\ref{zhizz}) (first line). There are two major differences from the left panels: (a) The distributions above the mode show {\em no significant variation} with \ppb\ centrality, and (b) shifts below the mode to lower \yt\ with {\em decreasing} centrality are accompanied by strong {\em increase} of distribution maxima. In effect, the procedure based on Eq.~(\ref{yold}) transformed the amplitude variation to an {\em apparently} uniform (and misleading) translation on \yt, thus demonstrating the need for a modified procedure.

For each hadron species fixed model $\hat H_{0i}(y_t)$ is determined by the most-central data distribution. That definition leads to more-precise determination of model parameters as reported in Table~\ref{pidparamsx}. The model descriptions follow the most-central data within statistical uncertainties as demonstrated in Figs.~\ref{pionaa4}, \ref{kaonaa4} and \ref{baryonaa4} (right).

%%%%%%%%%%%%
\section{Systematic uncertainties} \label{sys}

Four principal issues emerge in connection with a TCM description of PID \ppb\ spectra: 
(a) accuracy of the \ppb\ collision ``geometry,'' including \ppb\ impact parameter variation and \pn\ multiplicity \nch\ dependence;
(b) minimizing uncertainties due to proton detection inefficiencies and Lambda \pt\ acceptance limitations;
(c) uncertainties for  coefficients $z_{si}(n_s)$ and $z_{hi}(n_s)$ that form the central elements of the TCM description of \ppb\ PID spectra;
(d) uncertainties for hard-component shape evolution.

\subsection{TCM and collision geometry accuracy}

Accurate separation of hard and soft spectrum components via the TCM depends on precise determination of collision-system ``geometry'' (used broadly to refer to \mbox{A-B} centrality and/or N-N \nch\ dependence). The {\em method} for geometry determination is based on \pp\ spectrum analysis without {\em a priori assumptions}~\cite{ppprd} and analysis of ensemble-mean \mmpt\ data from \ppb\ collisions~\cite{tommpt}.

The ``geometry'' for \pp\ (\nn) collisions is determined solely by coefficient $\alpha$ in the expression $\bar \rho_h \approx \alpha(\sqrt{s}) \bar \rho_s^2$ with $\alpha(\sqrt{s})$ energy dependence as it appears in Fig.~20 (left) of Ref.~\cite{tomnewppspec} and its Table~II, where $\alpha$ for 5 TeV \nn\ is listed as 0.013 compared to 0.0127 employed in the present study. More details are provided in Ref.~\cite{tommpt} Sec.~III A. The uncertainty for $\alpha$ is 5\% at LHC energies.

The centrality dependence for 5 TeV \ppb\ collisions is derived from a TCM study of \mmpt\ data as reported in Ref.~\cite{tommpt} and as elaborated in Ref.~\cite{tomglauber}. The observed trends for \ppb\ collisions (especially since they differ strongly from a classical Glauber analysis~\cite{aliceglauber}) are explained in Ref.~\cite{tomexclude}. The critical parameters for \ppb\ centrality, aside from $\alpha$, are transition point $\bar \rho_{s0} \approx 15$ and slope $m_0 \approx 1/10$ that determine the function $x(n_s)$ as it appears for instance in Fig.~2 (left) of Ref.~\cite{ppbpid}. From $x(n_s)$ all other centrality parameters are determined in turn as reported in Ref.~\cite{tommpt}. Parameters $\bar \rho_{s0}$ and $m_0$ are determined via a TCM description of \mmpt\ data within data uncertainties out to $\bar \rho_0 \approx 115$. With respect to \ppb\ spectrum data for $\bar \rho_0 \approx 45$ or less the critical model parameters as reported in Table~\ref{rppbdata} are determined to a few percent. Accurate estimation of $x(n_s)$ then ensures that soft-component density $\bar \rho_s = (N_{part}/2)\bar \rho_{sNN}$ (with $N_{part} = N_{bin} + 1$) as it appears in Eq.~(\ref{zsi}) and $\bar \rho_h = N_{bin} \alpha \bar \rho_{sNN}^2$ as it appears in Eq.~(\ref{zhi}) are also accurate at the few percent level. That assessment is supported for $\bar \rho_s$ by invariance of kaon hard components below the mode in Fig.~\ref{kaonaa4} (c) and for $\bar \rho_h$ by invariance of Lambda hard components above the mode in Fig.~\ref{baryonaa4} (c).

\subsection{Minimizing proton and Lambda uncertainties}

Interpretation of proton and Lambda \pt\ spectra in the context of the TCM is compromised by apparent detection inefficiency for protons {\em above} 0.6 GeV/c and no acceptance for Lambdas {\em below} 0.65 GeV/c. As described in  the text, strategies have been introduced to minimize systematic uncertainties given those limitations.

From Fig.~\ref{protonxxx} (b,d) it is apparent that baryon hard components are negligible below \yt\ = 1.5 ($p_t \approx 0.3$ GeV/c). Thus, extrapolation of proton data/model ratio trends per Eq.~(\ref{zsi}) down to 0.2 GeV/c as in Fig.~\ref{protonxxx} (a) (left vertical line) leads to $z_{si}(n_s)$ values with few-percent accuracy. 
Correction of proton inefficiencies as described in Sec.~\ref{ineff} relies on estimation of a mean value for ratio parameter $\tilde z_i$ from proton $z_{si}(n_s)$ values presented in Table~\ref{zsixx} per Eq.~(\ref{zsinew}). The best-fit fixed value is $\tilde z_i = 5.8 \pm 0.1$. As described in Sec.~\ref{ineff} that value is used to generate TCM model spectra for protons per Eqs.~(\ref{rhosi}) (first line) and (\ref{pidhard}). Ratios of uncorrected data spectra to TCM models then lead to the correction model shown in Fig.~\ref{eppsproton} (left). Variation of $\tilde z_i$ over its range of uncertainties remains consistent with the bold correction curve within its width as plotted in Fig.~\ref{eppsproton}.

The lower-\pt\ cutoff for Lambda spectra at 0.65 GeV/c in Fig.~\ref{protonxxx} (c) precludes direct inference of $z_{s}(n_s)$. However, it is expected that the Lambda trend may be closely related to the proton trend. Detailed comparison of panels (a) and (c) of Fig.~\ref{protonxxx} provides reasonably accurate estimates of $z_{s}(n_s)$ for Lambdas. Proton and Lambda spectra are related at 0.85 GeV/c (vertical lines in Fig.~\ref{protonxxx} left) by factor 1.8. That factor provided an initial estimate of Lambda $z_{s}(n_s)$ values which, combined with measured $z_{h}(n_s)$ values, determined $\tilde z_i(n_s)$ trends used to generate TCM spectra compared directly with Lambda data in Fig.~\ref{piddata} (f). The comparison indicated that Lambda $z_{s}(n_s)$ might be reduced from that estimate by as much as 10\% to insure an optimal overall data description. The Lambda $z_{si}(n_s)$ values thus include a $\pm 5$\% uncertainty which propagates to Lambda $\tilde z_i(n_s)$ values in Fig.~\ref{zrat}.

\subsection{Table~\ref{zsixx} and \ref{zhixx} uncertainties}

As noted in Sec.~\ref{zcent} the overall scale of the $z_{si}(n_s)$ is determined by $\bar \rho_s = (N_{part}/2)\bar \rho_{sNN}$ in Eq.~(\ref{zsi}) which in turn depends on geometry parameters presented in Table~\ref{rppbdata} derived from \pp\ spectrum data as in Ref.~\cite{alicetomspec} and \mmpt\ data as in Ref.~\cite{tommpt}. $\bar \rho_s$ scaling applies to all $z_{si}(n_s)$ in common, not to  their point-to-point uncertainties, and is thus relevant to overall ratio $\tilde z_i(n_s)$ trends but not to their individual values. The overall scale of measured $z_{hi}(n_s)$ depends on the same parameters but in addition depends on the value of $\alpha(\sqrt{s})$ used to determine $\bar \rho_h = N_{bin} \alpha \bar \rho_{sNN}^2$ in Eq.~(\ref{zhi}).  $\alpha$ scaling applies to all $z_{hi}(n_s)$ in common, not to  their point-to-point uncertainties, and is likewise relevant only to overall ratio $\tilde z_i(n_s)$ trends.

The trends in Fig.~\ref{zrat} (right) imply that only two constants, those in the expression $\tilde z_i \approx 5.4 m_i(1 + 0.46 x \nu)$, determine the linear trends in the left panel that can, along with assigned values for the $z_{0i}$, reasonably be used to generate all $z_{si}(n_s)$ and $z_{hi}(n_s)$ values (see model curves in Fig.~\ref{zshx}). But that is basically all the elements unique to the PID TCM. Rather than making the \ppb\ PID TCM more complicated, introducing centrality dependent $\tilde z_{i}(n_s)$ and determining $z_{si}(n_s)$ and $z_{hi}(n_s)$ for each spectrum has greatly simplified the model. After accounting for the 12\% change from $\alpha = 0.0113$ in Ref.~\cite{ppbpid} to 0.0127 in the present study the $\tilde z_{i}(n_s)$ trends presented in Fig.~\ref{zrat} (left) are consistent within reported uncertainties with the fixed $z_h / z_s$ values in Table~\ref{otherparams} from Ref.~\cite{ppbpid}. 

\subsection{Hard-component shape evolution}

Determining spectrum hard components via Eq.~(\ref{yold}) on the one hand or Eqs.~(\ref{zhizz}) on the other may seem algebraically equivalent. However, the statistical consequences are quite different as discussed in Sec.~\ref{revisedspec}. In Eq.~(\ref{yold}) the normalized spectra $X_i(y_t,n_s)$ include hard component $H_i(y_t,n_s)$ that is therefore also normalized by product $z_{si}(n_s)\bar \rho_s$ which includes measurement uncertainties. Quantity $Y_i(y_t,n_s)$ includes a difference that may impose large relative errors below the hard-component mode because of systematic errors in $z_{si}(n_s)$. $Y_i(y_t,n_s)$ is further normalized by $\tilde z_i$ that may be a fixed quantity or centrality dependent but also includes measurement uncertainties. The result is hard component representation $H_i(y_t,n_s) / z_{hi}(n_s) \bar \rho_h$  that may be further biased (see the left panels of figures in Sec.~\ref{revisedspec}). Defining TCM model functions $\hat H_{0i}(y_t)$ is ambiguous as a result.

In Eqs.~(\ref{zhizz}) the difference $\bar \rho_{0i}(y_t,n_s) -  z_{si}(n_s)\bar \rho_s \hat S_{0i}(y_t)$ preserves data hard component $H_i(y_t,n_s)$ intact. Coefficients $z_{si}(n_s)$ are optimized as in Sec.~\ref{diffanal} to match as accurately as possible data soft components at low \pt, thus minimizing bias of hard component $H_i(y_t,n_s)$ below its mode. The difference is then normalized only by $\bar \rho_h = \bar \rho_s x\nu$ and the fixed number $\hat H_{0i}(\bar y_t)$. Isolated hard components $H_i(y_t,n_s)$ are thus unbiased except possibly at low \pt, and different hadron species are reliably comparable as is evident in the right panels of Sec.~\ref{revisedspec}.

Because hard components $H_i(y_t,n_s)$ remain intact as in Sec.~\ref{newdetails} left panels (except for numerical factors common to all hadron species) any factorization of \yt\ and $n_s$ dependence can be done after the fact. In particular, definition of TCM model function $\hat H_{0i}(y_t)$ can be based on precise data trends. It is apparent that for the most-central data all hard components appear symmetric about the mode. Fixed model functions $\hat H_{0i}(y_t)$ are defined on that basis (dashed curves in left panels). Based on those reference functions, quantities $z_{hi}(y_t,n_s)$ defined in Eq.~(\ref{zhiz}) and shown in right panels of Sec.~\ref{newdetails} represent all deviations of data spectra from the fixed TCM. The quality of the description is indicated by the $z_{hi}(y_t,n_s)$ curves for most-central events consistent with data within statistical uncertainties. Because of that differential format and the unique model definitions, the $\hat H_{0i}(y_t)$ parameters are determined with exceptional accuracy as indicated by the reduced uncertainty estimates in Table~\ref{pidparamsx}. Given the fixed TCM reference the $z_{hi}(y_t,n_s)$ ratio curves represent {\em all jet-related information} carried by \ppb\ PID spectrum data relative to nonPID data.

%%%%%%%%%%%%%%
\section{Discussion}  \label{disc}

Four issues from the present study warrant further discussion:
(a) The importance of accurate nonPID \ppb\ collision geometry (centrality) determination and implications therefrom,
(b) implications of the evolution with \nch\ of PID coefficients $z_{si}(n_s)$, $z_{hi}(n_s)$ and $\tilde z_{i}(n_s)$,
(c) interpretation of PID hard-component shape evolution with \nch\ and
(d) implications from the above for PID parton fragmentation to jets in \nn\ and \ppb\ collisions.

\subsection{PID $\bf p_t$ spectra and the $\bf p$-Pb geometry model} \label{ppbgeom2}

Effective analysis of \ppb\ PID \pt\ spectra requires an accurate  centrality description. Based on alternative critical assumptions major differences in inferred collision geometries may result as exemplified by columns in Table~\ref{rppbdata} for $\sigma' / \sigma_0$ and $N_{bin}'$ vs $\sigma / \sigma_0$ and $N_{bin}$, the former based on application of a classical Glauber model and the latter based on a precise TCM description of \ppb\ \mmpt\ data.

The TCM-based \ppb\ geometry developed in Ref.~\cite{ppbpid} was derived from nonPID \ppb\ \mmpt\ data reported in Ref.~\cite{alicempt} extending out to $\bar \rho_0 \approx 115$. The procedure was based on a TCM expression for \mmpt\ (with biased data denoted by $\bar p_t'$ below)
\bea
\frac{\bar P_t}{n_s} &=& [\xi + x(n_s) \nu(n_s)] \, \bar p_{t}' =  \bar p_{ts} + x(n_s) \nu(n_s)\bar p_{th}~~~
\eea 
assuming that $\bar p_{ts}$ and $\bar p_{th}$ are approximately constant and are derived from TCM model functions $\hat S_0(y_t)$ and $\hat H_0(y_t)$. $\bar P_t$ is the integrated \pt\ within some angular acceptance, $n_{ch}'$ is the integrated charge within the same acceptance subject to \pt\ constraints and $\bar p_{ts}' \approx \bar P_t / n_{ch}'$ is the conventional \mmpt\ as published in Ref.~\cite{alicempt}. $\xi \leq 1$ represents the fraction of particles surviving a \pt\ acceptance lower bound (bias). Given a model for $x(n_s)$ all other geometry elements are defined as described in Ref.~\cite{ppbpid} Sec.~5. The simple $x(n_s)$ model that best describes the Ref.~\cite{alicempt} \mmpt\ data is shown in Fig.~2 (left) of Ref.~\cite{ppbpid}. It follows a linear \pp\ trend with slope $\alpha$ up to transition point $\bar \rho_{s0}$ and then proceeds also linearly with one tenth the slope (via $m_0$) above that point. The reason for that empirical finding is explained in Refs.~\cite{tomglauber} and \cite{tomexclude}.

Based on analysis of \pp\ spectra~\cite{ppprd} the relation $x(n_s) \equiv \bar \rho_{hNN} / \bar \rho_{sNN} \approx \alpha \bar \rho_{sNN}$ is assumed for A-B collisions composed of {\em linear superpositions} of \nn\ collisions. The product $x(n_s)\nu(n_s)$ representing the hard/soft ratio $\bar \rho_h / \bar \rho_s$ plays a central role in the A-B TCM, with $\bar \rho_h =N_{bin} \bar \rho_{h_{NN}}$, $\bar \rho_s =(N_{part}/2) \bar \rho_{sNN}$ and $\nu \equiv 2 N_{bin} / N_{part}$. For \pp\ collisions $\bar \rho_{sNN} \equiv \bar \rho_s$ and $\nu \equiv 1$. These relations 
\bea \label{rhosrho0}
\bar \rho_s / \bar \rho_0&\approx & \frac{1}{1+ \alpha \bar \rho_{s}}~~~\text{p-p}
\\ \nonumber
&\approx& \frac{1}{1+ \alpha \rho_{sNN}\nu(n_s)}~~~\text{p-Pb}
\eea
and
\bea
\bar \rho_h / \bar \rho_s &\equiv & x(n_s) \nu(n_s) 
\\ \nonumber
&\approx &  \alpha \bar \rho_{s}~~~\text{p-p}
\\ \nonumber
&\approx& \alpha \bar \rho_{sNN} \nu~~~\text{p-Pb}.
\eea
are illustrated in the figure below.

Figure~\ref{680j} (left) shows TCM values for ratio $\bar \rho_s / \bar \rho_0$ and for \ppb\ (points) and \pp\ (curve) collisions. In either case the ratio can be expressed as $1/(1 + x\nu)$. Those trends on $\bar \rho_s$ can be compared with the trends for $z_{si}(n_s)$ on $x\nu$ in Fig.~\ref{zshx} (left) in the context of Eq.~(\ref{zsinew}). In either case the dominant parameter is the hard/soft ratio $x\nu$ which measures a jet contribution to hadron production relative to the nonjet component. For $z_{si}(n_s)$ the parameter $\tilde z_i$ in Eq.~(\ref{zsinew}) then determines the {\em specific importance} of jet production for hadron species $i$. That direct connection between the centrality trend of the spectrum soft component and jet production was first observed in Ref.~\cite{ppprd} Fig.~3 (left) and is the basis for argument in Secs.~\ref{pidfracdata} and \ref{ineff} supporting correction of proton detection inefficiencies based on low-\pt\ spectrum trends.

%%%%%%%%%%\bar y_t,
\begin{figure}[h]
	\includegraphics[width=3.3in]{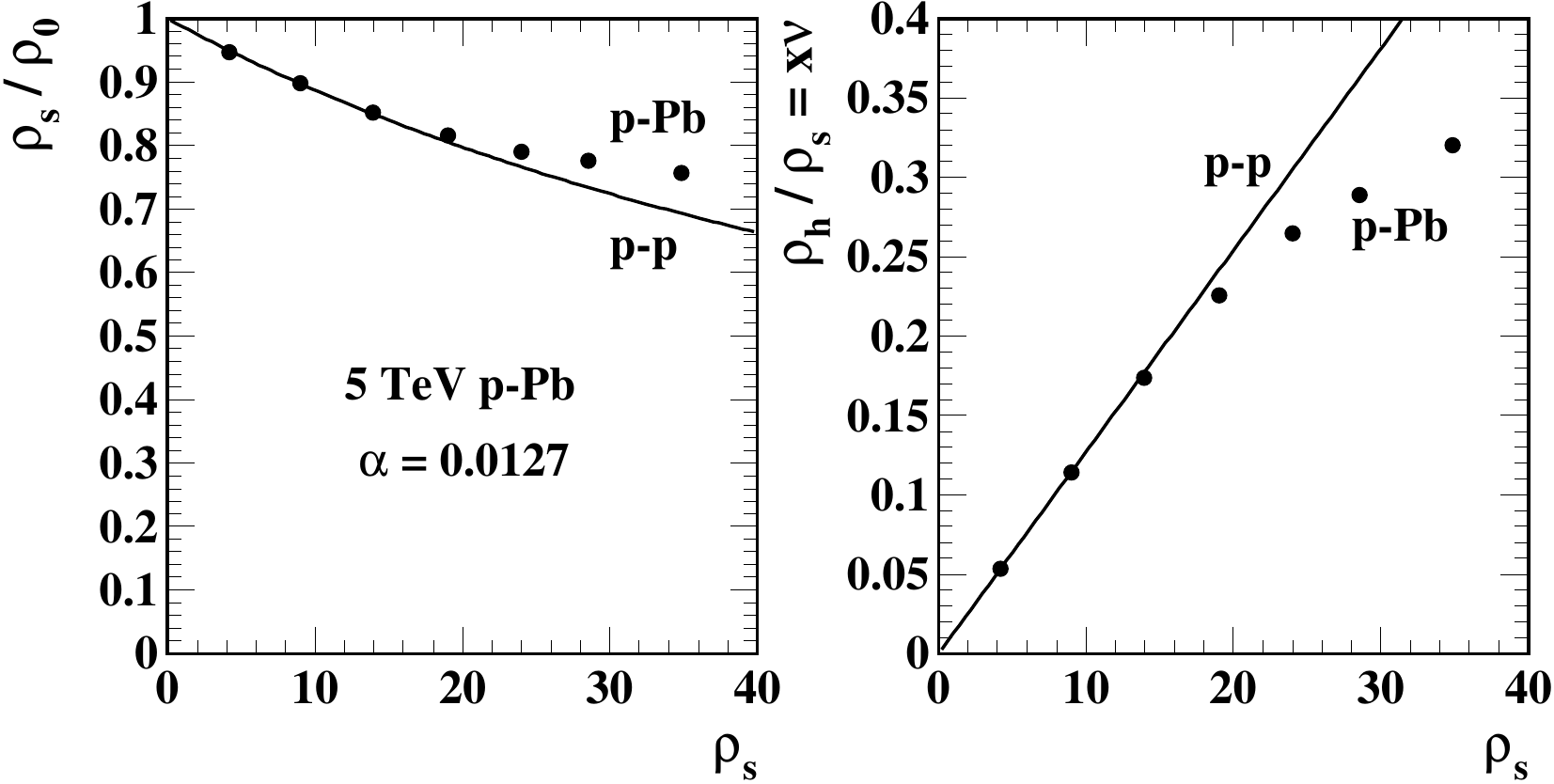}
	\caption{\label{680j}
		Left: Ratio $\bar \rho_s / \bar \rho_0$ vs $\bar \rho_s$ for \pp\ collisions (solid) and \ppb\ collisions (points) illustrating that jet production strongly affects the spectrum soft component at lower \pt.
		Right: Hard/soft ratio $\bar \rho_h / \bar \rho_s = x(n_s)\nu(n_s)$ for \pp\ collisions (solid) and \ppb\ collisions (points) illustrating the key role of parameter $\alpha$ (slope of the straight line for \pp\ collisions).
	}  %  alice680j
\end{figure}
%%%%%%%%%%%%

Figure~\ref{680j} (right) shows TCM values for ratio $\bar \rho_h / \bar \rho_s$ and for \ppb\ (points) and \pp\ (line) collisions. In either case the ratio is expressed as $x\nu$ with limiting cases described above. The \pp\ trend $x\nu \rightarrow x = \alpha \bar \rho_s$ (line) appears universal for $\sqrt{s} = $ 17 GeV to 13 TeV, $\alpha$ increasing logarithmically with $\sqrt{s}$ consistent with eventwise-reconstructed jet measurements (see Sec.~VI E of Ref.~\cite{alicetomspec}). 

Reduction of \ppb\ $x\nu$ below the \pp\ trend beyond $\bar \rho_s \approx 20$ is explained as follows: Below that  value demand for higher multiplicity \nch\ is provided by {\em single} peripheral \pn\ collisions with consequent increase of jet production according to $\bar \rho_{hNN} = \alpha \bar \rho_{sNN}^2$ (i.e.\ strong increase, line). Above $\bar \rho_s \approx 20$ {\em multiple} \pn\ collisions become competitive, i.e.\ \ppb\ centrality increases ($N_{part} > 2$). The rate of increase of $\bar \rho_{sNN}$ within $\bar \rho_s$ then decreases, reducing the mean \pn\ jet production rate relative to \pp\ collisions.

Given that context it is essential to compare this situation with the \mmpt\ data reported in Ref.~\cite{alicempt}. In Fig.~\ref{680j} four of seven \ppb\ centrality classes effectively coincide with single peripheral \pn\ collisions (curve and line) below $\bar \rho_s = 20$. The remaining three centrality classes deviate only modestly from isolated-\pn\ trends. The highest centrality class for Ref.~\cite{aliceppbpid} ($\bar \rho_0 \approx 45$) corresponding to $\bar \rho_s \approx 35$ should be compared with $\bar \rho_0 \approx 115$ and $\bar \rho_s \approx 75$ for the \mmpt\ data from Ref.~\cite{alicempt} that are described by the TCM in Ref.~\cite{tommpt} within point-to-point uncertainties.

The most-central \ppb\ event class for Ref.~\cite{aliceppbpid}, with $\bar \rho_0 \approx 45$, is reported there as representing 0-5\% of the total cross section and in Ref.~\cite{aliceppbglaub} is assigned $N_{part} \approx 16$. If that were correct the \pn\ mean charge density should be $\bar \rho_{0NN} \approx (2/N_{part})\bar \rho_0 \approx 5.5$, comparable to the 5 TeV \pp\ NSD value $\bar \rho_{0NSD} \approx  5 \approx \bar \rho_{sNSD}$. In that case jet production per \nn\ collision should be measured by $\alpha \bar \rho_{sNSD}^2$ and hard/soft ratio $x\nu \rightarrow \alpha \bar \rho_{sNSD}$ would be $\approx 0.06$ according to Fig.~\ref{680j} (right), not $> 0.3$ as indicated by the TCM that accurately describes \ppb\ \mmpt\ data (points). In contrast, the centrality trend in Fig.~\ref{piddata} (b) for the higher-\pt\ spectrum interval, with its large spacings for four more-peripheral spectra and close spacings for more-central spectra, matches the $x \nu$ trend in Fig.~\ref{680j} (right).

Given the availability of \ppb\ collision data from Ref.~\cite{alicempt} that extends out to $\bar \rho_0 \approx 115$ one may wonder why the PID spectrum analysis in  Ref.~\cite{aliceppbpid} was not extended over the same interval. The report in Ref.~\cite{aliceppbglaub} gives the impression that $\bar \rho_0 \approx 45$ is a limiting case (at least for V0A event selection -- see its Fig.~16). Then how to characterize the $\bar \rho_0 \approx 115$ data from Ref.~\cite{alicempt}? If one is concerned about the issue of collectivity in small systems a PID spectrum study including the highest possible particle densities ought to be desirable. Figure~\ref{680j} demonstrates that \ppb\ data as reported in Ref.~\cite{aliceppbpid} are not much different from isolated \pn\ collisions.

\subsection{Systematic evolution of $\bf z_{si}$, $\bf z_{hi}$ and $\bf \tilde z_i$ with $\bf n_{ch}$}

The new results of this study are presented in Secs.~\ref{zxspectra} [methods to obtain $z_{si}(n_s)$ and $z_{hi}(n_s)$],  \ref{zcentt} [``centrality'' trends for $z_{si}(n_s)$ and $z_{hi}(n_s)$] and \ref{newdetails} [shape evolution of PID hard components $H_i(y_t,n_s)$]. The term ``centrality'' should be used with caution in connection with \ppb\ collisions. One point of the previous subsection is that over a substantial range of \nch\ or $\bar \rho_0$ \ppb\ {\em centrality} may not change at all (i.e.\ $N_{part} \approx 2$). In what follows the expression ``\nch\ evolution'' is emphasized instead.

The $z_{si}(n_s)$ evolution in Fig.~\ref{zshx} (left) is directly related to Fig.~\ref{680j} (left). The expression for yield ratio $\bar \rho_{si} / \bar \rho_{0i}$ is just Eq.~(\ref{rhosrho0}) (second line) with factor $\tilde z_i$ added to $x\nu$. The expression for $z_{si}(n_s) / z_{0i}$ in Eq.~(\ref{zsinew}) is then $(\bar \rho_{si} / \bar \rho_{0i}) / (\bar \rho_s / \bar \rho_0)$. In each case the jet contribution at higher \pt\ controls the soft-component \nch\ evolution at lower \pt\ due to overall constraints by densities $\bar \rho_{0i}$ and $\bar \rho_{0}$ as noted. Factor $\tilde z_i$ measures the relative importance of jet production for hadron species $i$.

Figure~\ref{zrat} quantifies that relative importance. Ratio $\tilde z_i(n_s) = z_{hi}(n_s) / z_{si}(n_s)$ varies with hard/soft ratio $x\nu$ and with hadron mass $m_i$. The dependence on the former is modest, but the latter reveals a {\em dramatic increase in jet fragment production with fragment mass}. The description of Fig.~\ref{zrat} in Sec.~\ref{zcent} can be summarized by the relation $\tilde z_i \approx 5.4 m_i(1 + 0.46 x \nu)$ with $m_i$ the mass of species $i$. The relation applies equally well to mesons and baryons and has no significant dependence on strangeness. One should note that the same minimum-bias jet population contributes to fragments of each species $i$. It is the {\em relative} proportion of a given fragment species that varies.

\subsection{Hard-component shape evolution with $\bf n_{ch}$}

Section~\ref{zcentt} presents measured $z_{xi}(n_s)$ \nch\ trends that convey only part of the information associated with PID spectrum hard components -- variation of hard-component amplitudes evaluated at their modes -- which manifests as the trends summarized above. While the strong increase of jet fragment production with hadron mass seems universal the evolution of hard-component shapes appears to be unique for each fragment species.

Section~\ref{newdetails} demonstrates that evolution of hard components with \ppb\ \nch\ is substantial and systematic. For instance, Fig.~\ref{baryonaa4} demonstrates that baryon hard components exhibit no significant evolution above $y_t \approx 3.75$ ($p_t \approx 3$ GeV/c), but near and below the mode baryons show strong transport to {\em higher} \yt\ with increasing \nch. In contrast, kaon hard components in Fig.~\ref{kaonaa4} show significant evolution with \nch\ above the mode (transport to {\em lower} \yt\ with increasing \nch) but no significant variation below the mode (below $y_t \approx$ 2.7 or $p_t \approx 1$ GeV/c).

The pion hard component shows strong variation both above and below the mode, but the combination leads to very little variation {\em at} the mode as in Fig.~\ref{pionaa4}. For that reason pion $z_{hi}(n_s)$ is quite misleading (note Fig.~\ref{pionaa4}, right). Above the mode  meson (pion and kaon) evolution appears to be quantitatively quite similar. But unlike for kaons the pion hard component below the mode is strongly transported to lower \yt\ with increasing \nch. 

As noted in Sec.~\ref{ppbgeom2}, \ppb\ ``centrality'' changes little for the spectra reported in Ref.~\cite{aliceppbpid}. It is apparent in Sec.~\ref{newdetails} that the strongest hard-component changes correspond to the most-peripheral four \nch\ classes that are demonstrated in Sec.~\ref{ppbgeom2} to be nearly equivalent to single peripheral \pn\ collisions. The dominant control parameter is thus the mean \pn\ charge multiplicity. The relevant question then becomes how or why does jet formation vary in \nn\ collisions as a function of \nch?

\subsection{Jet production in $\bf N$-$\bf N$ collisions vs $\bf n_{ch}$}

Two degrees of freedom could vary with \nch\ in \nn\ collisions: (a) In the transverse plane \nn\ ``centrality'' (i.e.\ impact parameter) might vary leading to bias toward more or less ``hard'' (e.g.\ large-angle) parton-parton collisions~\cite{frankfurt}. (b) Along the collision axis demand for greater \nch\ could lead to deeper penetration of the event-wise PDF on momentum fraction $x$ and increased fraction of low-$x$ gluons which then scatter to large angles as jets. 

The relevance of \nn\ {\em centrality} to jet production as proposed in Ref.~\cite{frankfurt} can be questioned~\cite{tomue}. The observed relation $\bar \rho_h \approx \alpha \bar \rho_s^2$~\cite{ppprd} strongly suggests that each participant parton in one nucleon (i.e.\ included within an event-wise PDF represented by $\bar \rho_s$) may interact with {\em any} participant parton in the partner nucleon. If \nn\ centrality were relevant the parton-parton binary-collision trend should go as $\bar \rho_s^{4/3}$ as for \nn\ collisions within \aa\ collisions. The quadratic jet dependence on $\bar \rho_s$ is then directly related to \ppb\ centrality trends reported in Ref.~\cite{tomglauber} which in turn suggest that {\em exclusivity} is a property of \pn\ collisions within \pa\ collisions: A projectile proton may interact with only one target nucleon {\em at a time} (because all its participant partons are committed to that interaction) and \pn\ centrality is therefore irrelevant~\cite{tomexclude}.

Based on those observations hard-component shape evolution for \ppb\ collisions may be better understood. As previously demonstrated, spectrum hard components are {\em quantitatively} predicted by convolution of a measured jet energy spectrum with a measured ensemble of fragmentation functions~\cite{fragevo,hardspec,jetspec2,eeprd,mbdijets}. In connection with Fig.~7 of Ref.~\cite{ppbpid} it was pointed out that the trend of PID hard components from \ppb\ collisions is consistent with the trend of PID fragmentation functions (FFs)~\cite{eeprd}, particularly evolution of hard components below the mode where FF shapes have the greatest impact.

It is worth emphasizing again that data deviations from the TCM in Sec.~\ref{newdetails} are the {\em only} deviations of measured spectra from the fixed TCM and that those deviations offer important new insights into jet formation in A-B collisions not otherwise accessible. Models that claim to describe collision dynamics should be confronted with those details at the level of statistical uncertainties.

%%%%%%%%%%%
\section{Summary}\label{summ}

In a previous study an identified-hadron (PID)  two-component model (TCM) for \pt\ spectra from 5 TeV \ppb\ collisions was defined. The model was successful at separating jet-related (hard) and nonjet (soft) spectrum components for pions, kaons, protons and Lambdas. The TCM was used to predict ensemble-mean \mmpt\ trends for four hadron species and certain spectrum ratios (e.g.\ proton/pion and Lambda/kaon) to be compared with data. However, several issues remained unresolved.

In the present study the TCM has been further refined and remaining issues have been addressed. A correction based on spectrum behavior at low \pt\ is devised for an inferred proton detection inefficiency. A resonance contribution to spectra at low \pt\ is explicitly included in the pion soft-component model. And details of the PID TCM defined in the previous study are reviewed.

In a major upgrade for the present study centrality-dependent PID soft and hard coefficients $z_{si}(n_s)$ and $z_{hi}(n_s)$ are determined directly from spectra as precisely as possible. The coefficients represent PID charge densities in terms of nonPID quantities determine in previous studies. The coefficient centrality trends are described by simple TCM functions and tested for self consistency.

A new method is introduced to extract PID spectrum hard components that minimizes bias. Bias below the hard-component mode due to soft-component subtraction is greatly reduced.  Extracted hard components remain precisely as they were in intact spectra except for nonPID normalization common to all hadron species. Comparisons among species are thereby improved. In particular, coefficients $z_{hi}(n_s)$ are generalized to rapidity-dependent coefficients $z_{hi}(y_t,n_s)$ carrying {\em all data information} beyond fixed TCM reference structures. Detailed comparisons are made between old and new procedures.

A key result of the present study is demonstration that strangeness and baryon production are dominated by jets.  Jet-related production of pions, kaons, protons and Lambdas is all within the same order of magnitude in contrast to nonjet or soft-component production that ranges over two orders of magnitude. The new analysis procedure provides access to detailed hard-component evolution that reveals intriguing differences between jet-related meson and baryon trends. 

The \ppb\ analysis results to date -- reported here as Part I -- illustrate the precision achievable via the TCM applied as a {\em fixed} data reference across an ensemble of collision systems. The analysis also makes clear the large amount of information that is accessible within particle data given suitable application of {\em differential} methods relative to an appropriate fixed reference. In Part II the present results are extended to a PID study of yield and spectrum ratios and ensemble-mean \mmpt\ trends. Whereas the current analysis is based on the TCM as a fixed model, with deviations suggesting novel physics,  the follow-up procedure includes variation of the TCM to describe all data accurately within their statistical uncertainties as determined by standard statistical measures.

%%%%%%%%%%%%%%%%%%%%%%%%%%%%


\begin{thebibliography}{99}

\bibitem{ppbpid}  T.~A.~Trainor,
%``A two-component model for identified-hadron $\bf p_t$ spectra from 5 TeV p-Pb collisions,''
J. Phys. G \textbf{47}, no.4, 045104 (2020).
%doi:10.1088/1361-6471/ab5831
%[arXiv:1812.01151 [hep-ph]].
%6 citations counted in INSPIRE as of 27 Oct 2021
	
\bibitem{aliceppbpid}  B.~B.~Abelev {\it et al.} (ALICE Collaboration),
%``Multiplicity Dependence of Pion, Kaon, Proton and Lambda Production in p-Pb Collisions at $\sqrt{s_{NN}}$ = 5.02 TeV,''
Phys.\ Lett.\ B {\bf 728}, 25 (2014).
%doi:10.1016/j.physletb.2013.11.020
%[arXiv:1307.6796 [nucl-ex]].
%%CITATION = doi:10.1016/j.physletb.2013.11.020;%%
%336 citations counted in INSPIRE as of 29 Apr 2020
% THIS IS DATA FOR MY PPBPID PAPER IN JPHYSG

\bibitem{aliceppbpidnew}  J.~Adam {\it et al.} (ALICE Collaboration),
%``Multiplicity dependence of charged pion, kaon, and (anti)proton production at large transverse momentum in p-Pb collisions at $\mathbf{\sqrt{{\textit s}_{\rm NN}}}$ = 5.02 TeV,''
Phys.\ Lett.\ B {\bf 760}, 720 (2016).
%doi:10.1016/j.physletb.2016.07.050
%[arXiv:1601.03658 [nucl-ex]].
%%CITATION = doi:10.1016/j.physletb.2016.07.050;%%
%84 citations counted in INSPIRE as of 12 Apr 2020

\bibitem{ppbridge}  B.~Abelev {\it et al.} (ALICE Collaboration),
%``Long-range angular correlations on the near and away side in $p$-Pb collisions at $\sqrt{s_{NN}}=5.02$ TeV,''
Phys.\ Lett.\ B {\bf 719}, 29 (2013).
%doi:10.1016/j.physletb.2013.01.012
%[arXiv:1212.2001 [nucl-ex]].
%%CITATION = doi:10.1016/j.physletb.2013.01.012;%%
%577 citations counted in INSPIRE as of 20 Nov 2018

\bibitem{ppcms}  V.~Khachatryan {\it et al.} (CMS Collaboration),
%``Observation of Long-Range Near-Side Angular Correlations in Proton-Proton Collisions at the LHC,''
JHEP {\bf 1009}, 091 (2010).
%doi:10.1007/JHEP09(2010)091
%[arXiv:1009.4122 [hep-ex]].
%%CITATION = doi:10.1007/JHEP09(2010)091;%%
%712 citations counted in INSPIRE as of 20 Nov 2018

\bibitem{dusling} K.~Dusling, W.~Li and B.~Schenke,
%``Novel collective phenomena in high-energy proton–proton and proton–nucleus collisions,''
Int.\ J.\ Mod.\ Phys.\ E {\bf 25}, no. 01, 1630002 (2016).
%  doi:10.1142/S0218301316300022
%  [arXiv:1509.07939 [nucl-ex]].
%%CITATION = doi:10.1142/S0218301316300022;%%
%50 citations counted in INSPIRE as of 18 Jun 2017
% FLOW IS BIGGER IN SMALLER SYSTEMS: P. 15

\bibitem{tomnewppspec} T.~A.~Trainor,
%``A two-component model of hadron production applied to $\bf p_t$ spectra from 5 TeV and 13 TeV $\bf p$-$\bf p$ collisions at the large hadron collider,''
arXiv:2104.08423 [hep-ph].
%0 citations counted in INSPIRE as of 01 Aug 2021

\bibitem{tommodeltests} T.~A.~Trainor,
%``Statistical evaluation of fitted models applied to $\bf p_t$ spectrum data from 5 TeV and 13 TeV $\bf p$-$\bf p$ collisions at the large hadron collider,''
arXiv:2107.10899 [hep-ph].
%0 citations counted in INSPIRE as of 01 Aug 2021

\bibitem{aliceglauber} J.~Adam {\it et al.} (ALICE Collaboration),
%``Centrality dependence of particle production in p-Pb collisions at $\sqrt{s_{\rm NN} }$= 5.02 TeV,''
Phys.\ Rev.\ C {\bf 91}, no. 6, 064905 (2015).
%  doi:10.1103/PhysRevC.91.064905
%  [arXiv:1412.6828 [nucl-ex]].
%%CITATION = doi:10.1103/PhysRevC.91.064905;%%
%100 citations counted in INSPIRE as of 18 Jun 2017

\bibitem{tommpt} T.~A.~Trainor,
%``Ensemble-mean $\bf p_t$ and hadron production in high-energy nuclear collisions,''
arXiv:1708.09412.
%%CITATION = ARXIV:1708.09412;%%
%2 citations counted in INSPIRE as of 20 Oct 2017
	
\bibitem{tomglauber}  T.~A.~Trainor,
%''Glauber-model analysis of 5 TeV $\bf p$-Pb centrality compared to a two-component (soft + hard) model of hadron production in high-energy nuclear collisions''
arXiv:1801.05862.

\bibitem{tomexclude}  T.~A.~Trainor,
%``Exclusivity of p-N interactions within p-A collisions,''
arXiv:1801.06579.
%%CITATION = ARXIV:1801.06579;%%
%1 citations counted in INSPIRE as of 04 Nov 2018

\bibitem{alicetomspec}    T.~A.~Trainor,
%``Charge-multiplicity and collision-energy dependence of $p_t$ spectra from $p$-$p$ collisions at the relativistic heavy-ion collider and large hadron collider,''
J.\ Phys.\ G {\bf 44}, no. 7, 075008 (2017).
%  doi:10.1088/1361-6471/aa759e
%  [arXiv:1603.01337 [hep-ph]].
%%CITATION = doi:10.1088/1361-6471/aa759e;%%
%5 citations counted in INSPIRE as of 17 Dec 2017

\bibitem{ppquad}  T.~A.~Trainor and D.~J.~Prindle,
%``Charge-multiplicity dependence of single-particle transverse-rapidity $y_t$ and pseudorapidity η densities and 2D angular correlations from 200 GeV p-p collisions,''
Phys.\ Rev.\ D {\bf 93}, 014031 (2016).
%  [arXiv:1512.01599 [hep-ph]].

\bibitem{ppprd} J.~Adams {\it et al.}  (STAR Collaboration),
% ``The multiplicity dependence of inclusive p(t) spectra from p p  collisions
% at s**(1/2) = 200-GeV,''
Phys.\ Rev.\  D {\bf 74}, 032006 (2006).
%  [nucl-ex/0606028].
	
\bibitem{resonances} P.~M.~Lo,
%``Resonance decay dynamics and their effects on $p_T$-spectra of pions in heavy-ion collisions,''
Phys. Rev. C \textbf{97}, no.3, 035210 (2018).
%doi:10.1103/PhysRevC.97.035210
%[arXiv:1705.01514 [hep-ph]].
%12 citations counted in INSPIRE as of 29 Sep 2021
	
\bibitem{pbpbpid} T.~A.~Trainor,
%"Differential comparison of identified-hadron $\bf p_t$ spectra from high-energy A-B nuclear collisions based on a two-component model of hadron production",
arXiv:2001.03200.

\bibitem{alicepppid} S.~Acharya \textit{et al.} (ALICE Collaboration),
%``Multiplicity dependence of $\pi $, K, and p production in pp collisions at $\sqrt{s} = 13$ TeV,''
Eur. Phys. J. C \textbf{80}, no.8, 693 (2020).
%doi:10.1140/epjc/s10052-020-8125-1
%[arXiv:2003.02394 [nucl-ex]].
%10 citations counted in INSPIRE as of 19 May 2021

\bibitem{alicepbpbpidspec}  J.~Adam {\it et al.} (ALICE Collaboration),
%``Centrality dependence of the nuclear modification factor of charged pions, kaons, and protons in Pb-Pb collisions at $\sqrt{s_{\rm NN}}=2.76$ TeV,''
Phys.\ Rev.\ C {\bf 93}, no. 3, 034913 (2016).
%doi:10.1103/PhysRevC.93.034913
%[arXiv:1506.07287 [nucl-ex]].
%%CITATION = doi:10.1103/PhysRevC.93.034913;%%
%75 citations counted in INSPIRE as of 07 Jun 2019
%BIG PID SPECTRA
% THIS IS INSTRUCTION ON HOW TO SCREW UP PROTONS

\bibitem{alicepppidx} S.~Acharya \textit{et al.} (ALICE Collaboration),
%``Production of charged pions, kaons, and (anti-)protons in Pb-Pb and inelastic $pp$ collisions at $\sqrt {s_{NN}}$ = 5.02 TeV,''
Phys. Rev. C \textbf{101}, no.4, 044907 (2020).
%doi:10.1103/PhysRevC.101.044907
%[arXiv:1910.07678 [nucl-ex]].
%79 citations counted in INSPIRE as of 12 Dec 2021

\bibitem{stoeckerstatmodel} V.~Vovchenko, B.~D\"onigus and H.~Stoecker,
%``Canonical statistical model analysis of p-p , p -Pb, and Pb-Pb collisions at energies available at the CERN Large Hadron Collider,''
Phys. Rev. C \textbf{100}, no.5, 054906 (2019).
%doi:10.1103/PhysRevC.100.054906
%[arXiv:1906.03145 [hep-ph]].
%21 citations counted in INSPIRE as of 05 Aug 2021
% SHOWS PROTON INEFFICIENCY -- STAT MODEL NOT APPLIES TO JETS

\bibitem{thermalprotons} A.~Andronic, P.~Braun-Munzinger, B.~Friman, P.~M.~Lo, K.~Redlich and J.~Stachel,
%``The thermal proton yield anomaly in Pb-Pb collisions at the LHC and its resolution,''
Phys. Lett. B \textbf{792}, 304-309 (2019).
%doi:10.1016/j.physletb.2019.03.052
%[arXiv:1808.03102 [hep-ph]].
%45 citations counted in INSPIRE as of 29 Sep 2021

\bibitem{alicempt}  B.~B.~Abelev {\it et al.}  (ALICE Collaboration),
%``Multiplicity dependence of the average transverse momentum in pp, p-Pb, and Pb-Pb collisions at the LHC,''
Phys.\ Lett.\ B {\bf 727}, 371 (2013).
%  [arXiv:1307.1094 [nucl-ex]].
	
\bibitem{aliceppbglaub} J.~Adam {\it et al.} (ALICE Collaboration),
%``Centrality dependence of particle production in p-Pb collisions at $\sqrt{s_{\rm NN} }$= 5.02 TeV,''
Phys.\ Rev.\ C {\bf 91}, no. 6, 064905 (2015).
%  doi:10.1103/PhysRevC.91.064905
%  [arXiv:1412.6828 [nucl-ex]].
%%CITATION = doi:10.1103/PhysRevC.91.064905;%%
%100 citations counted in INSPIRE as of 18 Jun 2017

\bibitem{frankfurt} L.~Frankfurt, M.~Strikman and C.~Weiss,
%``Transverse nucleon structure and diagnostics of hard parton-parton processes at LHC,''
Phys. Rev. D \textbf{83}, 054012 (2011).
%doi:10.1103/PhysRevD.83.054012
%[arXiv:1009.2559 [hep-ph]].
%97 citations counted in INSPIRE as of 08 Oct 2021

\bibitem{tomue} T.~A.~Trainor,
%``Dijet production, collision centrality and backgrounds in high-energy p-p collisions,''
Phys. Rev. D \textbf{87}, no.5, 054005 (2013).
%doi:10.1103/PhysRevD.87.054005
%[arXiv:1210.5217 [hep-ph]].
%18 citations counted in INSPIRE as of 08 Oct 2021

\bibitem{fragevo}    T.~A.~Trainor,
%``Evolution of minimum-bias parton fragmentation in nuclear collisions,''
Phys.\ Rev.\  C {\bf 80}, 044901 (2009).
%[arXiv:0901.3387 [hep-ph]].

\bibitem{hardspec}  T.~A.~Trainor,
%``Centrality evolution of $p_t$ and $y_t$ spectra from Au-Au collisions at
%$\sqrt{s_{NN}} = 200$ GeV,''
Int.\ J.\ Mod.\ Phys.\  E {\bf 17}, 1499 (2008).
% arXiv:0710.4504.

\bibitem{eeprd}   T.~A.~Trainor and D.~T.~Kettler,
%``Extrapolating parton fragmentation to low $Q^{2}$ in $e^{+}$ - $e^{-}$ collisions,''
Phys.\ Rev.\ D {\bf 74}, 034012 (2006).
%  [hep-ph/0606249].

\bibitem{jetspec2}  T.~A.~Trainor,
%``Universal parametrization of jet production based on parton and fragment rapidities,''
Phys.\ Rev.\ D  {\bf 89}, 094011 (2014).
%  arXiv:1403.3685.

\bibitem{mbdijets} T.~A.~Trainor,
%``Manifestations of minimum-bias dijets in high-energy nuclear collisions,''
arXiv:1701.07866 [hep-ph].
%6 citations counted in INSPIRE as of 17 Dec 2021


\end{thebibliography}
\end{document}